\documentclass[12pt]{article}
\usepackage{amsmath,epsfig,graphicx}
\usepackage{textcomp}
\usepackage[T1]{fontenc} 

\input{epsf} 
\def\ltap{\raisebox{-.6ex}{\rlap{$\,\sim\,$}} \raisebox{.4ex}{$\,<\,$}} 
\def\gtap{\raisebox{-.6ex}{\rlap{$\,\sim\,$}} \raisebox{.4ex}{$\,>\,$}}

\newcommand\as{\alpha_{\mathrm{S}}} 
\newcommand\f[2]{\frac{#1}{#2}} 
\def\beq{\begin{equation}} 
\def\eeq{\end{equation}} 
\def\beeq{\begin{eqnarray}} 
\def\eeeq{\end{eqnarray}} 
 
\def\to{\rightarrow}

\def\msbar{{\overline {\rm MS}}}

\begin{document} 
\begin{titlepage}
\renewcommand{\thefootnote}{\fnsymbol{footnote}}
\vspace*{2cm}

\begin{center}
{\Large \bf W boson production at hadron colliders:}
\vskip 0.15cm
{\Large \bf the lepton charge asymmetry in NNLO QCD}
\end{center}

\par \vspace{2mm}
\begin{center}
{\bf Stefano Catani$^{(a)}$, Giancarlo Ferrera$^{(a)}$}
and
{\bf Massimiliano Grazzini$^{(a,b)}$}

\vspace{5mm}

$^{(a)}$INFN, Sezione di Firenze and
Dipartimento di Fisica e Astronomia,\\ 
Universit\`a di Firenze,
I-50019 Sesto Fiorentino, Florence, Italy\\

$^{(b)}$Institute for Theoretical Physics,
ETH-Zurich, 8093 Zurich, Switzerland

\vspace{5mm}

\end{center}

\par \vspace{2mm}
\begin{center} {\large \bf Abstract} \end{center}
\begin{quote}
\pretolerance 10000

We consider the production of $W^\pm$ bosons in hadron collisions, and the
subsequent leptonic decay $W\to l\nu_l$.
We study the asymmetry between the rapidity distributions of the 
charged leptons,
and we present its computation
up to the next-to-next-to-leading order (NNLO) in QCD perturbation theory.
Our calculation 
includes the dependence on the 
lepton kinematical cuts that are necessarily 
applied to select $W\to l\nu_l$ events in actual experimental analyses
at hadron colliders.
We illustrate the main differences between the $W$ and 
lepton charge asymmetry, and we discuss their physical origin and the
effect
of the QCD radiative corrections.
We show detailed numerical results on the charge asymmetry in  
$p{\bar p}$ collisions at the Tevatron, and we
discuss
the comparison with some of the available data.
Some illustrative results on the lepton charge asymmetry
in $pp$ collisions
at LHC energies are presented.

\end{quote}

\vspace*{\fill}
\begin{flushleft}
February 2010

\end{flushleft}
\end{titlepage}

\setcounter{footnote}{1}
\renewcommand{\thefootnote}{\fnsymbol{footnote}}

\section{Introduction}

The production of lepton pairs with high invariant mass, of $Z$ bosons and
of $W$ bosons, through the Drell--Yan (DY) mechanism
\cite{Drell:1970wh}, is the most `classical' hard-scattering process in 
hadron--hadron
collisions. 
The DY process provides important tests of the Standard 
Model (SM) and precise determinations of SM parameters, and it places stringent
constraints on many form of new physics.

At high-energy hadron colliders, the $W$ and $Z$ production processes 
have large production rates. 
They also
offer clean experimental signatures, 
because of the presence
of one high-$p_T$ lepton and large missing transverse energy 
(in the case of $W$ production), or of two high-$p_T$ leptons
of opposite charge (in the case of $Z$ production) in the final state.
Owing to these features, $W$ and $Z$ production are 
key processes for physics studies at the Tevatron and the LHC.

Single $W$ and $Z$ boson production is used at the 
Tevatron to precisely measure the mass and width of the $W$ boson
and to extract the electroweak (EW) mixing angle from measurements
of the forward--backward lepton asymmetry.

The production of DY lepton pairs gives important information on the parton
densities, or Parton Distribution Functions (PDFs), of the colliding hadrons.
DY production in low-energy proton--proton ($pp$) and proton--nucleon
collisions is more sensitive
to the sea quark densities of the proton than the structure functions measured 
in Deep Inelastic lepton--hadron Scattering (DIS).
$W$ and $Z$ production in proton--antiproton ($p{\bar p}$)
collisions is mainly sensitive
to the valence quarks of the proton, in combinations that are different
from those appearing in DIS structure functions.
The shape of the $W$ and $Z$ rapidity distributions in $pp$ collisions
at the LHC gives direct information on the quark and antiquark densities 
of the proton at high scale and small values of parton momentum fractions.

$W$ and $Z$ production at the Tevatron and the LHC represents important 
background
to other SM processes and signals of new physics. The production of DY lepton
pairs with invariant mass larger than the masses of the $W$ and $Z$ bosons
gives direct information and constraints on effective interactions between 
quarks and leptons that can originate from physics beyond the SM.

The importance of the DY process and the high precision achieved and achievable
by the experiments at the Tevatron and the LHC demand for theoretical 
predictions with corresponding accuracy. These theoretical predictions
require, in particular, the computation of QCD radiative corrections up to the
next-to-next-to-leading order (NNLO) in perturbation theory.

The DY process is one of the few processes for which 
NNLO QCD corrections are
known \cite{Hamberg:1990np}-\cite{Catani:2009sm}.
The important NNLO calculations of the total cross section and of the 
rapidity distribution
of the DY lepton pair were performed in Refs.~\cite{Hamberg:1990np} and 
\cite{Anastasiou:2003yy, Anastasiou:2003ds}, respectively. A limitation of these
NNLO calculations is that only the invariant mass and the rapidity of the
lepton pair are explicitly retained; the calculations are fully inclusive over
the hadronic (partonic) final state, and they are also inclusive over
the separate momenta of the two final-state leptons. This limitation is
particularly evident in the case of $W$ production, since the longitudinal
momentum of the 
neutrino from $W$ decay and, thus, the 
momentum of the lepton
pair are not measurable.
More generally, the 
identification and
selection of DY lepton pairs in actual experiments require the use of 
various kinematical cuts. Moreover, the explicit dependence on the
leptonic kinematical variables that are measurable by experiments gives
additional information on the dynamics of the DY process.  
The limitation of the inclusive calculations of 
Refs.~\cite{Hamberg:1990np}-\cite{Anastasiou:2003ds} is obviated by
considering fully-differential calculations of the DY process at the NNLO 
\cite{Melnikov:2006di,Catani:2009sm}.

The evaluation of higher-order QCD radiative corrections 
to hard-scattering processes 
is 
definitely a hard task.
The presence of infrared singularities at intermediate stages 
of the calculation does not permit
a straightforward implementation of numerical techniques.
In particular, {\em fully differential} calculations at the 
NNLO involve a substantial amount of conceptual, analytical and technical
complications \cite{Kosower}-\cite{ST}.
In $e^+e^-$ collisions, NNLO differential cross sections are 
known only for 
two~\cite{Anastasiou:2004qd, Weinzierl:2006ij} and three jet 
production~\cite{threejets, Weinzierl:2008iv}.
In hadron--hadron collisions, fully differential cross sections have 
been computed 
in the cases of Higgs production by gluon 
fusion \cite{Anastasiou:2004xq,Catani:2007vq}
and of the DY process \cite{Melnikov:2006di, Catani:2009sm}.

A significant observable in $W$ hadroproduction is the asymmetry
in the rapidity distribution of $W^+$ and $W^-$ bosons (see
Sect.~\ref{sec:pre}). In $p{\bar p}$ collisions, 
the $W^+$ and $W^-$ bosons are produced with equal rates; however, 
the $W^+$ is mainly produced 
in the proton direction, whereas the opposite happens for the $W^-$.
In $pp$ collisions, $W$ production is forward--backward symmetric;
however, the $W^-$ production rate is smaller than the $W^+$ production rate
and, moreover, the $W^-$ is mostly produced at central rapidities, while the
$W^+$ is mostly produced at larger rapidities.

These $W^+/W^-$ asymmetries are mainly due to the proton content of $u$ and 
$d$ quarks and, in particular, to the fact that $u$ quarks carry, on average,
more proton momentum fraction than $d$ quarks. Therefore, as already 
pointed out long ago \cite{earlyth}, the $W$ boson charge asymmetries
provide important quantitative information on the size and momentum fraction
distribution of the $u$ and $d$ parton densities of the proton. 

In hadron collisions, 
the produced $W$ bosons are identified by their
leptonic decay $W\to l\nu_l$.
Since the longitudinal component of the neutrino momentum is
unmeasured in experiments at hadron colliders, what is actually measured is
the rapidity of the charged lepton and the corresponding 
lepton charge asymmetry (rather than the $W$ charge 
asymmetry\footnote{A direct measurement of the $W$ charge asymmetry has 
recently been presented by the CDF collaboration \cite{cdfw}
(see Sect.~\ref{sec:tev}).}).

The first measurement \cite{cdfpre} 
of the lepton charge asymmetry in hadron collisions
was carried out by the CDF Collaboration at the Tevatron Run~I, 
using data from 
$p{\bar p}$ collisions at the
centre--of--mass energy  ${\sqrt s}=1.8$~TeV.
The final CDF measurement \cite{cdfrunI} at the Tevatron Run~I
is available since more than ten years. The recent data 
\cite{cdfe}-\cite{d0e}
from the CDF and D0 Collaborations at the Tevatron Run~II
(${\sqrt s}=1.96$~TeV) are more precise
than Run~I data, extend at larger rapidities
and give additional information on the dependence on the lepton
transverse energy. 

The CDF Run~I measurement \cite{cdfrunI} of the lepton charge 
asymmetry has played (and still plays) a relevant role in testing and
constraining the quark parton densities of the proton. In particular,
these data are continuously used in the 
PDF
global fits 
of the MRST and CTEQ Collaborations since their 
MRST1998 \cite{Martin:1998sq} and
CTEQ5 \cite{Lai:1999wy} analyses, respectively.
The most recent 
PDF analysis of the MRST/MSTW Collaboration
\cite{Martin:2009iq, Martin:2009bu} already includes Run~II measurements
\cite{cdfe, d0m}
in the global fit.

Future LHC measurements \cite{CMSnote}
will study the lepton charge asymmetry at energies larger than Tevatron energies.
The lepton charge asymmetry at the LHC is sensitive to PDFs with parton momentum
fractions that are smaller (up to about a factor of seven)
than those probed at the Tevatron.

The lepton charge asymmetry is a typical observable that depends on various
kinematical selection cuts,
such as those that are necessary and applied in actual experimental 
configurations
to identify $W\to l\nu_l$
events. This dependence has to properly
be taken into account in the corresponding theoretical results.
QCD studies of the lepton charge asymmetry has so far been performed
by considering radiative corrections up to next-to-leading order (NLO).
For instance, the QCD calculations of Refs.~\cite{cdfpre}-\cite{Lai:1999wy}
use the NLO code
DYRAD \cite{Giele:1993dj} and the RESBOS 
code \cite{Balazs:1997xd} (which includes higher-order soft-gluon contributions 
to the transverse-momentum spectrum of the~$W$).

In this paper we present the 
calculation of the lepton charge asymmetry in NNLO QCD.
We use the numerical program of Ref.~\cite{Catani:2009sm}, which
encodes the NNLO radiative corrections to the DY process at the
fully-differential level. This allows us 
to compute the lepton charge asymmetry by including the kinematical
cuts applied in experimental analyses.

The paper is organized as follows. In Sect.~\ref{sec:pre} we introduce the
charge asymmetry and the setup of our calculation.
In Sect.~\ref{sec:tev} we consider $p{\bar p}$ collisions and
present the results of our calculation at the Tevatron Run II.
We also discuss the comparison with some of the available Tevatron data.
In Sect.~\ref{sec:lhc} we consider the charge asymmetry in $pp$ collisions
and present some illustrative results at LHC energies.
We briefly summarize our results in Sect.~\ref{sec:sum}.

\section{Preliminaries}
\label{sec:pre}

We consider the process
\beq
\label{pro}
h_1(p_1) + h_2(p_2) \longrightarrow \; W+X 
\longrightarrow \; l\nu_l +X \;\;.
\eeq
The $W$ boson is produced by the collision 
of the two incoming hadrons 
$h_1$ and $h_2$ (with momenta $p_1$ and $p_2$), 
and then it decays leptonically. The accompanying final state
is denoted by $X$. 

In the centre--of--mass frame of the colliding hadrons,
$y_W$ and $y_l$ denote the rapidities of the $W$ boson and of the charged 
lepton $l$, respectively. We consider the $W$ decay in electrons or muons 
($l=e,\mu$) and we treat the charged leptons in the massless approximation;
thus, $y_l$ coincides with the lepton pseudorapidity $\eta_l$,
where $\eta_l= - \ln (\tan \f{\theta_l}{2})$ and $\theta_l$ is the lepton
scattering angle.
The $W$ production cross section at fixed $y_W$ is denoted by
$d\sigma_{h_1h_2}(W^{\pm})/dy_W$. The analogous rapidity cross section
of the decaying charged lepton is denoted by 
$d\sigma_{h_1h_2}(l^{\pm})/dy_l$. The $W$ and lepton charge asymmetries are
denoted by $A_{h_1h_2}(y_W)$ and $A_{h_1h_2}(y_l)$, respectively. They are
defined as
\begin{equation}
\label{was}
A_{h_1h_2}(y_W)=\frac{d\sigma_{h_1h_2}(W^+)/dy_W \;-\; 
d\sigma_{h_1h_2}(W^-)/dy_W}
{d\sigma_{h_1h_2}(W^+)/dy_W \;+\; d\sigma_{h_1h_2}(W^-)/dy_W} \;\;,
\end{equation}
\begin{equation}
\label{las}
A_{h_1h_2}(y_l)=\frac{d\sigma_{h_1h_2}(l^+)/dy_l \;-\; 
d\sigma_{h_1h_2}(l^-)/dy_l}
{d\sigma_{h_1h_2}(l^+)/dy_l \;+\; d\sigma_{h_1h_2}(l^-)/dy_l} \;\;.
\end{equation}

We also introduce the following notation for the kinematical variables
of the decaying leptons: $E_T$ is the transverse energy of the charged lepton, 
$E_T^\nu$ is the neutrino transverse energy (which corresponds to the missing
transverse energy, \,$\slash \!\!\!\!E_T$, measured by experiments),
$\phi_{l\nu}$ is the azimuthal angle between the lepton and neutrino transverse
momenta,
and $M_T= \sqrt{2 E_T E_T^\nu (1- \cos \phi_{l\nu})}$ is the leptonic
transverse mass.

In actual experimental determinations, the measured rapidity cross sections and 
charge asymmetries depend on the kinematical cuts (e.g., lepton isolation
requirements, minimum $M_T$) and the kinematical variables (e.g., $E_T$)
that are used to identify and select the observed  
$W \to l {\nu}_l$ events.
This dependence, which is not explicitly denoted in Eqs.~(\ref{was}) and 
(\ref{las}), is fully taken into account in our calculation,
as briefly mentioned below and shown in Sects.~\ref{sec:tev} 
and \ref{sec:lhc}.

In this work we study the rapidity cross sections and the charge asymmetries in
QCD perturbation theory. A generic differential cross section $d\sigma$ for the
process in Eq.~(\ref{pro}) is given as
\beq
\label{dsigma}
d\sigma(p_1,p_2) = \sum_{a,b} \int_0^1 dx_1 \int_0^1 dx_2 
\;f_{a/h_1}(x_1, \mu_F^2) \;f_{b/h_2}(x_2, \mu_F^2) 
\;d{\hat \sigma}_{ab}(x_1p_1,x_2p_2;\mu_F^2) \;\;,
\eeq
where $f_{a/h}(x, \mu_F^2)$ ($a=q,{\bar q},g$) are the parton distributions
of the colliding hadron $h$, $\mu_F$ is the corresponding factorization 
scale, and we use the $\msbar$ factorization scheme. 
The partonic cross section $d{\hat \sigma}$ is computed up to include its 
NNLO contribution in perturbative QCD. We have:
\beq
\label{psigma}
d{\hat \sigma}(p_1,p_2;\mu_F^2) = d{\hat \sigma}^{(0)}(p_1,p_2)
+ \as(\mu_R^2) \;d{\hat \sigma}^{(1)}(p_1,p_2;\mu_F^2) + \as^2(\mu_R^2) \;
d{\hat \sigma}^{(2)}(p_1,p_2;\mu_F^2,\mu_R^2)
+ {\cal O}(\as^3) \;\;,
\eeq
where $\as(\mu_R^2)$ is the QCD running coupling, $\mu_R$ is the
renormalization scale, and we use the $\msbar$ renormalization scheme. 
The leading order (LO) partonic cross section is 
$d{\hat \sigma}^{(0)}$, while $d{\hat \sigma}^{(1)}$ and 
$d{\hat \sigma}^{(2)}$ are the NLO and NNLO corrections,
respectively.

Our computation of the partonic cross section is carried out by using the NNLO
numerical program of Ref.~\cite{Catani:2009sm}. This NNLO QCD calculation is 
organised at the fully differential level, and it is encoded in a 
partonic Monte Carlo program.
It allows the user to compute
differential cross sections and observables with arbitrary kinematical 
requirements and acceptance cuts on the produced $W$, leptons and 
accompanying final state $X$. The only essential restriction is that the
constraints applied to the final state have to be IR safe\footnote{The observable
must be independent of the presence of arbitrarily-soft partons and independent 
of the individual-parton momenta of a bunch of collinear partons.} at 
the partonic level. 

The hadronic cross sections at the LO, NLO and NNLO are evaluated according to
Eq.~(\ref{dsigma}). The N$^n$LO (with $n=0,1,2$) partonic cross sections in
Eq.~(\ref{psigma})
use the expression of $\as(\mu_R^2)$ at the $n$-th order 
(i.e., we use the $\mu_R$ dependence at the level of $(n+1)$ loops), and they
are consistently convoluted with parton densities at each corresponding order.
The reference value of $\as(M_Z)$ is fixed at the actual
value used in the corresponding set of parton densities. We consider $N_f=5$
flavours of (effectively) massless quarks, and thus we have two `up-type'
quarks $(u, c)$ and three `down-type' quarks $(d, s, b)$.

Recent sets of parton densities, which are obtained by analyses of various
collaborations, are presented in 
Refs.~\cite{Martin:2009iq, Martin:2009bu, cteqpdf, nnpdf, Alekhin:2005gq, 
Alekhin:2009ni, GJRpdf, JimenezDelgado:2009tv, herapdf}.
Among these sets, only those of Refs.~\cite{Martin:2009iq}, 
\cite{Alekhin:2009ni}
and \cite{JimenezDelgado:2009tv} include NNLO parton densities with 
$N_f=5$ (effectively) massless quarks.
Since the main purpose of our work is the study of rapidity cross sections and
asymmetries at the NNLO, we consider only the parton density sets of  
Refs.~\cite{Martin:2009iq}, 
\cite{Alekhin:2009ni}
and \cite{JimenezDelgado:2009tv}. Moreover, to avoid multiple presentations of
similar results, we mostly use the parton densities
of Ref.~\cite{Martin:2009iq}. The global fit of Ref.~\cite{Martin:2009iq}
includes also some data on the lepton charge asymmetry at the Tevatron: 
the ensuing parton densities are thus expected to produce better agreement
with available measurements of charge asymmetries.

In our calculation, the $W$ boson is treated off shell, thus including 
finite-width effects, and its leptonic decay retains the corresponding spin
correlations. The values of the mass and total width of the $W$ boson
are $M_W = 80.398$~GeV and $\Gamma_W=2.141$~GeV. The Fermi constant is set 
to the value $G_F = 1.16637\times 10^{-5}$~GeV$^{-2}$, and we use the following
(unitarity constrained) values of the CKM matrix elements:
$V_{ud}=0.97419$, $V_{us}=0.2257$, $V_{ub}=0.00359$,
$V_{cd}=0.2256$, $V_{cs}=0.97334$, $V_{cb}=0.0415$. All these values of 
EW parameters are taken from the PDG 2008 
\cite{Amsler:2008zzb}. The $W$ boson EW couplings to quarks and leptons
are treated at the tree level, so that the above parameters are sufficient to
fully specify the EW content of our calculation. 
In particular, the tree-level leptonic width
of the $W$ boson implies the value $BR(W\to l{\nu})=10.62$~\% of the
leptonic branching ratio\footnote{For comparison, the electron and muon
branching ratios of the PDG \cite{Amsler:2008zzb} are 
$BR(W\to e {\nu}_e)=(10.75 \pm 0.13)$~\% and
$BR(W\to \mu {\nu}_\mu)=(10.57 \pm 0.15)$~\%, respectively.}.

We note that our calculation is invariant under CP transformations.
Therefore, in $p{\bar p}$ collisions, the rapidity cross sections 
$d\sigma_{p{\bar p}}/dy$ of $W^+ (l^+)$ and $W^- (l^-)$ are simply related
by the replacement $y_W (y_l) \leftrightarrow -y_W (-y_l)$, and the charge
asymmetry fulfils $A_{p{\bar p}}(-y)= -A_{p{\bar p}}(y)$. Analogously,
in $pp$ collisions, the rapidity cross sections $d\sigma_{p{p}}/dy$ 
of $W^\pm (l^\pm)$ and the charge
asymmetry $A_{pp}(y)$ are invariant with respect to 
the replacement $y \leftrightarrow -y$.

\section{Rapidity 
cross section 
and asymmetry at the Tevatron}
\label{sec:tev}

In this section we consider $p{\bar p}$ collisions. We recall the 
main features of the production mechanism of the $W^{\pm}$ bosons
and their decaying leptons. Then, we present the results of our
QCD calculation at the centre--of--mass energy ${\sqrt s}=1.96$~TeV.
Unless otherwise stated, throughout the paper
we use the MSTW2008 sets \cite{Martin:2009iq} 
of parton densities, and we fix the
renormalization and factorization scales at the value $\mu_R=\mu_F=M_W$.

\setcounter{footnote}{1}

\subsection{$W$ rapidity distribution and asymmetry}
\label{sec:wtev}

The LO cross section is controlled by the partonic subprocesses
\beeq
\label{pro+}
U + {\overline D} &\longrightarrow& \; W^{+}  
\longrightarrow \; l^{+} \;{\nu}_l \;\;, \\
\label{pro-}
D + {\overline U} &\longrightarrow& \; W^{-}  
\longrightarrow \; l^{-} \;{\bar \nu}_l \;\;,
\eeeq
where $U (D)$ generically denote an `up-type' (`down-type') quark. 
Most of the proton momentum is carried by $u$ quarks (and gluons), while
most of the antiproton momentum is carried by $\bar u$ antiquark (and gluons).
Therefore, owing to the flavour structure of the processes in 
Eqs.~(\ref{pro+}) and (\ref{pro-}), the produced $W^+$ tends to follow the
direction of the colliding proton, while the $W^-$ tends to follows the 
direction of the colliding antiproton.

\begin{figure}[htb]
\begin{center}
\begin{tabular}{c}
\epsfxsize=11truecm
\epsffile{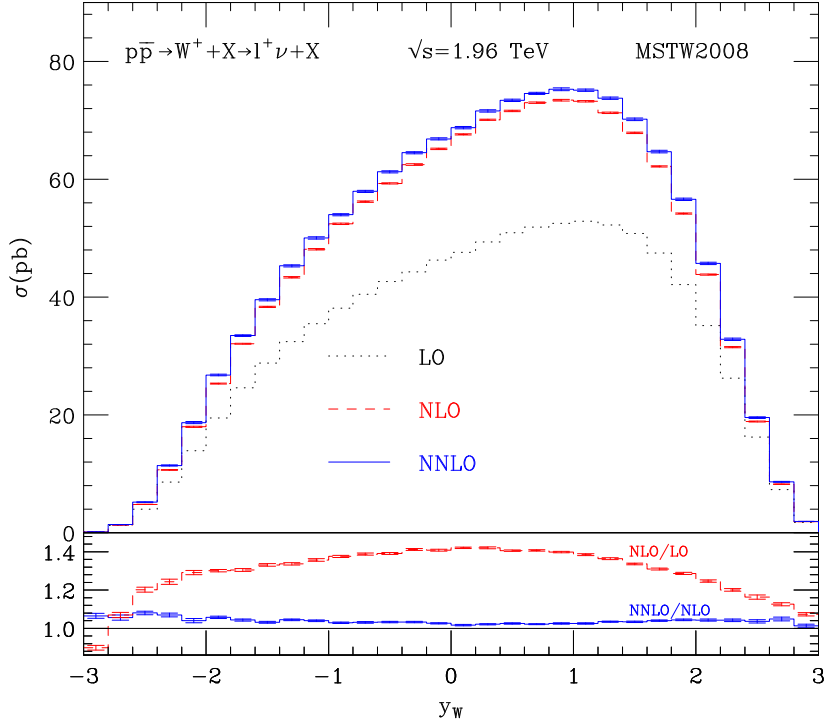}\\
\end{tabular}
\end{center}
\caption{\label{fig:etaW}
{\em Rapidity distribution of an on-shell $W^+$ boson at the Tevatron Run~II
in LO (black dotted), NLO (red dashed) and NNLO (blue solid) QCD. 
No cuts are applied on the leptons and on their accompanying final state.
The height of each histogram bin gives the value (in $pb$) of the cross section
in the corresponding rapidity bin.
The lower panel shows the ratios NLO/LO (red dashed) and 
NNLO/NLO (blue solid) of the cross section results in the upper panel.
}}
\end{figure}

\begin{figure}[htb]
\begin{center}
\begin{tabular}{c}
\epsfxsize=15truecm
\epsffile{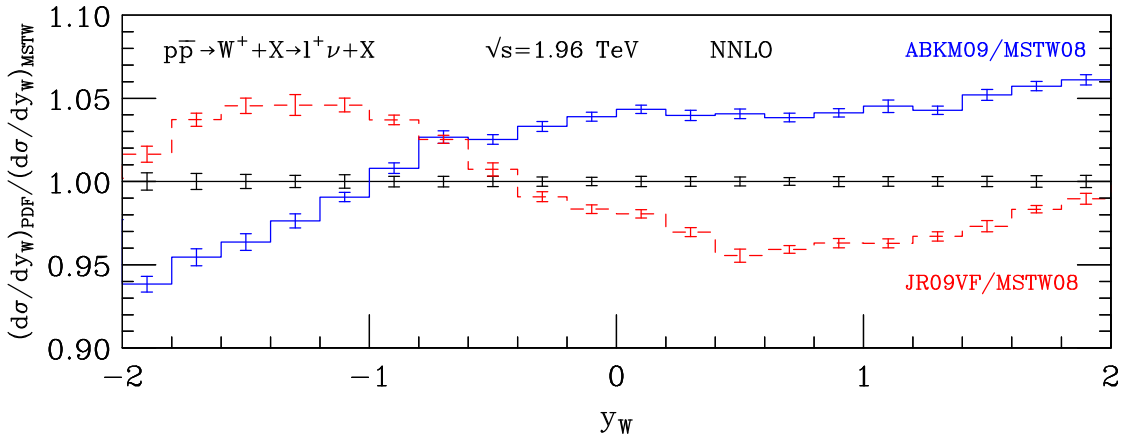}\\
\end{tabular}
\end{center}
\caption{
\label{fig:KW}
{\em On-shell $W^+$ boson production at the Tevatron Run~II in NNLO QCD.
The rapidity cross sections computed with the ABKM09 (blue solid)
and JR09VF (red dashed)
parton densities are rescaled by the corresponding MSTW2008 result.
}}
\end{figure}

We define the sign of the rapidity $y$ so that the forward region $(y>0)$ 
corresponds to the direction of the momentum of the incoming proton.
In Fig.~\ref{fig:etaW} we present the rapidity distribution for the
inclusive production of an on-shell $W^+$. The $W^+$ bosons are mostly produced
forward, and the rapidity distribution is peaked at $y_W\sim 1$.
The three histograms give the results of our LO, NLO and NNLO calculation 
of the cross section $\sigma=\sigma(W^+) BR(W\to l{\nu})$. 
The height of each histogram bin gives the value of the cross section
in the corresponding rapidity bin\footnote{The average value
$(d\sigma/dy)_i$ of the rapidity cross section in the $i$-th bin is obtained
by rescaling the corresponding bin cross section $\sigma_i$ by the bin size
$\Delta y=0.2$, i.e. $(d\sigma/dy)_i=\;\sigma_i/0.2=5 \,\sigma_i$.}. 
The error bars\footnote{Unless otherwise stated, 
hereafter the error bars in the 
histograms of our QCD computations always refer to the 
numerical error in the Monte Carlo integration
carried out by our program.}
reported in the histograms refer to an estimate of the numerical error in the
Monte Carlo integration carried out by our program. These error bars are small
and hardly visible in the plot of Fig.~\ref{fig:etaW} 
(see Fig.~\ref{fig:KW} and related comments).

Having computed the rapidity cross section at the first three orders 
in QCD perturbation theory, we can examine the quantitative convergence of the
perturbative expansion. To this purpose,
we introduce $y$ dependent ratios of the results at two
successive perturbative orders. More precisely, we define the
following NLO and NNLO `K factors':
\beq
\label{wkfators}
K_{NLO}(y) = \f{ \;\; \bigl[ d\sigma/dy \bigr]_{NLO}}{\bigl[ d\sigma/dy 
\bigr]_{LO}}
\;\;, \quad \quad
K_{NNLO}(y) = 
\f{ \;\; \bigl[ d\sigma/dy \bigr]_{NNLO}}{\bigl[ d\sigma/dy \bigr]_{NLO}}
 \;\;,
\eeq
where $\bigl[ d\sigma/dy \bigr]_{LO}$, 
$\bigl[ d\sigma/dy \bigr]_{NLO}$ and $\bigl[ d\sigma/dy \bigr]_{NNLO}$
are the LO, NLO and NNLO cross sections.

The K factors for on-shell $W^+$ production
are shown in the lower panel of Fig.~\ref{fig:etaW}. 
The NLO effects are bigger than the NNLO effects on both the normalization
and the shape of the rapidity cross section.
In the rapidity interval
$|y_W| \ltap 2$, the NLO K factor varies in the range
$K_{NLO}(y_W) \sim~$1.3--1.4, while the NNLO K factor
varies in the range $K_{NNLO}(y_W) \sim~$1.02--1.04.
We recall \cite{Martin:2009iq} that the K factors computed from the ratio of 
the total (i.e. integrated over $y_W$)
cross sections are  $K_{NLO}=1.35$ and $K_{NNLO}=1.03$.
The fact that $K_{NNLO}(y_W)$ is much closer to unity than $K_{NLO}(y_W)$ 
indicates a
very good quantitative convergence of the truncated perturbative expansion,
as first found in Ref.~\cite{Anastasiou:2003ds}.
In particular, the value of $K_{NNLO}-1$ can consistently be used as a measure
of the theoretical uncertainty due to the uncalculated contributions
from higher orders (i.e., beyond NNLO).

We have repeated our QCD calculation of the $W^+$ rapidity cross section 
by using 
the parton density sets of the ABKM Collaboration
\cite{Alekhin:2005gq, Alekhin:2009ni} and of the Dortmund Group
\cite{GJRpdf, JimenezDelgado:2009tv}. Here we limit ourselves to
presenting the results at the NNLO. In Fig.~\ref{fig:KW} we show the NNLO
ratios $(d\sigma/dy_W)_{PDF}/(d\sigma/dy_W)_{MSTW}$, where
the cross section in the numerator is computed by using either the
ABKM09 set \cite{Alekhin:2009ni} or the 
JR09VF set \cite{JimenezDelgado:2009tv}, while the cross section in the 
denominator uses the MSTW2008 set. We see that the NNLO
ratios are different from unity and depend on $y_W$; varying $y_W$
in the rapidity interval
$|y_W| \ltap 2$, 
the differences can reach the level of about 5\%.
The $W^+$ bosons from the ABKM09 (JR09VF) partons are produced slightly
more (less) forward than those from the MSTW2008 partons.
Considering the NNLO total cross sections, the ABKM09 (JR09VF) result
is about 3\% higher (1\% lower) than the MSTW2008 result:
we find the values\footnote{The errors on the values of the cross sections
are those from the Monte Carlo integration in our calculation.}
$\sigma_{NNLO}= 1.391 \pm 0.002$~nb,
$1.349 \pm 0.002$~nb and
$1.338 \pm 0.002$~nb, which agree with the corresponding results in 
Refs.~\cite{Alekhin:2009ni}, 
\cite{Martin:2009iq}\footnote{The authors of Ref.~\cite{Martin:2009iq}
use the leptonic branching ratio $B_{l\nu}=10.80$\%,
which is the PDG 2008 value that is obtained by averaging the
electron, muon and tau branching ratios.}
and \cite{JimenezDelgado:2009tv}.
This quantitative agreement is a numerical check of our calculation,
since the results of 
Refs.~\cite{Martin:2009iq,Alekhin:2009ni,JimenezDelgado:2009tv}
are obtained by using the direct calculation of the NNLO total cross section 
\cite{Hamberg:1990np}.

The error bars reported in Fig.~\ref{fig:KW} are the numerical errors of our 
Monte Carlo computations: they are the relative errors of the MSTW08
result presented in Fig.~\ref{fig:etaW} and the analogous errors of the
corresponding ABKM09 and JR09VF results.
In the rapidity range $|y_W| \ltap 2$, these errors are smaller than
$\pm 1$\% and, typically, at the level 
of about $\pm 5$\textperthousand. 

The PDF analyses of 
Refs.~\cite{Martin:2009iq, Alekhin:2009ni, JimenezDelgado:2009tv}
include estimates of the PDF uncertainties that originate from the experimental
errors of the data used in the corresponding global fits.
These PDF uncertainties can be used to evaluate the ensuing errors on the
theoretical computation of physical observables. We do not explicitly show
the PDF errors on the $W^+$ rapidity distribution.
Considering the region of small and medium rapidities (say, $|y_W| \ltap 2$),
these errors depend
slightly on $y_W$ and, therefore, their sizes can be argued from
the PDF error on the corresponding total cross sections. The quoted PDF errors
at the one-sigma level (68\% C.L.) on the NNLO total cross section
are about $\pm 1.7$\% (MSTW2008) \cite{Martin:2009iq},
$\pm 1$\% (ABKM09) \cite{Alekhin:2009ni} and
$\pm 1.2$\% (JR09VF) \cite{JimenezDelgado:2009tv}. 
The differences between the values of the NNLO total cross section 
obtained by the three
PDF sets are (almost) covered by these PDF errors, but 
these PDF errors do not fully
cover the PDF differences shown by the $y_W$ dependent ratios 
in Fig.~\ref{fig:KW}.

The CDF Collaboration 
has recently presented the first direct measurement of the $W$ asymmetry
\cite{cdfw}.
The experimental
difficulty in the determination of the neutrino's longitudinal momentum
is resolved \cite{Bodek:2007cz}
by means of a Monte Carlo driven extrapolation of the lepton rapidity
distribution, and the final CDF results refer to the charge asymmetry
of inclusive production of on-shell $W$ bosons.

\begin{figure}[htb]
\begin{center}
\begin{tabular}{c}
\epsfxsize=12truecm
\epsffile{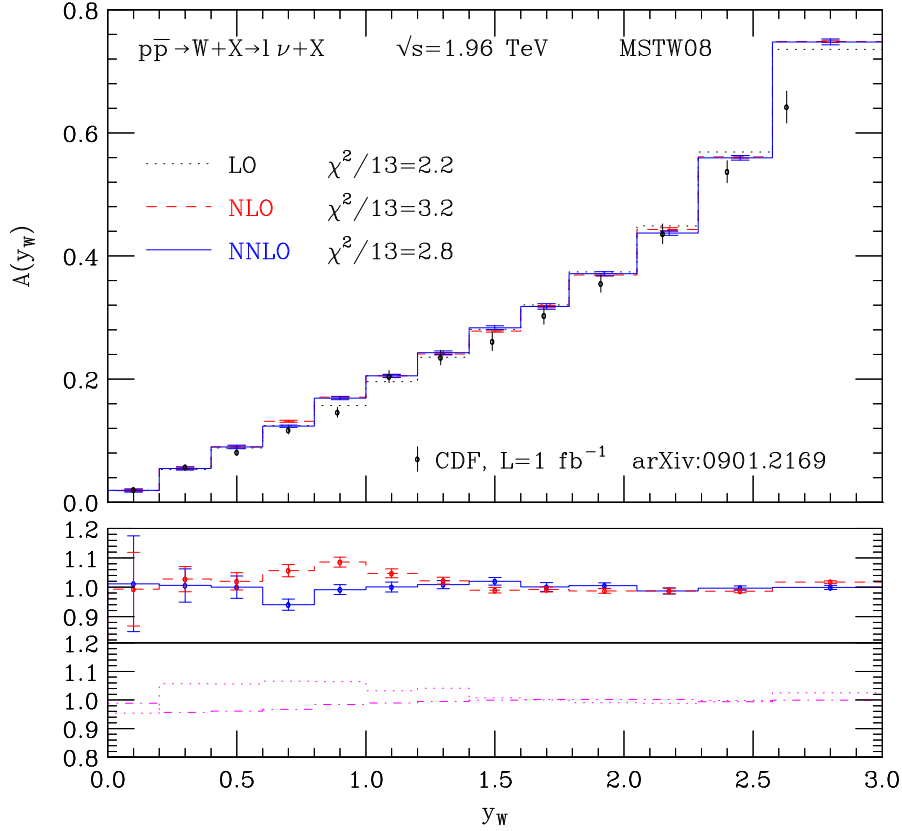}\\
\end{tabular}
\end{center}
\caption{\label{fig:asyW-nocuts}
{\em The charge asymmetry of on-shell $W$ production at the Tevatron Run~II. 
Upper panel: the CDF data \cite{cdfw} are compared with the QCD calculations
at the LO (black dotted), NLO (red dashed) and NNLO (blue solid).
Lower panel: the ratios NLO/LO (red dashed) and 
NNLO/NLO (blue solid) of the QCD results in the upper panel, and the 
corresponding ratios $({NLO/LO})_{LO}$ (magenta dotted) and 
$({NNLO/NLO})_{LO}$ (magenta dot-dashed) computed by using LO partonic cross
sections.}
}
\end{figure}

In Fig.~\ref{fig:asyW-nocuts} we present our 
perturbative results for the
asymmetry of an on-shell $W$ boson at LO (dotted histogram), 
NLO (dashed histogram) and NNLO (solid histogram),
and we compare them with the CDF data
\footnote{Here and in the following calculations of rapidity distributions
at the Tevatron Run~II, the sizes of the rapidity bins are 
always fixed to be equal to those used in the corresponding measurements
of the CDF and D0 Collaborations.
}.
The results in Fig.~\ref{fig:asyW-nocuts} confirm previous 
findings \cite{Anastasiou:2003ds}:
the perturbative corrections to the $W$ asymmetry are 
small. In particular, the corrections to the asymmetry are smaller than the
corresponding corrections to the rapidity cross section 
(see Fig.~\ref{fig:etaW}): this is somehow expected since the asymmetry involves
ratio of cross sections. More precisely, the smallness of the QCD radiative
corrections to the asymmetry could have been argued by the direct inspection
of the K factors in Fig.~\ref{fig:etaW}, since they have a high degree of
symmetry with respect to the exchange $y_W \leftrightarrow - y_W$
(note that a forward--backward symmetric K factor for the rapidity 
cross section would imply no radiative corrections to the asymmetry).


The effect of the radiative corrections is quantified by the NLO (dashed
histogram) and NNLO (solid histogram) asymmetry K factors, which are shown
in the lower panel of Fig.~\ref{fig:asyW-nocuts}. These K factors are computed
analogously to Eq.~(\ref{wkfators}) by simply replacing $d\sigma/dy$ with 
the asymmetry $A(y)$. Both K factors are close to unity, with small deviations
in the region of small to medium $y_W$.
The K factors vary in the ranges $K_{NLO}(y_W)\sim$~0.98--1.08 and 
$K_{NNLO}(y_W)\sim$~0.94--1.02, 
and the NNLO effects tend to be smaller than the NLO effects.
To understand the origin of these effects, we have recomputed the NLO and
NNLO asymmetry K factors
by always using the LO partonic cross sections, though still using the PDFs
at LO, NLO and NNLO. This procedure removes the effect of the 
radiative
corrections in the partonic cross sections, 
so that the NLO (NNLO) K factor is directly sensitive
to the NLO/LO (NNLO/NLO) ratio of PDFs. These `PDF driven' NLO (dotted histogram)
and NNLO (dot-dashed histogram) K factors are also displayed in the lower panel
of Fig.~\ref{fig:asyW-nocuts}. Since each `PDF driven' 
K factor closely follows the quantitative behaviour of the
corresponding K factor in the upper part 
of the lower panel, we conclude that the bulk of the radiative
effects is produced by the variation of the PDFs in going from LO to NLO and to
NNLO.

We have studied the factorization and renormalization 
scale dependence of the QCD results by varying
$\mu_F=\mu_R$ between $M_W/2$ and $2M_W$.
We find that the scale dependence of the $W$ asymmetry at NNLO 
is at the level of the 
numerical errors from our NNLO Monte Carlo computation (i.e. the error bars
of the NNLO result in the upper panel of Fig.~\ref{fig:asyW-nocuts}). 

The theoretical predictions in Fig.~\ref{fig:asyW-nocuts} 
agree well
with the CDF data,
except
in the last (i.e. highest-rapidity) bin.
To measure the consistency between the theoretical and experimental results,
we consider the $\chi^2$ probability function:
\beq
\label{chi2}
\f{\chi^2}{N_{\rm pts.}}
= \f{1}{N_{\rm pts.}}
\sum_{i=1} \frac{({\rm th}_i-{\rm exp}_i)^2}{\Delta^2_{i, {\rm exp}}} \;\;,
\eeq
where $N_{\rm pts.}$ is the number of data points (rapidity bins),
${\rm th}_i$ and ${\rm exp}_i$ denote the theoretical and 
experimental values of the $i$-th data point (i.e. in the $i$-th rapidity bin),
respectively, and $\Delta_{i, {\rm exp}}$ is the corresponding experimental error.
The values of $\chi^2/N_{\rm pts.}$ at each perturbative order
are reported in Fig.~\ref{fig:asyW-nocuts}, 
and are dominated by the contribution of the last bin
(removing the contribution of the last bin from the computation
of $\chi^2$, at the NNLO we obtain $\chi^2/12 = 1.6$).

\begin{figure}[htb]
\begin{center}
\begin{tabular}{c}
\epsfxsize=10truecm
\epsffile{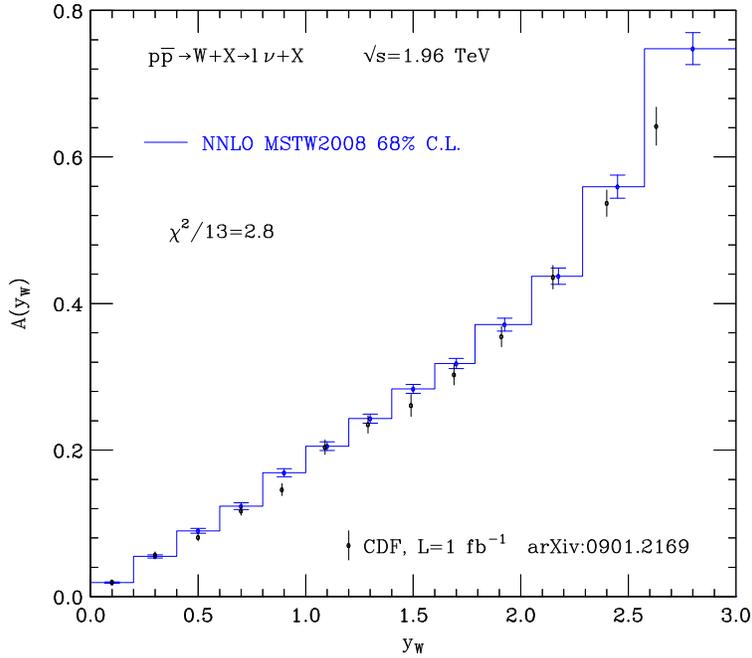}\\
\end{tabular}
\end{center}
\caption{\label{fig:wmstw}
{\em The charge asymmetry of on-shell $W$ production at the Tevatron Run~II.
The CDF data \cite{cdfw} are compared with the NNLO MSTW2008 result,
including the corresponding PDF errors at the 
$1\sigma$
level \cite{Martin:2009iq}.
}}
\end{figure}

\begin{figure}[ht]
\begin{center}
\begin{tabular}{cc}
\includegraphics[width=0.47\textwidth]{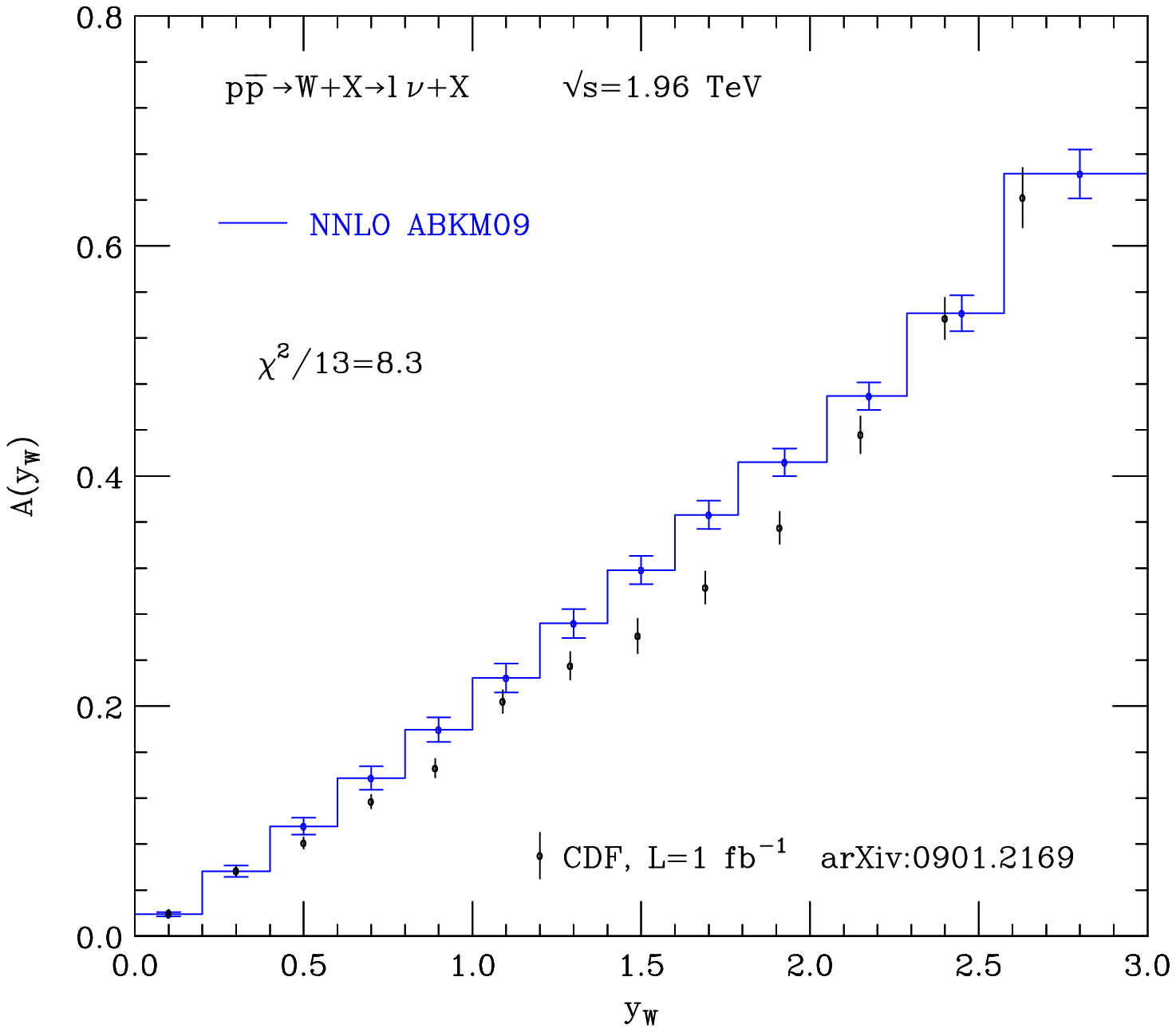} & 
\includegraphics[width=0.47\textwidth]{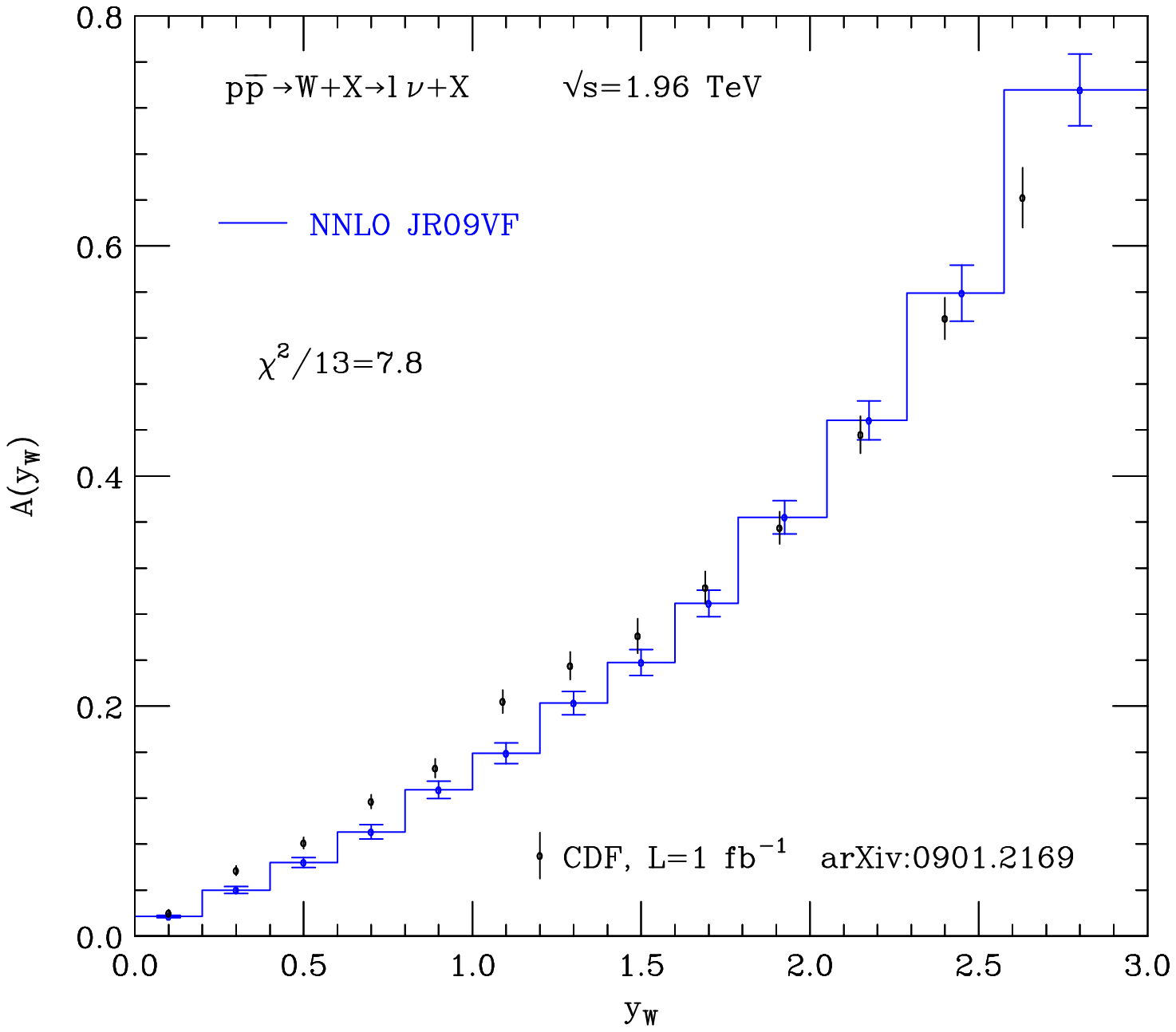}\\
\end{tabular}
\end{center}
\caption{\label{fig:wasycdfaljr}
{\em The charge asymmetry of on-shell $W$ production at the Tevatron Run~II.
The CDF data \cite{cdfw} are compared with 
the NNLO ABKM09 result \cite{Alekhin:2009ni} (left)
and the NNLO JR09VF result \cite{JimenezDelgado:2009tv} (right),
including the corresponding PDF errors.}}
\end{figure}

The NNLO result (solid histogram) of Fig.~\ref{fig:asyW-nocuts} is reported in 
Fig.~\ref{fig:wmstw} by including the PDF errors from the 
MSTW2008 parton densities \cite{Martin:2009iq}. The inclusion of the PDF errors
increases the consistency between the CDF data and the MSTW2008 prediction.
We have repeated the NNLO calculation of $W$ asymmetry at the Tevatron Run~II
by using the ABKM09 \cite{Alekhin:2009ni} and 
JR09VF \cite{JimenezDelgado:2009tv} partons: the results (solid histograms), 
with the corresponding PDF errors, are presented in Fig.~\ref{fig:wasycdfaljr}.
Considering the different slopes of the cross section ratios in 
Fig.~\ref{fig:KW}, we expect that the $W$ charge asymmetry 
from
the ABKM09 (JR09VF) partons is typically larger (smaller) than the asymmetry
from the MSTW2008 partons. This expectation is confirmed by the results in 
Figs.~\ref{fig:wmstw} and \ref{fig:wasycdfaljr}. We see that the 
ABKM09 prediction tends to overshoot the CDF data, while the 
JR09VF prediction tends to undershoot the CDF data in the region of small and
medium rapidities. The values of $\chi^2/N_{\rm pts.}$ computed from
the ABKM09 and JR09VF partons are reported in the corresponding plots of 
Fig.~\ref{fig:wasycdfaljr}.
Unlike the case of the MSTW2008 partons, the contribution of the
highest-rapidity bin to the values of $\chi^2$ is not dominant. 
Removing the contribution of the last bin from the computation
of $\chi^2$, we obtain $\chi^2/12 = 8.9$ (ABKM09 partons) and
$\chi^2/12 = 7.4$ (JR09VF partons).

We add some comments on the CDF measurement of the $W$ asymmetry. Since the
longitudinal momentum of the neutrino from $W$ decay is not determined
experimentally, the rapidity of the $W$ and its asymmetry cannot directly be
measured. The results of Ref.~\cite{cdfw} regard 
the first `direct determination', rather than `direct measurement', 
of the $W$ asymmetry.

The CDF Collaboration \cite{cdfw} selects $W \to e \nu_e$ events and measures
the momentum of the electrons and positrons and the missing (i.e. neutrino)
transverse momentum. Assuming that the decaying $W$ is on-shell
(or, more generally, that the invariant mass of the $W$ is fixed),
this kinematical
information determines the neutrino's longitudinal momentum (and, thus, $y_W$)
within a twofold ambiguity (see, e.g., Eqs.~(\ref{loraddif}) or (\ref{hokin})).
This twofold ambiguity is resolved on a statistical (event--by--event) basis
by using information of theoretical calculations. This information regards,
for instance, the rapidity distributions\footnote{The relation between 
the rapidity distributions of the $W$ and the charged lepton is discussed in
Sect.~\ref{sec:ltev}.} of the $W$ boson 
(Ref.~\cite{cdfw} uses the NNLO calculation of Ref.~\cite{Anastasiou:2003ds}) 
and of the charged lepton and the related boson--lepton momentum correlations
(Ref.~\cite{cdfw} uses results from ${\rm MC@NLO}$ \cite{mcnlo}).
The results of the theoretical calculations depend on the input PDFs
(Ref.~\cite{cdfw} explicitly considers the MRST2006 NNLO set \cite{MRST2006}
and the CTEQ6.1 NLO set \cite{CTEQ6.1}). The use of this amount of theoretical
information requires a consistent (and reliable) estimate of the theoretical
uncertainty that eventually affects the determination of the $W$ asymmetry.
For instance, Ref.~\cite{cdfw} considers the effect of the PDF errors of the
CTEQ6.1 NLO set (though these PDF errors are known 
not to always match the differences
between various available sets of PDFs).
Another theoretically-driven corrections to be taken into account \cite{cdfw}
regards the off-shell $W$ effects in the experimentally selected 
$W \to e \nu_e$ events. These effects depend on the finite width of the 
$W$ and on the experimentally accepted $W$ mass range, and they are more
important at high rapidities. 

Reference \cite{cdfw}
reports the result of the $W$ asymmetry in each $y_W$ bin, and it also presents
the value, $\langle y_W \rangle$, of the `average' bin center in each bin.
The `average' bin centers include the estimated corrections 
(assuming a fixed $W$ mass of
80.403~GeV) \cite{cdfw} to the
$W$ asymmetry from off-shell $W$ effects. In our computation of $\chi^2$
(whose values are reported in 
Figs.~\ref{fig:asyW-nocuts}--\ref{fig:wasycdfaljr}), 
we have compared the theoretical calculations of the $W$ asymmetry with the 
CDF data at the level of histogram bins.
As pointed out to us by the MSTW Group \cite{james}, the value of $\chi^2$
can vary if the comparison 
between the theoretical calculations and the CDF data is performed 
by considering
the values of the $W$ asymmetry at the `average' bin centers.
The NNLO calculation of the asymmetry at the `average' bin centers can be
carried out by using the numerical program of 
Ref.~\cite{Anastasiou:2003ds} (we could also use our Monte Carlo program 
and compute the asymmetry in a small rapidity bin around each 
`average' bin center). 
We have constructed an approximated functional form of
$A(y_W)$ by fitting our histogram results, and we have used this 
approximation
to compute $\chi^2$ from the comparison with the CDF data at 
the `average' bin centers. In the case of the MSTW2008 partons, we find values
of $\chi^2/N_{\rm pts.}$ that are relatively close to unity; the main
difference with respect to the values of $\chi^2/N_{\rm pts.}$ in 
Fig.~\ref{fig:asyW-nocuts} and
\ref{fig:wmstw} is due to the reduced contribution of the last 
(i.e. highest-rapidity) bin. In the cases of the ABKM09 and JR09VF partons,
we find values of $\chi^2/N_{\rm pts.}$ that are slightly reduced (about
10\%) with respect to those reported in Fig.~\ref{fig:wasycdfaljr}.

\subsection{Charged lepton rapidity 
distribution and asymmetry}
\label{sec:ltev}

We now move to our study of the rapidity distributions
of the charged lepton from $W$ decay. There are important differences between 
the rapidity distributions of the $W$ boson and of its decaying lepton.
These differences originate from 
the underlying short-distance dynamics and kinematics, as briefly described
below.

\noindent {\em EW dynamics correlations.}\\
\noindent Owing to the spin 1 nature of the $W$ boson, its production and 
decay mechanisms are correlated. The actual correlation depends on the
$V-A$ 
coupling of the $W$ boson to both the annihilating $q{\bar q}$ pair 
and the decaying lepton pair. The main effect of the correlation
can be understood by simply considering the LO partonic subprocesses in 
Eqs.~(\ref{pro+}) and (\ref{pro-}). The corresponding angular distribution
of the charged lepton is (see, e.g., Ref.~\cite{book})
\beq
\label{stardist}
\f{1}{{\hat \sigma}_{U{\overline D}}^{(0)}} \;
\f{d{\hat \sigma}_{U{\overline D}}^{(0)}}{d \cos\theta_{lD}^*} = 
\f{1}{{\hat \sigma}_{D{\overline U}}^{(0)}} \;
\f{d{\hat \sigma}_{D{\overline U}}^{(0)}}{d \cos\theta_{lD}^*}
= \f{3}{8} 
\left( 1+\cos\theta_{lD}^* \right)^2
 \;\;,
\eeq
where $\theta^*$ is the scattering angle of the charged lepton
in the centre--of--mass system of the colliding quark and antiquark.
More precisely, $\theta_{lD}^*$ is the lepton scattering angle with respect
to the direction of the `down-type' quark or antiquark.
Therefore, the form of the angular distribution on the right-hand side of
Eq.~(\ref{stardist}) implies that at the partonic level the
charged lepton tends to follow the direction of the colliding ${\overline D}$
(see Eq.~(\ref{pro+})) or $D$  (see Eq.~(\ref{pro-})).

The rapidity (angular) distribution of the charged lepton
at the hadronic level results from the combined effect
of Eq.~(\ref{stardist}) and of the parton densities of the colliding hadrons. 
More precisely, since the $W$ and lepton rapidities are related by
\begin{equation}
\label{yrap}
y_l=y_W+\f{1}{2}\ln\f{1+\cos\theta^*}{1-\cos\theta^*}\; ,
\end{equation}
the lepton rapidity distribution arises from the convolution of 
Eq.~(\ref{stardist}) with the $W$ rapidity distribution.

In the case of $p{\bar p}$ collisions, the $W$ boson tends to follow the
direction of the colliding `up-type' quark or antiquark and, therefore,
the forward--backward asymmetry 
produced by the EW distribution in Eq.~(\ref{stardist}) exactly acts in the
opposite direction. As a consequence, 
the rapidity distribution of the positive (negative) charged lepton is shifted
backward (forward) with respect to the distribution of the parent $W^+$ ($W^-$).
The relative weight of the two competitive effects depends on the
detailed  kinematical correlations between the lepton and the boson, and it can
be controlled by varying, for instance, the lepton $E_T$. 

\noindent {\em Kinematics correlations.} \\
\noindent In hadron collisions $W$ boson events are selected by requiring
a lower limit on the invariant mass (or, typically, on the leptonic transverse
mass $M_T$) of the $W$ boson. Provided this lower limit is close to $M_W$
(though smaller than $M_W$) the kinematics of the $W$ and its decaying leptons
is well described by using the narrow-width approximation (NWA), i.e. by assuming
that the $W$ is on-shell.
At the LO in perturbative QCD, the $W$ boson is produced with a vanishing
transverse momentum $q_T$; therefore, within the NWA the kinematical variables
of the $W$ and charged lepton fulfil the relation
(the symbol `$\sim$' denotes the use of the NWA)
\beq
\label{lokin}
M_W \simeq  2 E_T \cosh (y_W-y_l) \;\;, \quad \quad E_T \ltap M_W/2 \;\;,
\eeq
or, equivalently, $1-\cos^2\theta^*=4E_T^2/M_W^2$.
At fixed $E_T$, the rapidities of the $W$ and the charged lepton are thus
directly correlated:
\beq
\label{loraddif}
|y_W-y_l| \simeq \ln \left[ \f{M_W}{2E_T} + 
\sqrt{\left( \f{M_W}{2E_T} \right)^2 - 1} \;\right] \;\;.
\eeq
In particular, by increasing $E_T$, $y_l$ is forced to be close to $y_W$
and the rapidity distribution of the charged lepton tends to follow
the rapidity distribution of the $W$, thus minimizing the impact of the 
rapidity asymmetry produced by the EW dynamics (i.e. by Eq.~(\ref{stardist})). 

Incidentally, we also note that, at the LO and within the NWA, the leptonic
variables $E_T^\nu, M_T$ and $E_T$ are not independent. We have
\beq
\label{loconstr}
E_T^\nu \simeq E_T \;\;, \quad \quad M_T \simeq 2 E_T \;\;,
\eeq
so that fixing, for instance, $E_T$ fully specifies both $E_T^\nu$ and $M_T$.

Having discussed the main differences between 
the rapidity distributions of the $W$ boson and of the charged lepton $l$,
we add a comment on a direct consequence of theses differences.
The $W$ and $l$ rapidity distributions
have a different dependence on 
the parton densities of the
colliding hadrons. 

In $p{\bar p}$ collisions, for instance, owing to the effect
of the angular distribution in Eq.~(\ref{stardist}), 
the $y_l$ distribution of the $l^+$ at positive (negative) rapidity is more
(less) sensitive to the antiquark densities of the proton
than the $y_W$ distribution of the $W^+$ at positive (negative) rapidity.
This point is quantitatively illustrated in Sect.~11.1 of 
Ref.~\cite{Martin:2009iq}.

Moreover, the boson and lepton rapidity distributions probe the parton densities
at different typical values of parton momentum fractions $x_1$ and $x_2$ (see
Eq.~(\ref{dsigma})). In LO QCD and using the NWA, the production
of $W^{\pm}$ bosons (and their decaying $l^{\pm}$ leptons)
is mostly sensitive to the region of parton momentum fractions with
$x_1 x_2 \simeq M_W^2/s$. The typical value of the ratio $x_1/x_2$ is instead
controlled by the boson or lepton rapidity. Fixing the rapidity of the 
$W$, we have
\beq
\label{xratiow}
\frac{1}{2}\ln (x_1/x_2) \simeq \pm y_W \;\;,
\eeq
and different values of the ratio $x_1/x_2$ are explored by varying $y_W$,
Owing to the kinematical relation (\ref{loraddif}), we thus have
\beq
\label{xratiol}
\frac{1}{2}\,| \ln (x_1/x_2) | \simeq \left|  y_l \pm \ln \left( \f{M_W}{2E_T} + 
\sqrt{ \f{M_W^2}{4E_T^2}  - 1} \;\right) 
 \right| \;\;.
\eeq
We see that equal values of $y_W$ and $y_l$ actually probe different values of
the ratio $x_1/x_2$, the difference being controlled by the value of
the lepton $E_T$.

The differences between the rapidity distributions of the $W$ boson
of the charged lepton $l$ have a direct impact on any observables 
that are measured, or computed, by applying restrictions on the kinematics 
of the leptons from $W$ decay. This comment is valid also for 
observables that directly refer to kinematical variables of the $W$ boson.
For instance, 
it is interesting to study how the results for the charge asymmetry of 
fully-inclusive $W$ production (Fig.~\ref{fig:asyW-nocuts})
are affected by typical selection cuts that are used in experiments at the
Tevatron Run~II. 

\begin{figure}[htb]
\begin{center}
\begin{tabular}{cc}
\includegraphics[width=0.48\textwidth]{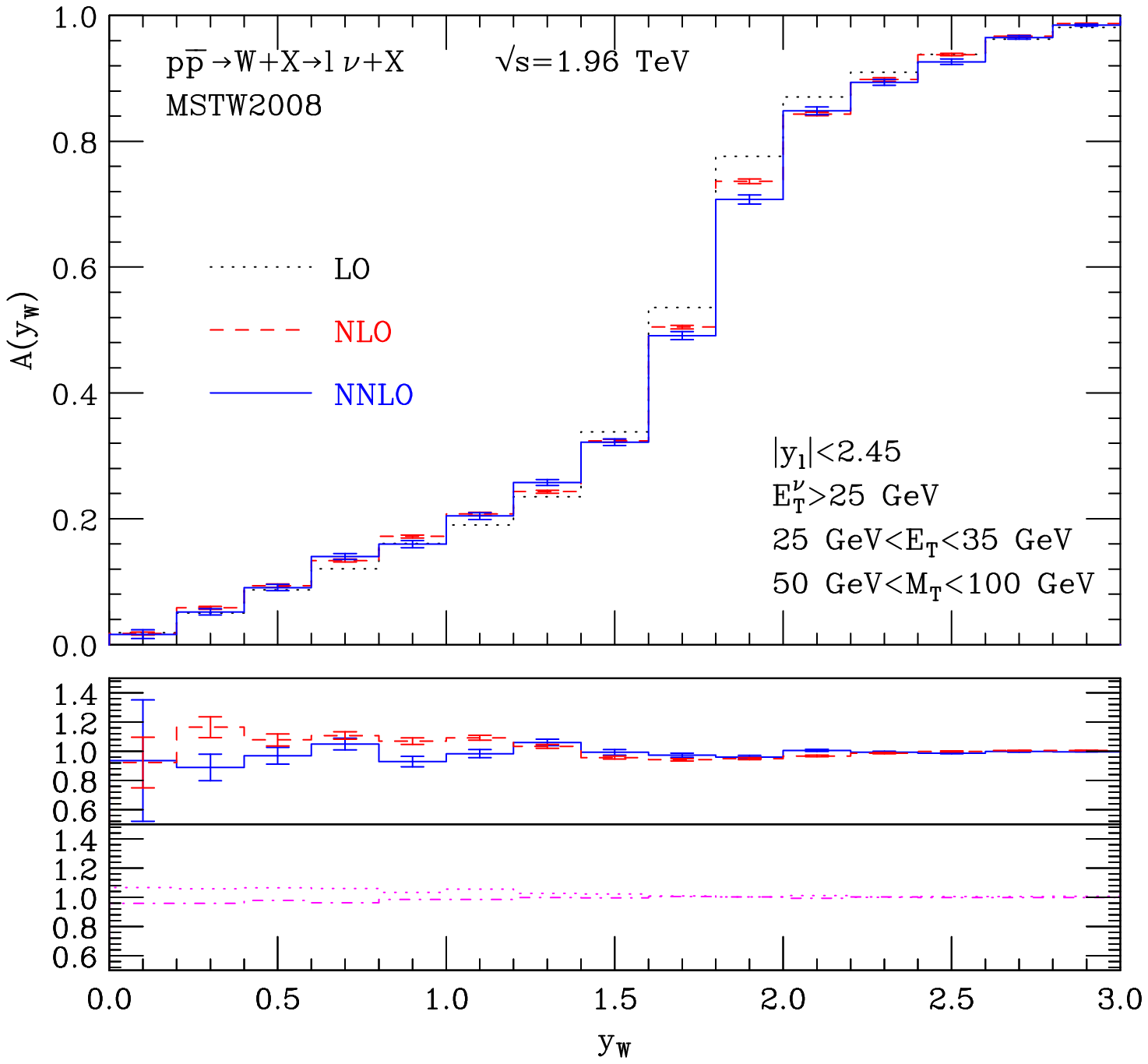} 
& \includegraphics[width=0.48\textwidth]{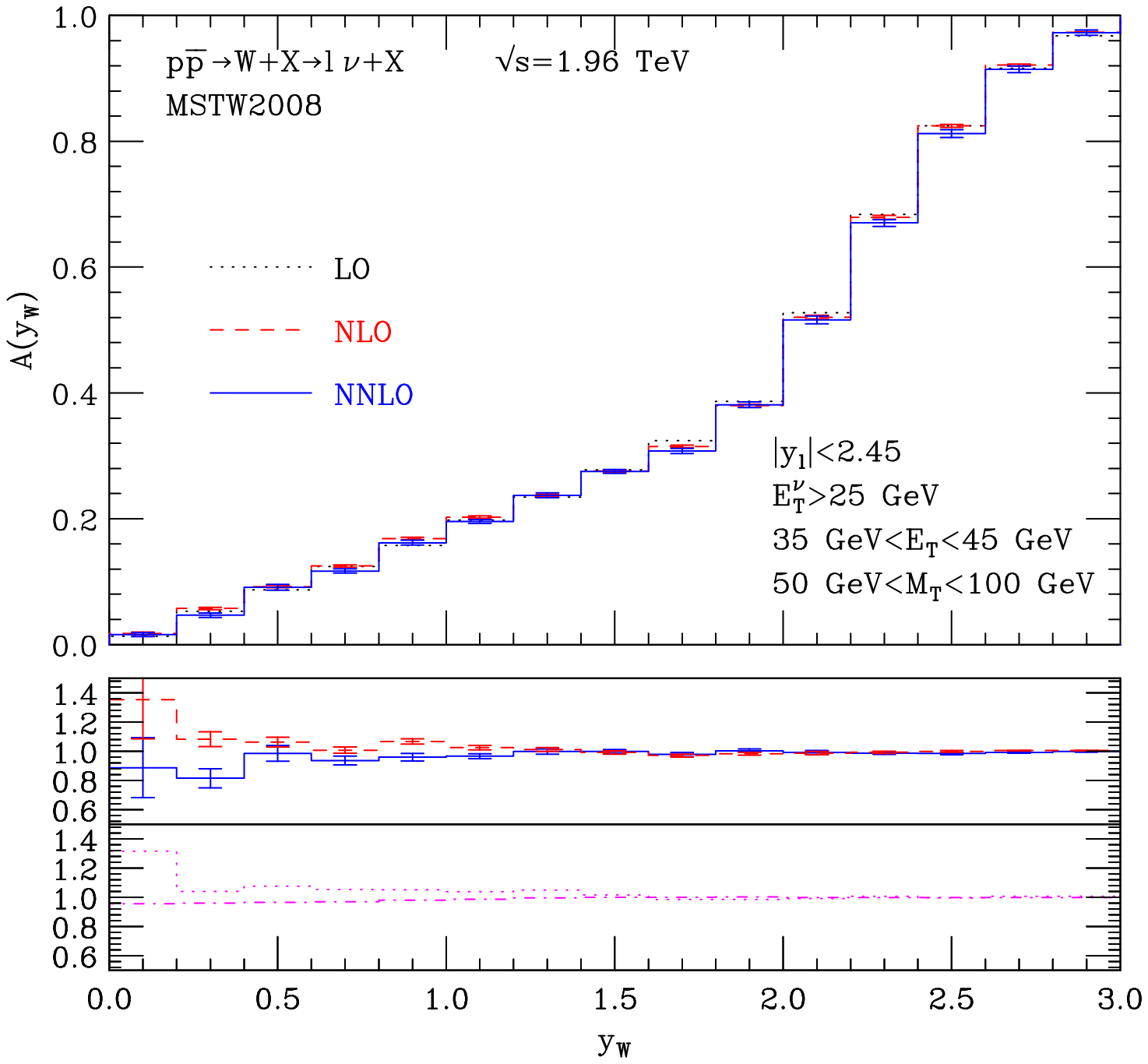}\\
\end{tabular}
\end{center}
\caption{\label{fig:wasycdf}
{\em $W$ production at the Tevatron Run~II with lepton selection cuts.
The rapidity of the charged lepton is constrained to be in the region 
$|y_l|\leq 2.45$, and its transverse energy is in the intervals 
25~GeV $< E_T<$ 35~GeV (left) and
35~GeV $< E_T<$ 45~GeV (right). Upper panels: the $W$ charge asymmetry
at the LO (black dotted), NLO (red dashed) and NNLO (blue solid).
Lower panels: the ratios NLO/LO (red dashed) and 
NNLO/NLO (blue solid) of the QCD results in the upper panels, and the 
corresponding ratios $({NLO/LO})_{LO}$ (magenta dotted) and 
$({NNLO/NLO})_{LO}$ (magenta dot-dashed) computed by using LO partonic cross
sections.}
}
\end{figure}

For illustrative purpose, we consider the lepton selection cuts
used by the CDF Collaboration in Ref.~\cite{cdfe}:
the observed charged lepton must be produced in the 
rapidity region $|y_l|\leq y_{l, \,{\rm MAX}}$ 
with $y_{l, \,{\rm MAX}}=2.45$, and it has to be isolated 
from hadronic activity (the CDF isolation criterion is described later in
this subsection); 
the selected $W \to l \nu$ events
are required to have $E_T^\nu>25$~GeV
and 50~GeV~$<~M_T~<$ 100~GeV.
In Fig.~\ref{fig:wasycdf} we show the QCD results on the
$W$ charge asymmetry after the
implementation of the lepton selection cuts.
Following the experimental analysis \cite{cdfe},
we consider two bins of the charged
lepton $E_T$: 
25~GeV $< E_T<$ 35~GeV (left panels in Fig.~\ref{fig:wasycdf}) and
35~GeV $< E_T<$ 45~GeV (right panels in Fig.~\ref{fig:wasycdf}).

Comparing the histogram in Fig.~\ref{fig:asyW-nocuts} with those in 
Fig.~\ref{fig:wasycdf}, we see that the shape (also the size) of the 
$W$ charge asymmetry is 
changed by the introduction of the lepton
selection cuts.

As in Fig.~\ref{fig:asyW-nocuts}, the lower panels in Fig.~\ref{fig:wasycdf}
show the asymmetry K factors at NLO (dashed) and NNLO (solid), and the
K factors computed with the corresponding NLO (dotted) and NNLO (dot-dashed)
PDFs by using the LO partonic cross sections.
The NLO and NNLO K factors in Fig.~\ref{fig:wasycdf} are 
similar
to those in Fig.~\ref{fig:asyW-nocuts}: the radiative corrections to the
asymmetry remain small after introducing the selection cuts, and the main
effect of these corrections acts in the region where
$y_W\ltap 1.5$. As in the case of the inclusive-$W$ asymmetry,
the effect of the radiative corrections is mostly produced by the 
variation of the PDFs in going from 
the LO PDF set to the NLO set and to NNLO set.

To explain the origin of the differences between the charge asymmetry results
in Figs.~\ref{fig:asyW-nocuts} and \ref{fig:wasycdf}, we have to understand
which are the most significant lepton selection cuts.
The charge asymmetry in Figs.~\ref{fig:asyW-nocuts} refers to on-shell $W$
production. The results in Fig.~\ref{fig:wasycdf} are obtained without fixing
the $W$ invariant mass to its on-shell value; the off-shell effects have,
however, a little impact on the results since the 
minimum value of 
$M_T$ is smaller than  $M_W$ and close to it.
The charged lepton isolation has also a minor impact, since it is not effective
at the LO and the radiative corrections are small.
At the LO and within the NWA, the constraints 
$E_T^\nu>25$~GeV and 50~GeV~$<~M_T~<$ 100~GeV are superseded by the 
constraints on $E_T$: using Eq.~(\ref{loconstr}), 
the cut 25~GeV $< E_T<$ 35~GeV
implies $E_T^\nu \gtap 25$~GeV and 50~GeV~$\ltap~M_T~\ltap$ 70~GeV, and the cut 
35~GeV $< E_T<$ 45~GeV implies $E_T^\nu \gtap 35$~GeV and 
70~GeV~$\ltap~M_T~\ltap$ 90~GeV. We can conclude that the 
mainly-relevant parameters that control the results in Fig.~\ref{fig:wasycdf}
are the maximum rapidity $y_{l, \,{\rm MAX}}$ and the $E_T$
of the charged lepton.

The cut on the charged lepton rapidity selects a fraction of the
inclusively-produced $W$, and this fraction is not uniform with respect to 
$y_W$. Owing to the EW dynamics correlations 
discussed at the beginning of this subsection, the $l^+ \,(l^-)$ originates
from a parent $W^+ \,(W^-)$ that is preferably produced at $y_W > y_l 
\,(y_W < y_l)$. Therefore, the forward--backward symmetric cut, 
$- y_{l, \,{\rm MAX}} < y_l < y_{l, \,{\rm MAX}}$, on $y_l$ selects
an event sample with an enriched component of forward-produced $W^+$
(backward-produced $W^-$):
the $y_W$ charge asymmetry of this sample is thus 
higher
than the 
asymmetry of the fully-inclusive sample. 
The increase of the (absolute value of the)
asymmetry is larger in the region of high values of $|y_W|$, 
where the selection cut 
$|y_l| < y_{l, \,{\rm MAX}}$ is more effective (the constraint 
$|y_l| < y_{l, \,{\rm MAX}}$ has a relatively-small effect on the
$W$ bosons at $y_W \sim 0$). 
Moreover, the region of high-$|y_W|$ values
where the asymmetry increases depends on $E_T$.
Increasing $E_T$ this region
moves to larger values of $|y_W|$ since, 
owing to the lepton--boson kinematic correlations, 
by increasing $E_T$ the lepton and boson rapidities get closer 
(see Eq.~(\ref{loraddif})) and, consequently, the 
cut $|y_l| < y_{l, \,{\rm MAX}}$ selects the parent $W$ bosons more uniformly
over an extended region of $|y_W|$.

The qualitative features that we have just discussed are 
confirmed by the quantitative results in Fig.~\ref{fig:wasycdf}.
In the low-$E_T$ bin, the $W$ charge asymmetry (left panel in
Fig.~\ref{fig:wasycdf}) closely follows the inclusive-$W$ asymmetry
(Fig.~\ref{fig:asyW-nocuts}) in the region $y_W \ltap 1.5$, and it sharply
increases at larger values of $y_W$.
In the high-$E_T$ bin, the $W$ charge asymmetry (right panel in
Fig.~\ref{fig:wasycdf}) closely follows the inclusive-$W$ asymmetry
(Fig.~\ref{fig:asyW-nocuts}) up to $y_W \sim 2$; then, it increases and,
at large $y_W$ ($y_W \sim 3$), it reaches the values of 
the charge asymmetry in the low-$E_T$ bin.

In summary, the differences between the charge asymmetry results in 
Figs.~\ref{fig:asyW-nocuts} and \ref{fig:wasycdf} are due to the fact
that the rapidity distribution of the lepton is, on average, more 
forward--backward symmetric than the rapidity distribution of the
parent $W$. More precisely, as discussed at the beginning of this subsection,
the rapidity distribution of the $l^+ (l^-)$ lepton is slightly bent
backward (forward) with respect to the distribution of the $W^+$ ($W^-$)
boson.

\begin{figure}[htb]
\begin{center}
\begin{tabular}{cc}
\includegraphics[width=0.48\textwidth]{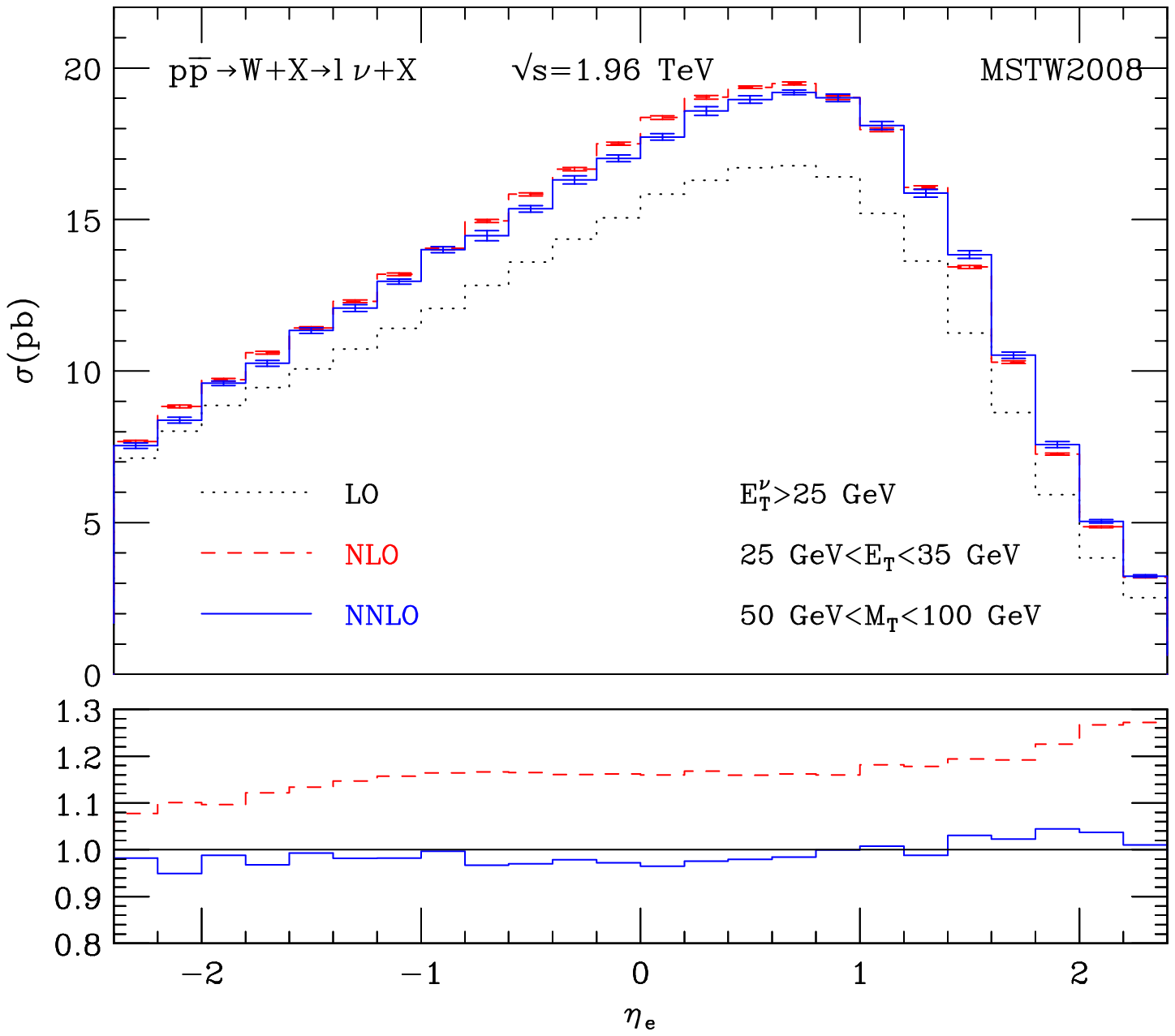} & 
\includegraphics[width=0.48\textwidth]{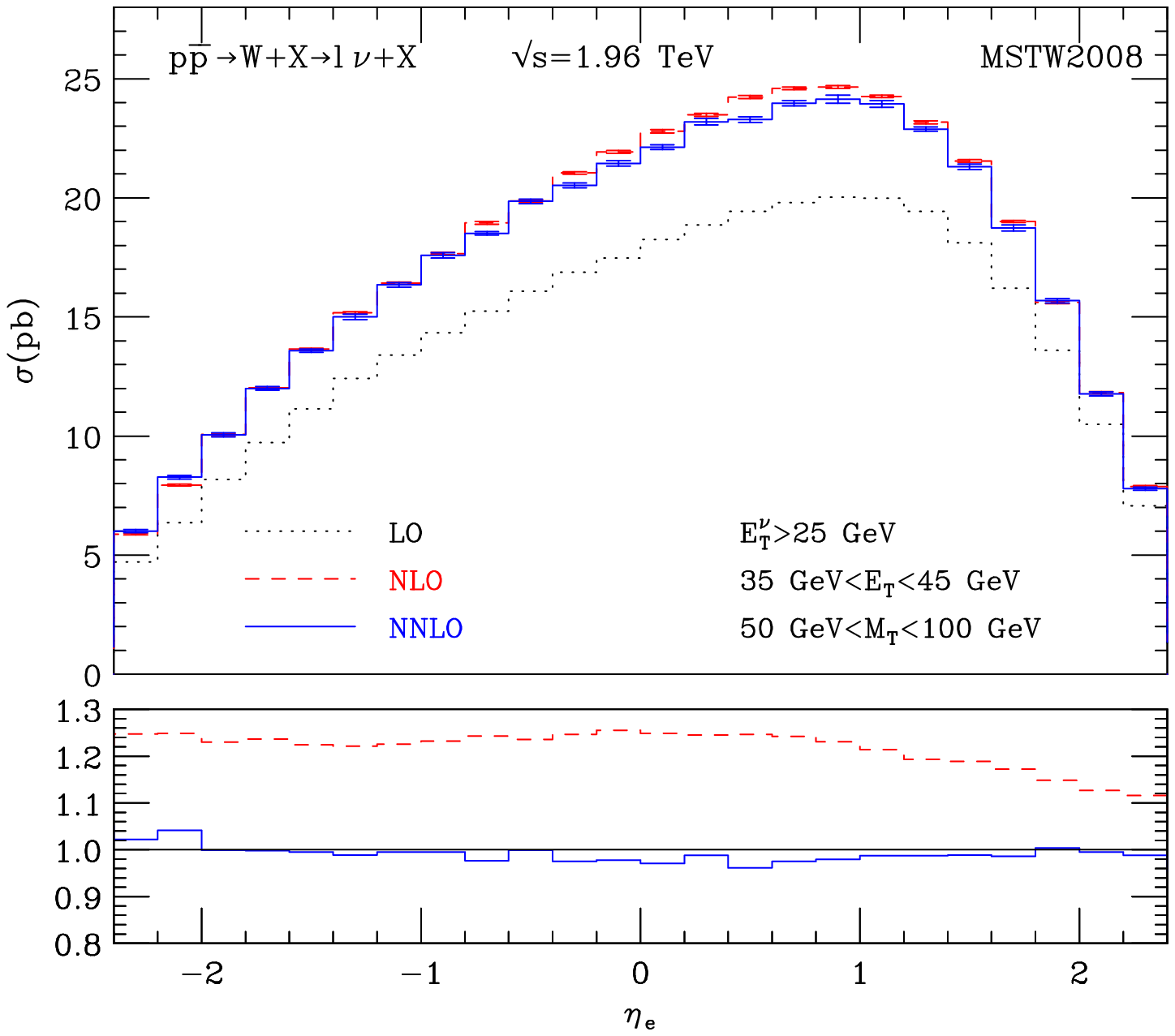}\\
\end{tabular}
\end{center}
\caption{\label{fig:etacdf}
{\em Rapidity distribution of the charged lepton from $W^+$ decay up to NNLO
QCD: low-$E_T$ bin (left), high-$E_T$ bin (right). In the lower plots the NLO/LO (dashed) and NNLO/NLO (solid) ratios are shown.}}
\end{figure}

\setcounter{footnote}{1}

The typical behaviour of the $l^+$ rapidity\footnote{Hadron collider experiments
actually measure the lepton pseudorapity $\eta_l$, and we use the label 
$\eta_l$ in the figures of the paper. Since in our QCD calculations with
massless leptons we have $\eta_l = y_l$, in the text we always refer
to the lepton rapidity and we equivalently use the
labels $\eta_l$ and $y_l$.} cross section  
is shown in Fig.~\ref{fig:etacdf}. 
As in Fig.~\ref{fig:wasycdf}, we consider 
the lepton selection cuts of Ref.~\cite{cdfe} and two $E_T$ bins.
The three histograms in each upper panel of Fig.~\ref{fig:etacdf}
give the results of our LO, NLO and NNLO calculation 
of the lepton rapidity cross section in each corresponding $E_T$ bin.
As in the case of the $W$ rapidity distribution (see Fig.~\ref{fig:etaW}),
the height of each histogram bin gives the value of the cross section
in the corresponding rapidity bin.

In the lower panels of Fig.~\ref{fig:etacdf}, we present the NLO (dashed) 
and NNLO (solid) K factors, computed from the rapidity cross sections
in the upper panels (see Eq.~(\ref{wkfators})).
The NLO effects are bigger than the NNLO effects on both the normalization
and the shape of the lepton rapidity cross section, analogously to the case 
of the $W$ rapidity cross section (see Fig.~\ref{fig:etaW}).
In the low-$E_T$ bin, the NLO K factor varies in the range
$K_{NLO}(y_l) \sim~$1.08--1.28
(considering  $|y_l| \ltap 2$, the range is $K_{NLO}(y_l) \sim~$1.10--1.22),
while the NNLO K factor
varies in the range $K_{NNLO}(y_l) \sim~$0.94--1.04.
In the high-$E_T$ bin, the NLO K factor varies in the range
$K_{NLO}(y_l) \sim~$1.12--1.26
(considering  $|y_l| \ltap 2$, the range is $K_{NLO}(y_l) \sim~$1.16--1.26),
while the NNLO K factor
varies in the range $K_{NNLO}(y_l) \sim~$0.96--1.04.
These lepton K factors tend to be closer to unity than the corresponding 
K factors for inclusive $W$ production. The same tendency is observed by
considering the K factors computed from the ratio of
the total accepted cross sections (i.e. integrated over the rapidity range
$|y_l|\le 2.45$): 
in the low-$E_T$ bin, we find $K_{NLO}=1.16$ and $K_{NNLO}=0.99$;
in the high-$E_T$ bin, we find $K_{NLO}=1.21$ and $K_{NNLO}=0.99$.

Although the QCD radiative corrections to the lepton rapidity cross section
tend, on average, to be smaller than those to the $W$ rapidity cross section, the 
lepton K factors in Fig.~\ref{fig:etacdf} and the inclusive K factors in 
Fig.~\ref{fig:etaW} have definitely different shapes. The difference is
particularly evident at NLO. In particular, the lepton K factors
are certainly less forward--backward symmetric than the inclusive K factors.
Therefore, we can expect that the effect of the radiative corrections
on the lepton charge asymmetry is larger than the effect on the $W$ charge
asymmetry, especially a high values of $y_l$.

In Figs.~\ref{fig:etaW} and \ref{fig:etacdf} we see that both the $W^+$ 
and the $l^+$ are mainly produced in the forward region.
Comparing the shapes of the $W^+$ and $l^+$ rapidity distributions, 
we can also see the effect of the $V-A$ structure of the EW interactions:
in going from the $W^+$ to the  $l^+$, the peak of the distribution is shifted
and the overall distribution is bent toward the backward direction.
In agreement with Eqs.~(\ref{lokin}) and (\ref{loraddif}), 
the effect is less evident when higher values of $E_T$ are selected.
These features are expected from a basic LO analysis of boson--lepton
correlations (as discussed at the beginning of this subsection).
The impact of higher-order QCD corrections on this LO picture can be seen
by a careful inspection of the K factors in Fig.~\ref{fig:etacdf}:
the QCD corrections {\em partly compensate} 
for the boson--lepton differences that
occur at the LO.
To a good approximation, 
the K factors in the low-$E_T$ (high-$E_T$) bin monotonically increase 
(decrease) as $y_l$ increases. Thus, 
the effect of the NLO (and NNLO) corrections in the lower-$E_T$ bin is 
to (slightly) shift the lepton rapidity distribution forward, whereas the 
opposite happens in the higher-$E_T$ bin.
As a consequence, in the lower (higher) $E_T$ bin we expect 
the effect of QCD corrections on the lepton charge asymmetry 
to be positive (negative). Moreover, owing to the monotonic behaviour of the 
K factors, the absolute size of this effect is expected to increase at high 
$|y_l|$. 

The qualitative effect of the QCD radiative corrections deserves some comments.
Our physical interpretation is that QCD corrections produce
dynamical and kinematical decorrelation effects of the LO correlations between 
the rapidity distributions of the $W$ boson and of its decaying lepton.

The QCD corrections to the LO partonic cross section in Eq.~(\ref{stardist})
are produced by multiparton radiation from the initial-state (colliding) 
partons. Since the QCD couplings are purely vector-like and flavour blind 
(insensitive
to the difference between 'up-type' and 'down-type' quarks and antiquarks),
QCD radiation dynamically dilutes the forward--backward asymmetry 
(see Eq.~(\ref{stardist}))
that is produced at the LO by the $V-A$ structure of the EW couplings.
After convolution with the PDFs of the initial-state partons,
this partonic diluted asymmetry reduces the impact of the lepton decay on the
shape of the (PDF-driven) rapidity distribution of the parent $W$. 

The QCD corrections also affect the kinematics of the $W$. In particular, 
the $W$ boson is no longer produced with a vanishing transverse momentum $q_T$
and, thus, Eq.~(\ref{lokin}) can  be modified as follows:
\beq
\label{hokin}
M_W \simeq 2 E_T \,
\bigl[ \, \cosh (y_W-y_l) + {\cal O}( q_T/M_W ) \, \bigr] \;\;.
\eeq
Since the typical values of the transverse momentum are $q_T \ltap \as M_W$, 
the correction term ${\cal O}( q_T/M_W )$ is not large.
Nonetheless, the effect of this correction term reduces the LO correlation
between the rapidities of the $W$ and the charged lepton 
(see Eq.~(\ref{loraddif})). In particular, the right-hand side of 
Eq.~(\ref{hokin}) shows that
the kinematical decorrelation produced by the correction term 
becomes more important by increasing $E_T$ 
(i.e. when $y_l$ is closer to $y_W$).

In the following we concentrate our attention on the lepton charge asymmetry.
We present our perturbative QCD calculations up to NNLO, and their comparison
with some of the published Tevatron data.

At the Tevatron Run~II, the CDF and D0 Collaborations have performed 
measurements of the lepton charge asymmetry by analyzing data samples with
increasing integrated luminosity $L$ (statistics), namely $L=170~{\rm pb}^{-1}$
\cite{cdfe}, 300~${\rm pb}^{-1}$ \cite{d0m} and 750~${\rm pb}^{-1}$ \cite{d0e}.
We list the lepton selection cuts that are used in these measurements and
implemented in our corresponding QCD calculations. As already mentioned, the
electron (and positron) event cuts of the CDF Collaboration \cite{cdfe}
are $E_T^\nu>25$~GeV and 50~GeV~$< M_T <$ 100~GeV, 
and the electron charge
asymmetry is measured 
in three different $E_T$ bins: 
$E_T >25$~GeV, 25~GeV $< E_T <$ 35~GeV and 35~GeV $< E_T<$ 45~GeV.
The D0 muon charge asymmetry \cite{d0m} is measured in the region 
where\footnote{The D0 measurement uses $p_T >20$~GeV, where $p_T$ is the
transverse momentum of the muon. In our QCD calculations with massless leptons,
$p_T$ and $E_T$ are equivalent.} 
$E_T >20$~GeV, with the selection cuts $E_T^\nu>20$~GeV and $M_T >$ 40~GeV.
The electron (and positron) event cuts of the D0 Collaboration \cite{d0e}
are $E_T^\nu>25$~GeV and $M_T >$ 50~GeV, 
and the electron charge
asymmetry is measured 
in three different $E_T$ bins: 
$E_T >25$~GeV, 25~GeV $< E_T <$ 35~GeV and $E_T >$ 35~GeV.
In all these measurements the charged leptons are required to be isolated.

The charged lepton 
isolation is defined by considering a cone along the 
direction $\{\eta_l, \phi_l\}$ of the lepton momentum in 
pseudorapidity--azimuth $(\eta$--$\phi)$ space. The cone radius is 
$R=\sqrt{(\eta-\eta_l)^2 + (\phi-\phi_l)^2}$ and the hadronic (partonic)
transverse energy in the cone is denoted by $E_T^{\rm iso}$.
The CDF and D0 isolation criterion for electrons and positrons fixes $R=0.4$:
the CDF Collaboration requires $E_T^{\rm iso}/E_T < 0.1$ \cite{cdfe}, while 
D0 requires $E_T^{\rm iso}/E_T < 0.15$ \cite{d0e}.
The D0 isolation criterion for $\mu^{\pm}$ \cite{d0m}
requires $E_T^{\rm iso} < 2.5$~GeV,
where $E_T^{\rm iso}$ is the hadronic transverse energy in a hollow cone 
of inner radius $R=0.1$ and outer radius $R=0.4$.

The MSTW Group has analyzed the Tevatron Run~II data on the lepton charge
asymmetry in the context of his global fit of PDFs \cite{Martin:2009iq}.
Their QCD calculation of the lepton charge asymmetry uses the NNLO code
FEWZ \cite{Melnikov:2006di}, although the partonic cross sections 
are computed only up to NLO
(see Sect.~11.1 in Ref.~\cite{Martin:2009iq}). The main conclusions of the 
MSTW study are as follows \cite{Martin:2009iq}.
The CDF electron asymmetry (low-$E_T$ and high-$E_T$ bins) \cite{cdfe} and
D0 muon asymmetry \cite{d0m}
data are reasonably well fitted at NNLO (at high $y_l$, some evidence for a
systematic discrepancy between the NNLO fit and the data is reported).
The D0 electron asymmetry data \cite{d0e} are not included in the 
MSTW PDF fit: their inclusion in the global analysis
does not permit to obtain a good quality NNLO
fit, with significant tension between the D0 electron asymmetry data
and other 
data (DIS structure functions and low-mass DY production).

\begin{figure}[htb]
\begin{center}
\begin{tabular}{cc}
\includegraphics[width=0.46\textwidth]{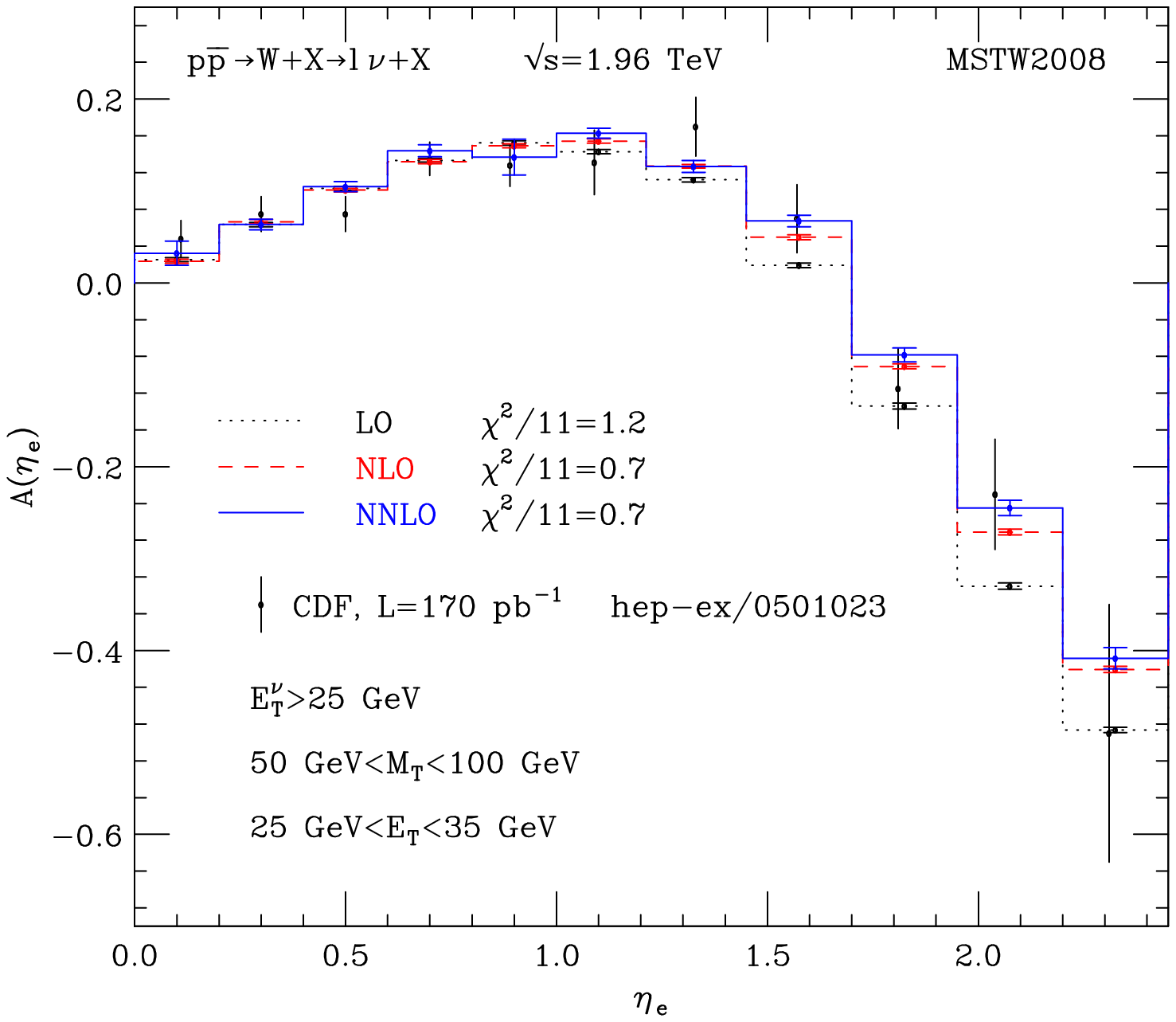} 
& \includegraphics[width=0.46\textwidth]{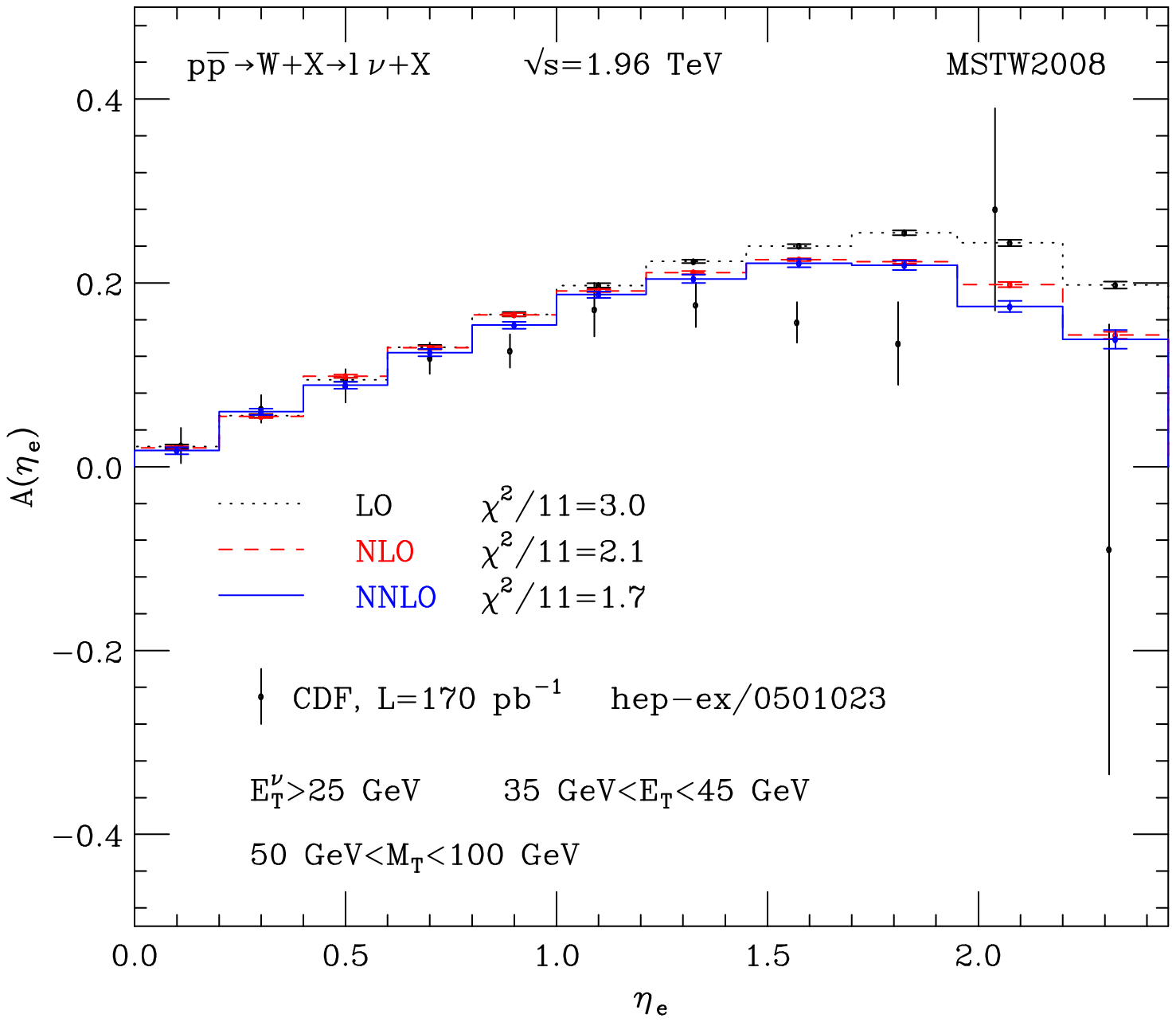}\\
\includegraphics[width=0.46\textwidth]{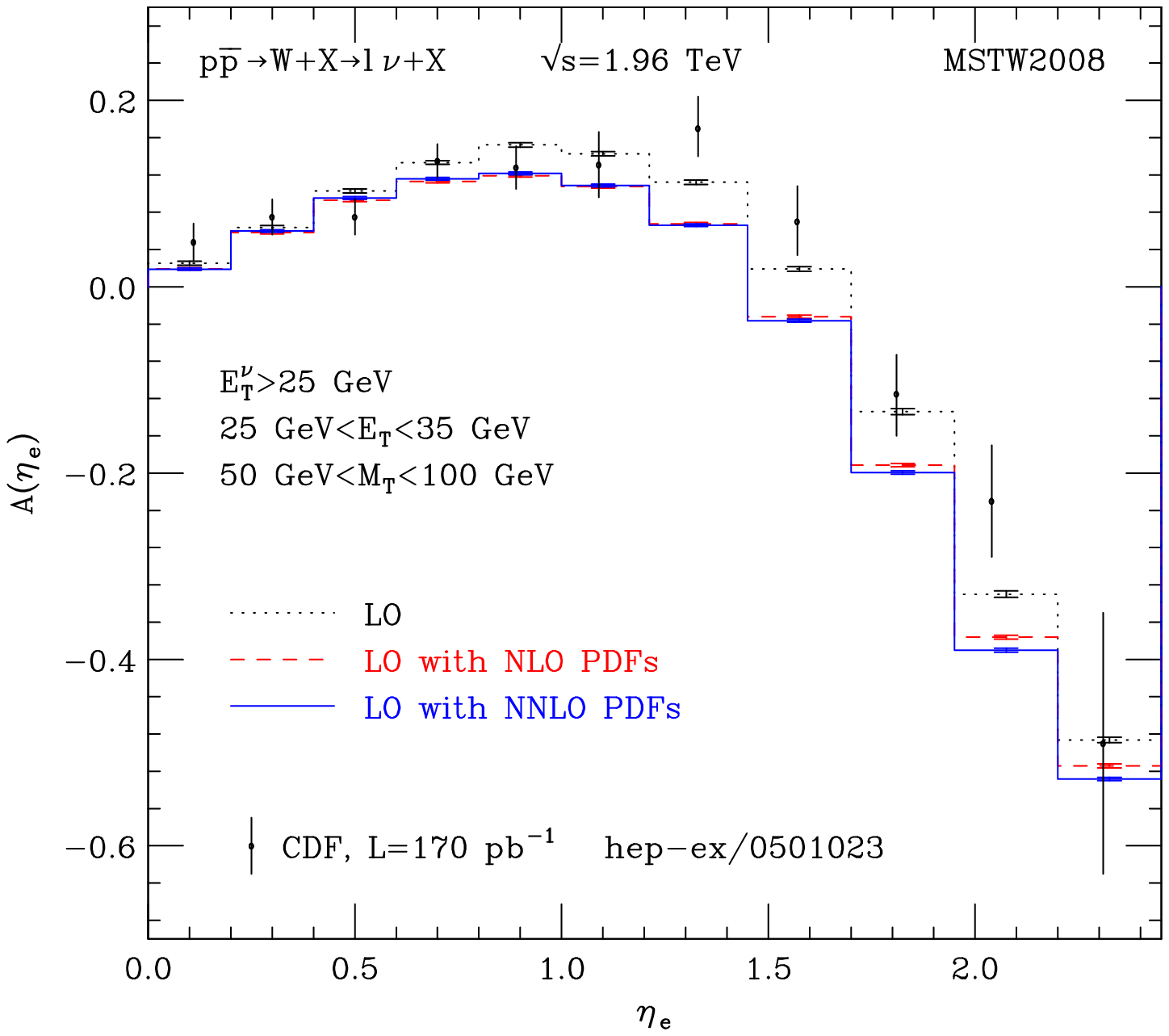} 
& \includegraphics[width=0.46\textwidth]{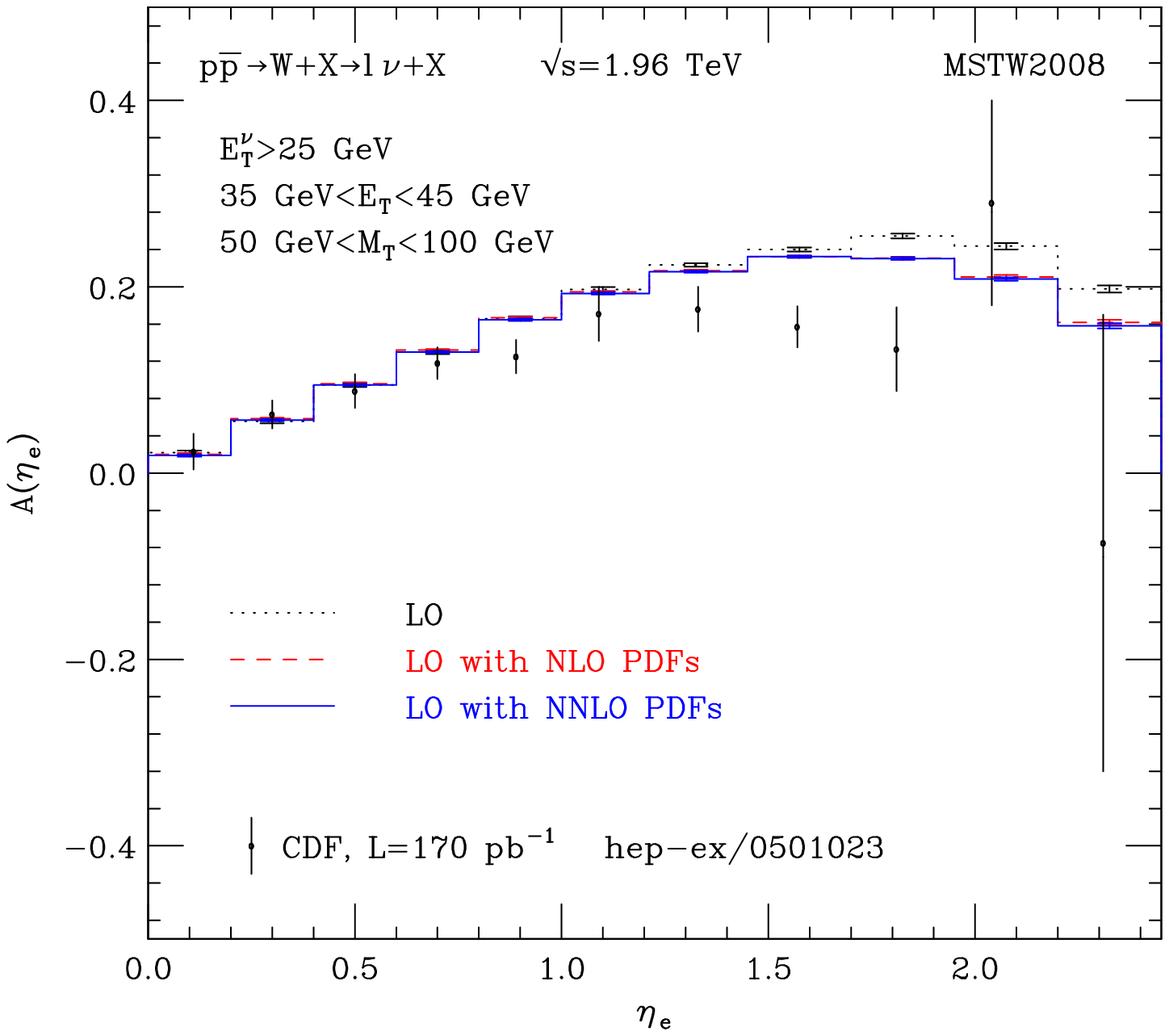}\\
\end{tabular}
\end{center}
\caption{\label{fig:asycdf}
{\em Electron charge asymmetry up to NNLO QCD compared to the CDF 
data of Ref.~\cite{cdfe}: a) lower $E_T$ bin; b) higher $E_T$ bin; c) lower
$E_T$ bin with LO partonic cross section; d) higher $E_T$ bin with LO partonic cross section.}}
\end{figure}

We first consider the electron charge asymmetry in the experimental 
configuration of the CDF data \cite{cdfe}, which cover the rapidity region
$|\eta_e| \leq 2.45$. We examine the low-$E_T$ and the high-$E_T$ bins 
(the same bins as in Figs.~\ref{fig:wasycdf} and \ref{fig:etacdf}), whose data
are included in the MSTW2008 fit. 
In Fig.~\ref{fig:asycdf} we report the CDF data\footnote{The size of the
rapidity bins is not explicitly reported in Table~I of Ref.~\cite{cdfe}. 
The bin size is (from correspondence with C.~Issever) $\Delta \eta_e=0.2$ 
in the central region $(|\eta_e| < 1.2)$ and $\Delta \eta_e=0.25$ 
in the forward region $(1.2 < |\eta_e| < 2.45)$.}
and present the results of our calculations.
 
The results of the LO, NLO and NNLO calculations in the low-$E_T$ and 
high-$E_T$ bins are shown in Figs.~\ref{fig:asycdf}(a) and \ref{fig:asycdf}(b),
respectively. At small values of $\eta_e$, the radiative corrections lead to
little effects. In the low-$E_T$ bin (Fig.~\ref{fig:asycdf}(a)),
as $\eta_e$ increases, both the NLO and NNLO effects slightly increase the 
value of the asymmetry; the NNLO effect is smaller than the experimental 
uncertainties. These small and positive radiative corrections are   
consistent with the corresponding shift of the lepton rapidity 
distribution observed in the left-side plot of Fig.~\ref{fig:etacdf}.
In the high-$E_T$ bin (Fig.~\ref{fig:asycdf}(b)),
as $\eta_e$ increases, the effect of the QCD radiative corrections is negative and 
the asymmetry slightly decreases; the NNLO effect is definitely smaller than
the experimental uncertainties. These negative contributions of the radiative
corrections are also consistent with the effects already seen in the
right-side plot of Fig.~\ref{fig:etacdf}.

\begin{figure}[ht]
\begin{center}
\begin{tabular}{cc}
\includegraphics[width=0.47\textwidth]{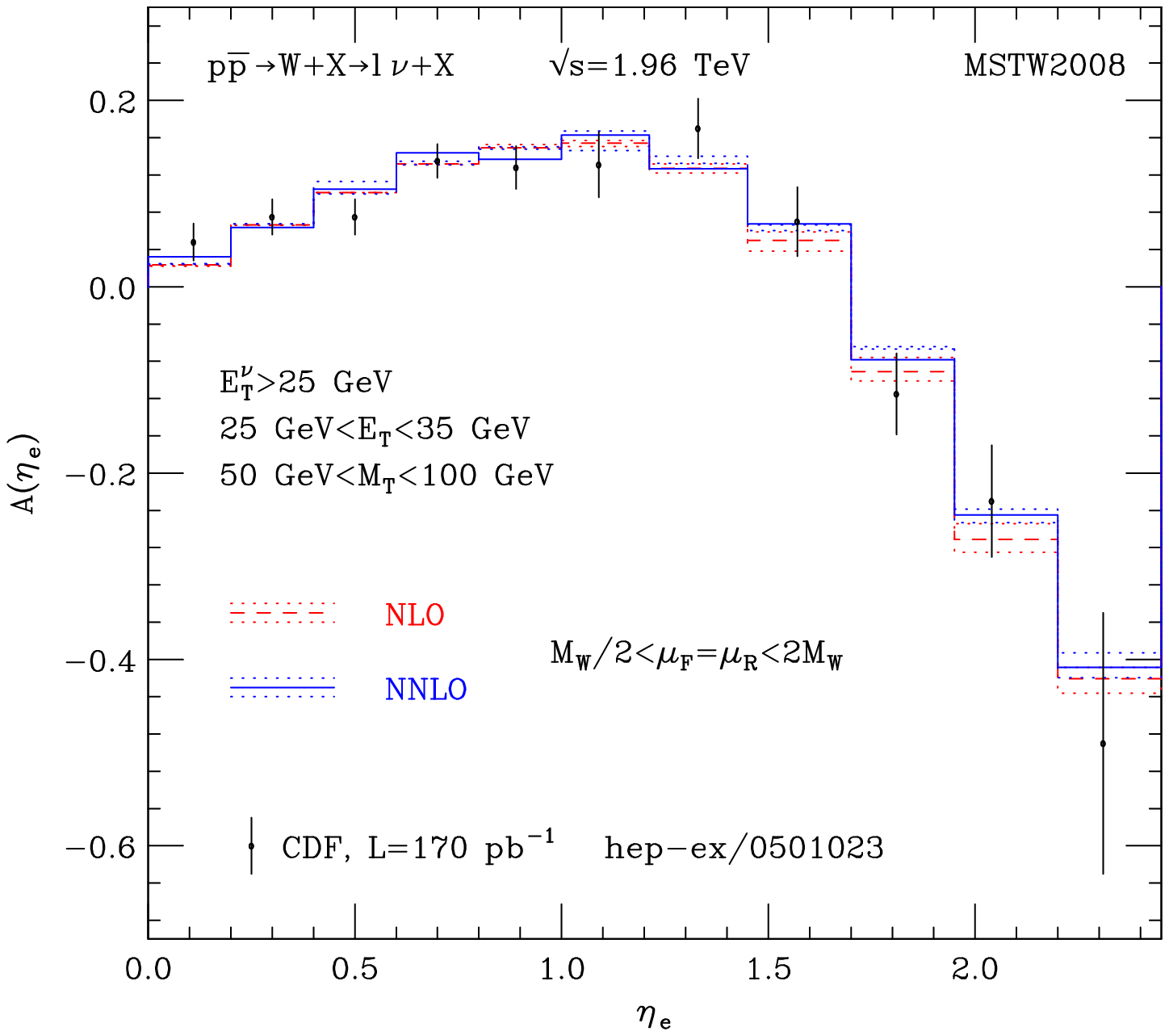} & 
\includegraphics[width=0.47\textwidth]{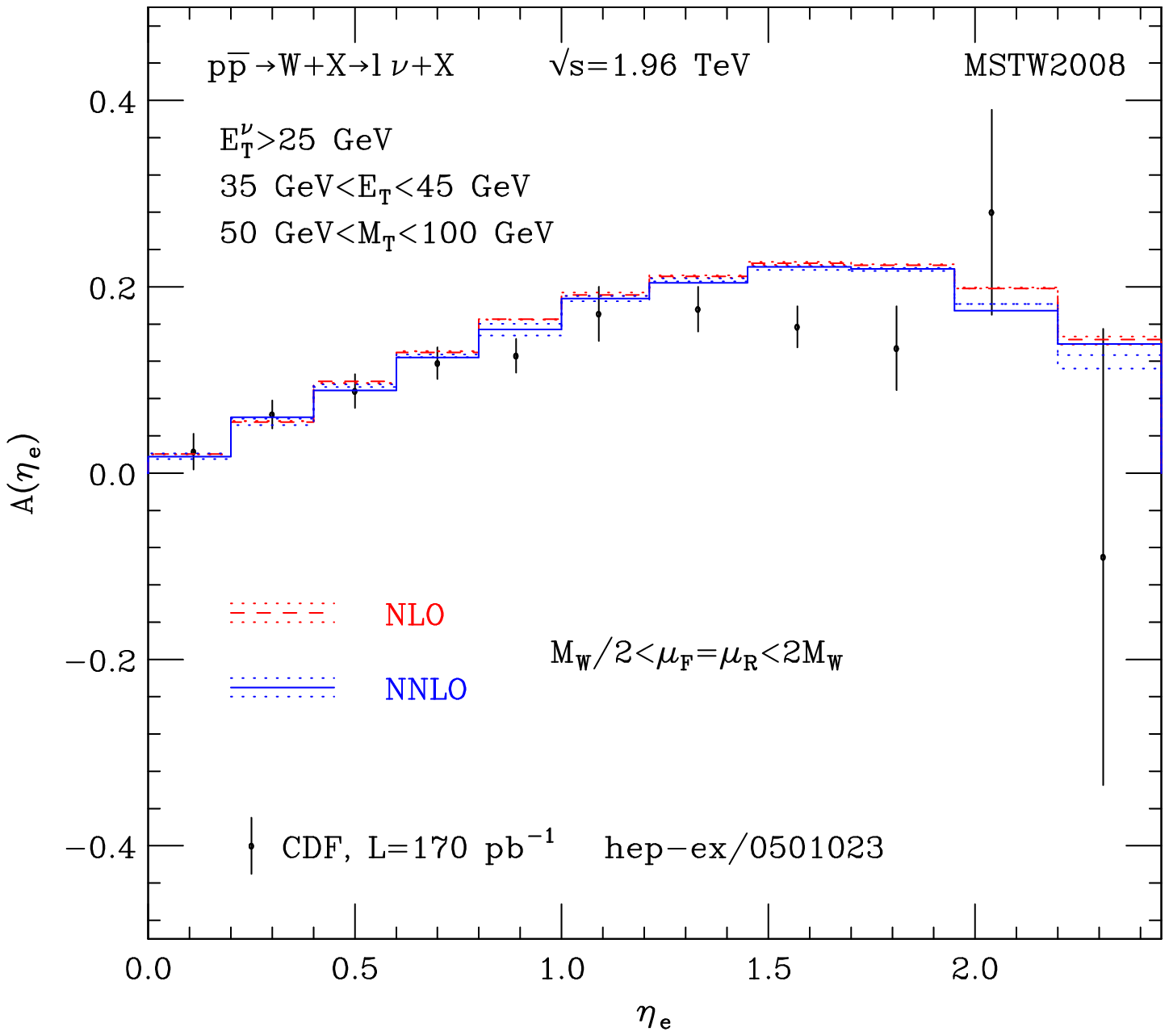}\\
\end{tabular}
\end{center}
\caption{\label{fig:asycdfscale}
{\em Scale uncertainty bands for the electron asymmetry: low-$E_T$ bin (left),
high-$E_T$ bin (right).}}
\end{figure}

In Figs.~\ref{fig:asycdf}(c) and \ref{fig:asycdf}(d) we present the results
obtained by convolution of the LO, NLO and NNLO PDFs with the LO partonic cross
sections. Comparing the histograms in Fig.~\ref{fig:asycdf}(a) with the
corresponding (in the sense of corresponding PDFs)
histograms in Fig.~\ref{fig:asycdf}(c), we observe a very different pattern of
perturbative corrections: in the low-$E_T$ bin, the radiative corrections to the
lepton asymmetry are definitely dominated by the radiative contribution to the
partonic cross sections. In the high-$E_T$ bin we observe a different behaviour.
The histograms in Fig.~\ref{fig:asycdf}(b) are qualitatively similar to the
corresponding histograms in Fig.~\ref{fig:asycdf}(d), and the NLO and NNLO
effects on the asymmetry are thus mostly driven by the corresponding PDFs.

We add some overall qualitative comments on the results in Fig.~\ref{fig:asycdf} 
and on
their comparison with the corresponding results for the $W$ charge asymmetry
(see Fig.~\ref{fig:asyW-nocuts}). 
Owing to the invariance under CP transformations, 
in the following comments we always refer to the rapidity region $y > 0$.
In $p{\bar p}$ collisions,
the lepton asymmetry is definitely different 
from the $W$ asymmetry. The lepton
asymmetry is smaller than the $W$ asymmetry and the difference increases 
as the rapidity increases.
The difference is mostly due to the LO effect
produced by the $V-A$ structure of the EW couplings in the underlying partonic
process (see Eq.~(\ref{stardist})).
By increasing the lepton $E_T$, the difference is reduced because of the 
increased LO kinematical correlation between the $W$ and the lepton
(see Eq.~(\ref{loraddif})). The QCD radiative corrections to the charge asymmetry
are small in both the $W$ and lepton cases. Their effect is relatively
larger in the lepton case.
Since the lepton charge asymmetry depends on the EW asymmetry generated by the
partonic subprocesses, it is more sensitive to the QCD radiative corrections
to these partonic subprocesses. As previously mentioned, at low values of $E_T$,
QCD radiation dynamically reduces the partonic EW asymmetry and, therefore, the
value of the lepton charge asymmetry slightly increases with respect to its
LO result (i.e. the difference between the lepton and $W$ asymmetry tends to be 
slightly reduced). 
This dynamical effect is 
compensated by a QCD kinematical effect (see Eq.~(\ref{hokin})) that weakens
the LO kinematical correlation between the $W$ and the lepton at high values
of $E_T$. 
By increasing $E_T$,  the dynamical effect is eventually over-compensated
by the kinematical effect and, therefore, the
value of the lepton charge asymmetry slightly decreases with respect to its
LO result (i.e. the difference between the lepton and $W$ asymmetry tends to be 
slightly increased).

In Fig.~\ref{fig:asycdfscale} we consider the scale dependence of the electron
charge asymmetry at NLO and NNLO. The scale dependence bands 
are obtained by fixing $\mu_F=\mu_R=\mu$ and considering the values 
$\mu=M_W/2$ and $2M_W$
in the calculations at NLO and NNLO. 
The scale dependence is small at
both NLO and NNLO
(at NNLO, the scale dependence is comparable to the numerical
errors of our NNLO Monte Carlo computation). 
In particular, the NLO and NNLO bands tend to overlap
in both the low-$E_T$ and high-$E_T$ bins.

The values of $\chi^2$ from the comparison of the CDF electron data with the 
QCD
calculations are reported in Figs.~\ref{fig:asycdf}(a) and \ref{fig:asycdf}(b).
The data are well fitted by the MRSTW2008 NNLO partons, especially in the 
low-$E_T$ bin. Considering the high-$E_T$ bin, we see that the QCD results tend
to overshoot the data in the intermediate region of lepton rapidities; the value
of $\chi^2$ is dominated by the contributions in the region 
$1.2 \ltap \eta_e \ltap 1.9$. The NNLO effects reduce the lepton asymmetry at
intermediate and large rapidities, but their impact is not yet quantitatively
relevant in comparison with the size of the experimental uncertainties.

We now move to briefly consider the lepton selection cuts used in the D0
measurement of the muon charge asymmetry \cite{d0m}. 
\begin{figure}[htb]
\begin{center}
\begin{tabular}{c}
\epsfxsize=12truecm
\epsffile{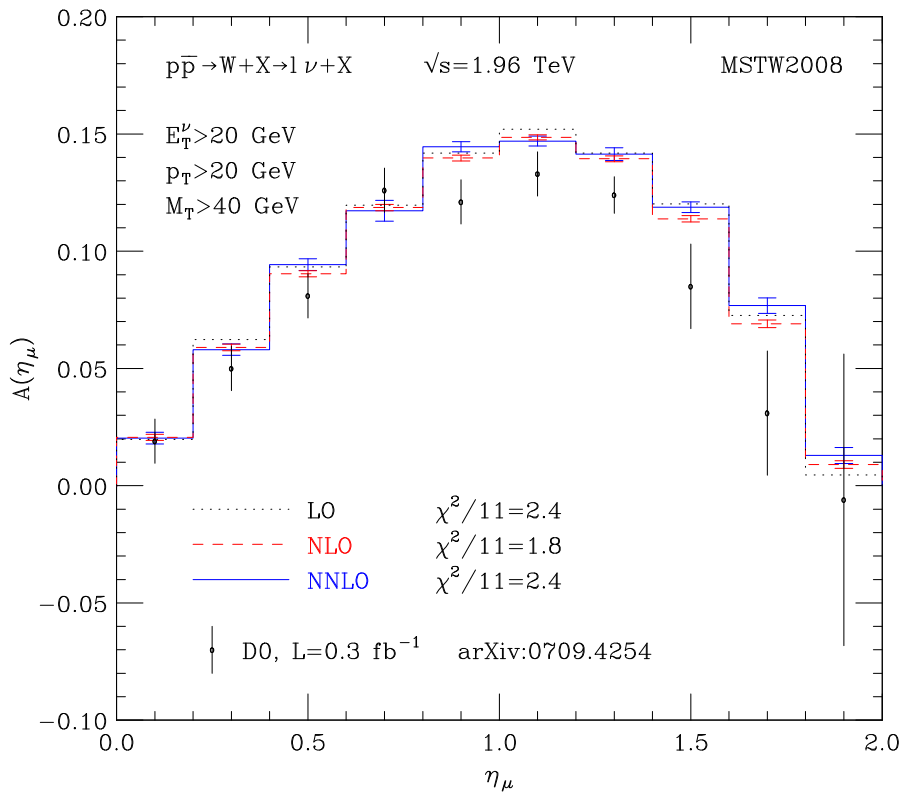}\\
\end{tabular}
\end{center}
\caption{\label{fig:asyd0muon}
{\em Lepton charge asymmetry up to NNLO QCD compared to the D0 data of 
Ref.~\cite{d0m}.}}
\end{figure}
The D0 data cover the
rapidity region where $|\eta_\mu| \leq 2$. 
In Fig.~\ref{fig:asyd0muon} we display 
the data and present the results of our corresponding calculation at LO, NLO and
NNLO. Note that the lepton $E_T$ is selected in the entire region where 
$E_T > 20$~GeV. This region includes both the low-$E_T$ and high-$E_T$ regions
that are examined separately in Fig.~\ref{fig:asycdf}. The main features of the
results in Fig.~\ref{fig:asyd0muon} are thus intermediate between those of 
Figs.~\ref{fig:asycdf}(a) and \ref{fig:asycdf}(b). The QCD radiative corrections
are small. In particular, at high $\eta_\mu$, the small NNLO effect is positive
and slightly increases the NLO charge asymmetry.

In Fig.~\ref{fig:asyd0muon} we can see that the size of the QCD radiative
corrections is certainly smaller than the experimental errors of the D0 data.
The values of $\chi^2$ from our QCD calculations are reported in the figure.
The D0 data are reasonably well fitted at NNLO by the MSTW2008 PDFs, and
the conclusions of the MSTW analysis \cite{Martin:2009iq}
are unchanged by the effect of the NNLO
corrections to the partonic cross sections. 
A systematic difference between the data and the QCD result appears starting
from the region of intermediate rapidities $(\eta_\mu \gtap 0.9)$.

\begin{figure}[ht]
\begin{center}
\begin{tabular}{cc}
\includegraphics[width=0.48\textwidth]{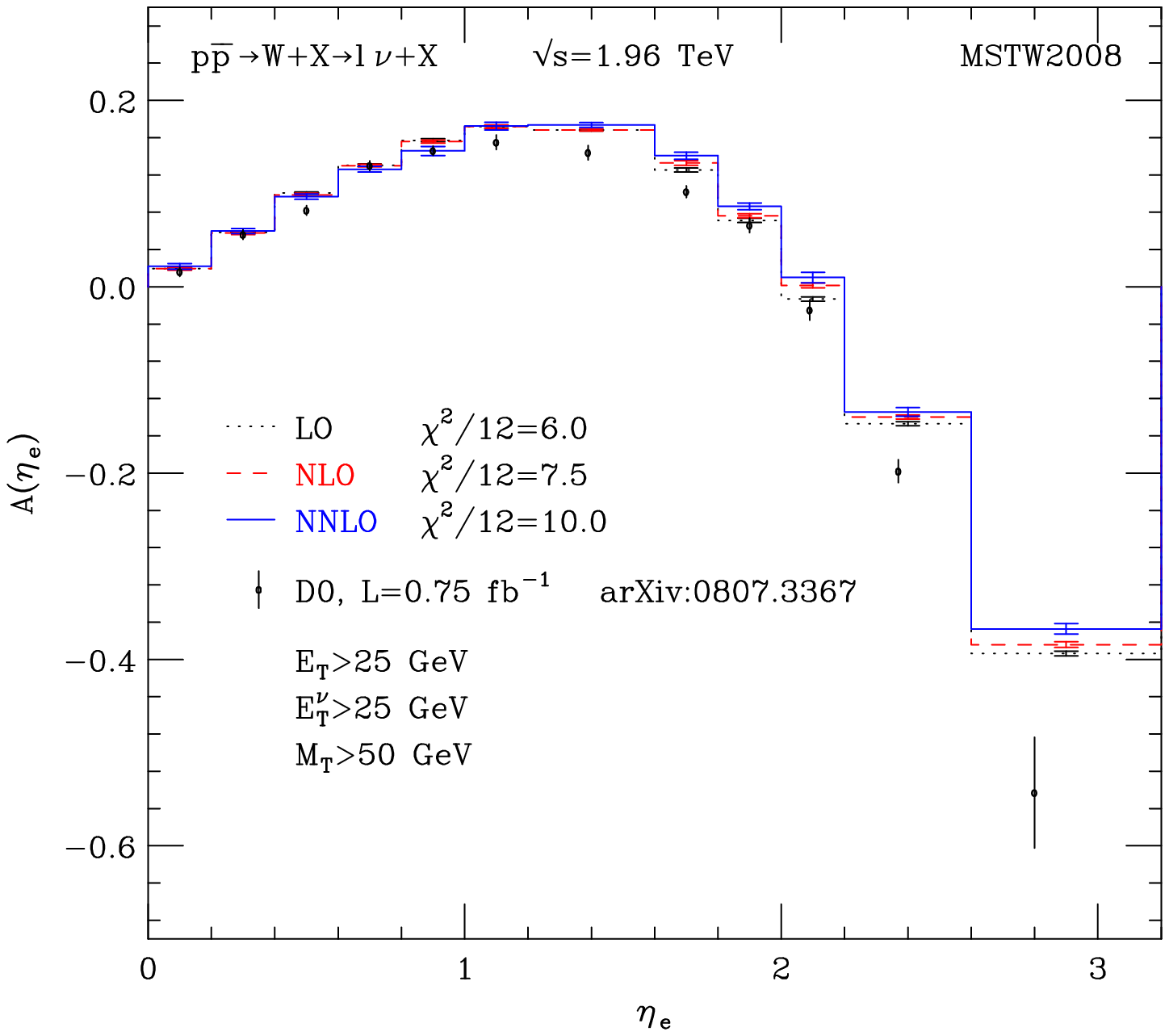} & 
\includegraphics[width=0.48\textwidth]{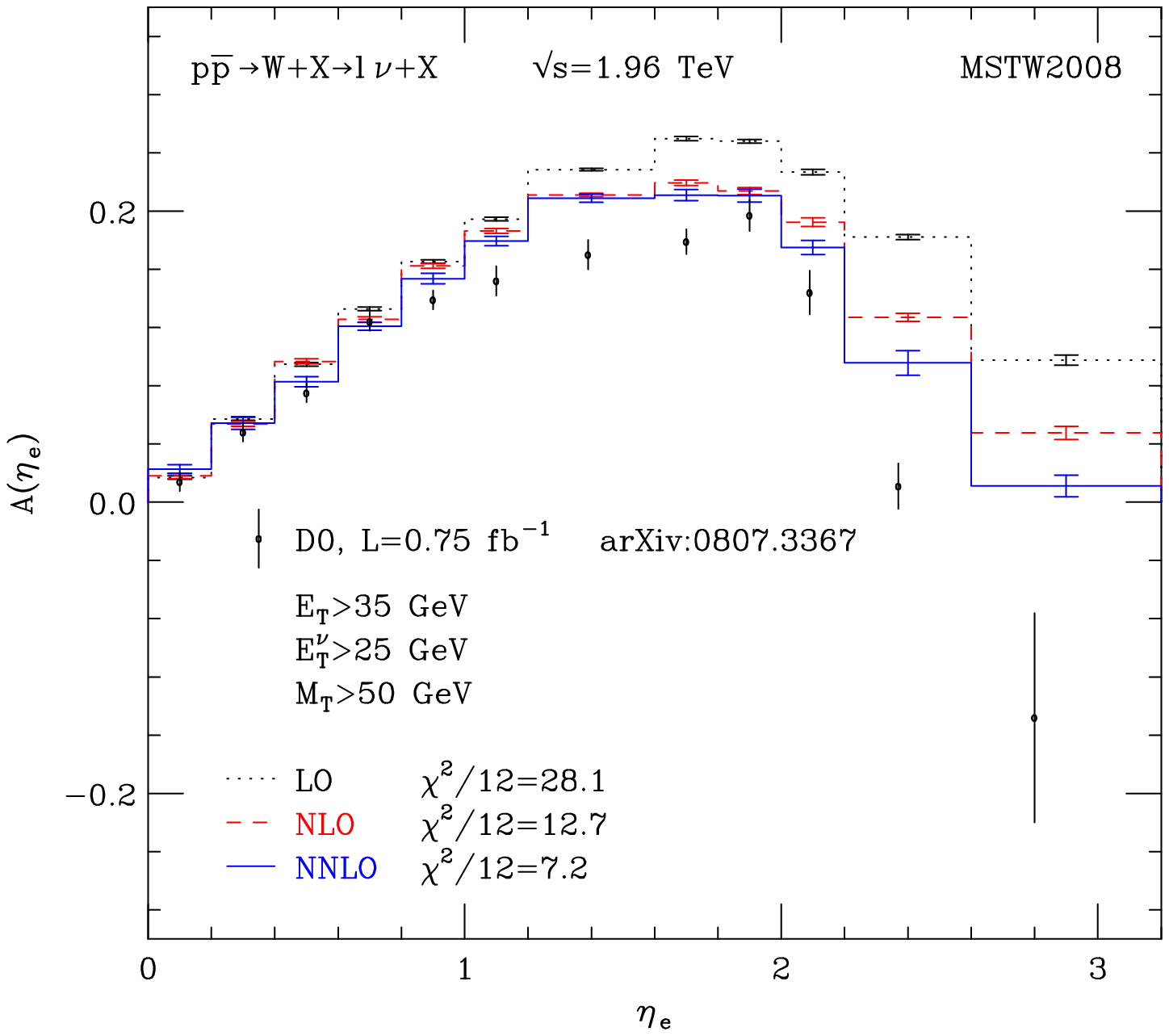}\\
\end{tabular}
\end{center}
\caption{\label{fig:asyD0}
{\em Electron charge asymmetry up to NNLO QCD compared to the 
D0 data of Ref.~\cite{d0e}: $E_T>25$ GeV (left), $E_T>35$ GeV (right).}}
\end{figure}

We finally consider the D0 electron charge asymmetry \cite{d0e}. The D0
measurement extends to high rapidities, $|\eta_e| \leq 3.2$. 
We recall that
the D0 data are not included in the PDF fit of the MSTW 
Group \cite{Martin:2009iq}.
We examine two regions of $E_T$: $E_T>25$~GeV (wide $E_T$ region) and 
$E_T>35$~GeV (high-$E_T$ region).

In the left panel of Fig.~\ref{fig:asyD0}, we consider the wide $E_T$ region:
we display the D0 electron data and present the corresponding QCD results at 
LO, NLO and NNLO. The QCD results behave similarly to those in 
Fig.~\ref{fig:asyd0muon}. This is not unexpected since, at the LO and within
the NWA (see Eq.~(\ref{loconstr})), the relevant lepton selection cuts
used in these two cases are quite similar: we have $E_T \gtap 25$~GeV and
$E_T \gtap 20$~GeV in Fig.~\ref{fig:asyD0} (left) and Fig.~\ref{fig:asyd0muon},
respectively. At high values of the electron rapidity, 
the effect of the NLO and NNLO corrections is positive and increases the
deviation of the QCD results from the D0 data.
We see that (consistently with the comments in Ref.~\cite{Martin:2009iq})
the agreement between the QCD calculations and the data is poor. This is
quantitatively confirmed by the values of $\chi^2$, which are reported in 
Fig.~\ref{fig:asyD0} (left). We note that the value of $\chi^2$ increases
in going from LO to NLO and to NNLO.

In the right panel of Fig.~\ref{fig:asyD0}, we consider the high-$E_T$ region.
In this region (as in the case of the high-$E_T$ bin in Fig.~\ref{fig:asycdf})
the effect of the NLO and NNLO corrections is negative as the lepton
rapidity increases toward high values. This effect reduces the difference
between the QCD calculations and the D0 data,  although a 
substantial disagreement between them still persists at the NNLO.
The value of $\chi^2$ is reported in Fig.~\ref{fig:asyD0} (right);
it decreases in going from LO to NLO and to NNLO.

\begin{figure}[ht]
\begin{center}
\begin{tabular}{cc}
\includegraphics[width=0.48\textwidth]{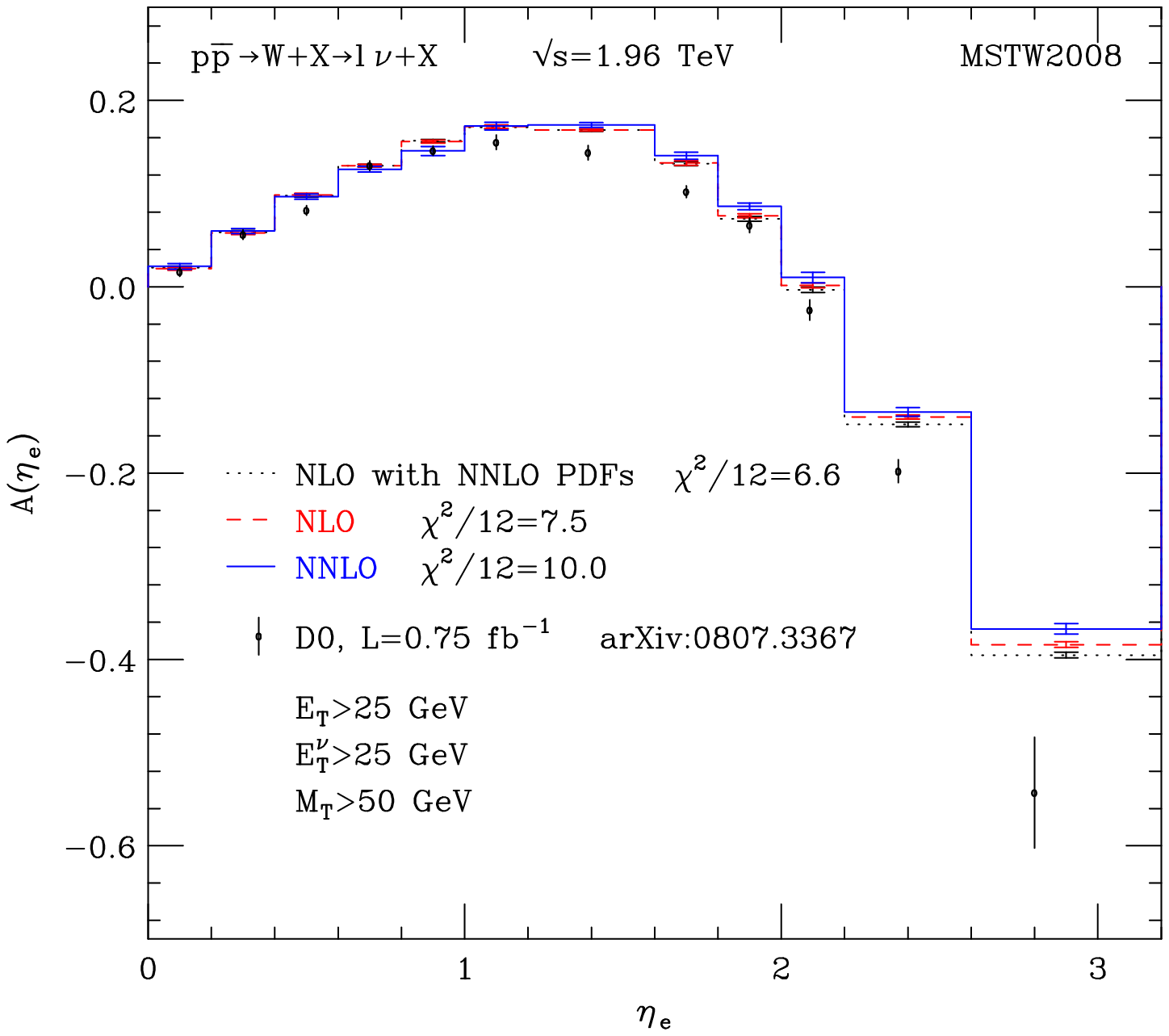} & 
\includegraphics[width=0.48\textwidth]{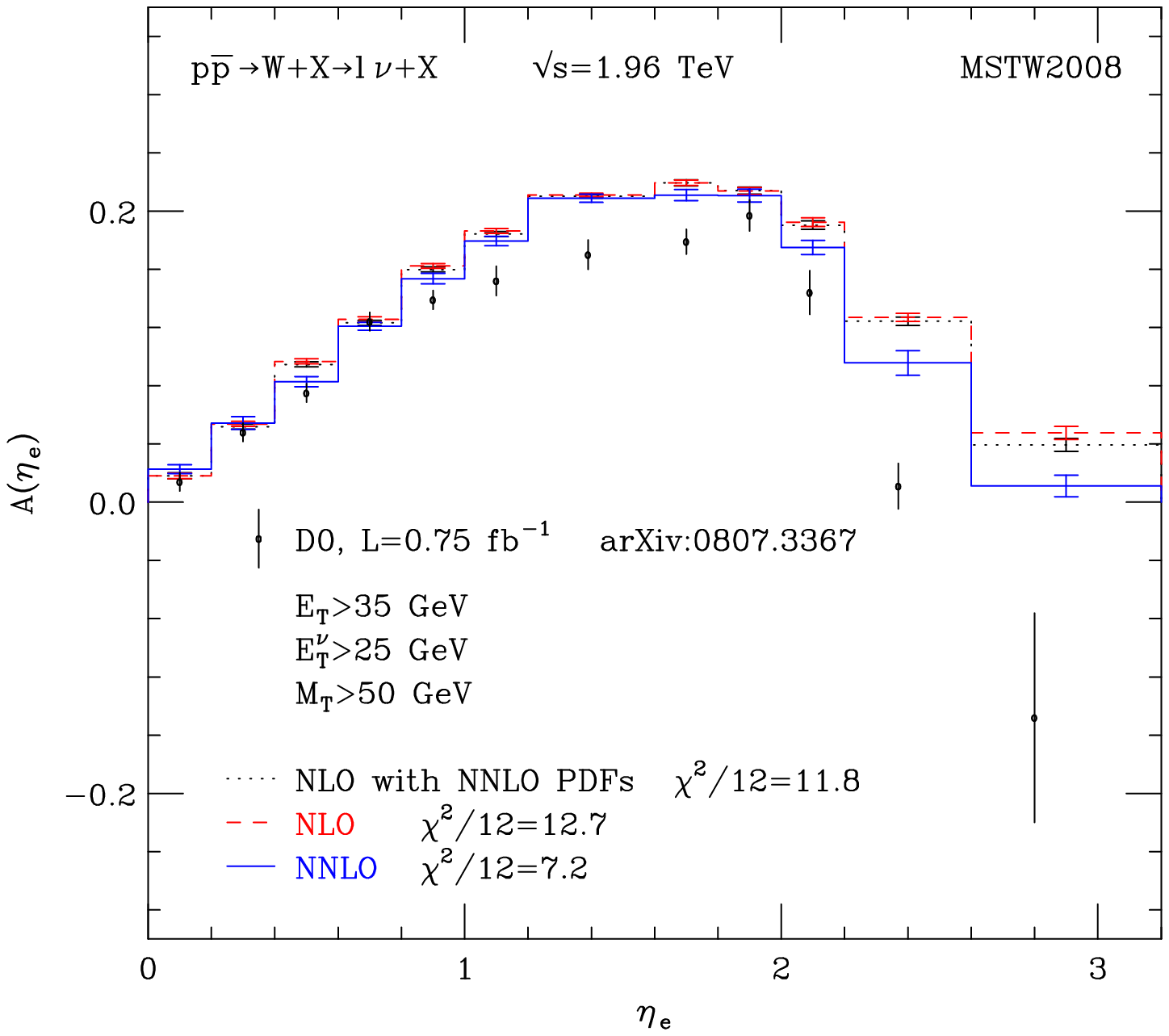}\\
\end{tabular}
\end{center}
\caption{\label{fig:asyD0b}
{\em Electron charge asymmetry up to NNLO and D0 
data of Ref.~\cite{d0e}: $E_T>25$ GeV (left), $E_T>35$ GeV (right). The NNLO prediction (solid) is compared to the NLO result (dashed), and to the NLO result with NNLO PDFs (dots).}}
\end{figure}
\begin{figure}[htb]
\begin{center}
\begin{tabular}{cc}
\includegraphics[width=0.47\textwidth]{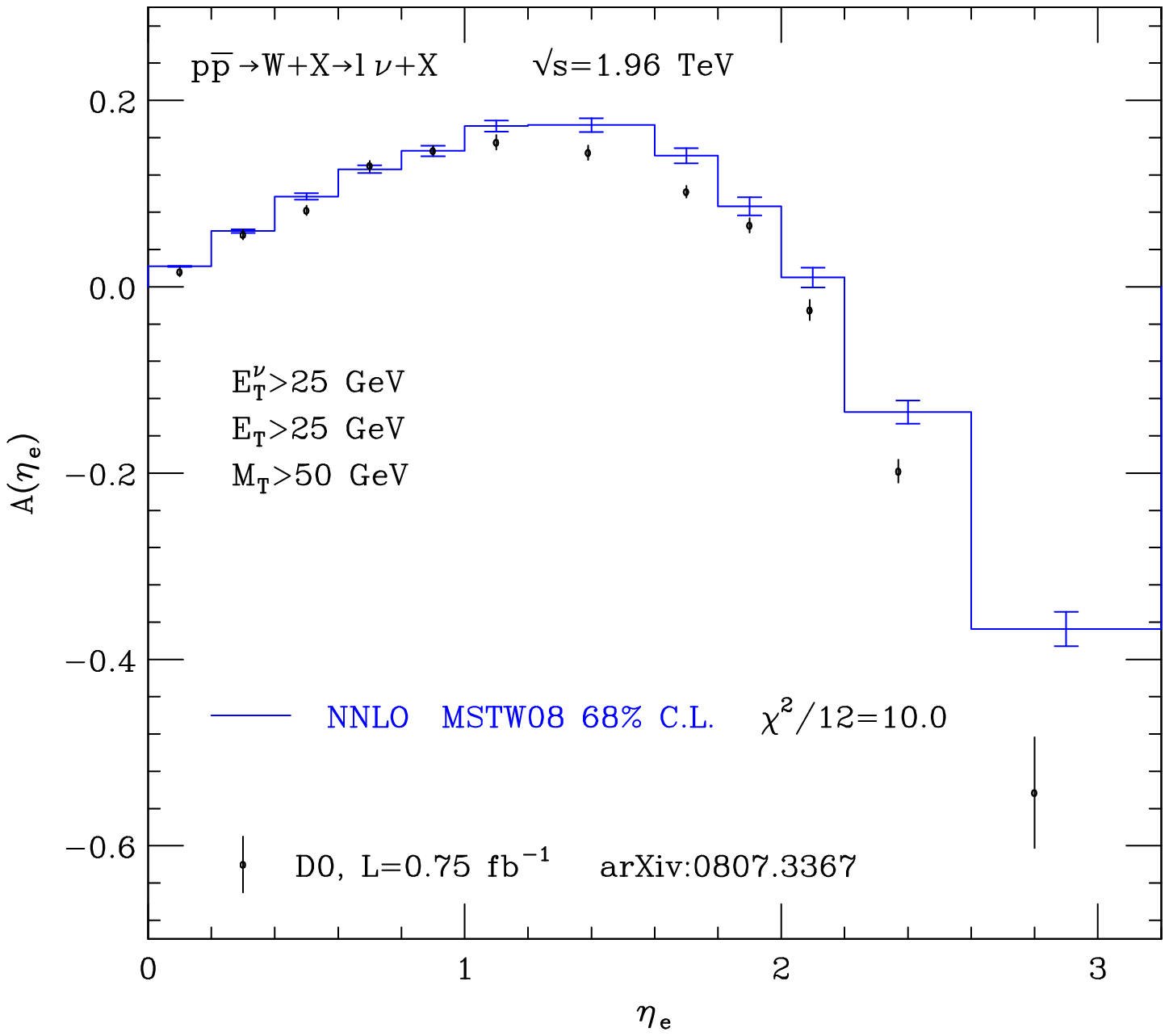} & 
\includegraphics[width=0.47\textwidth]{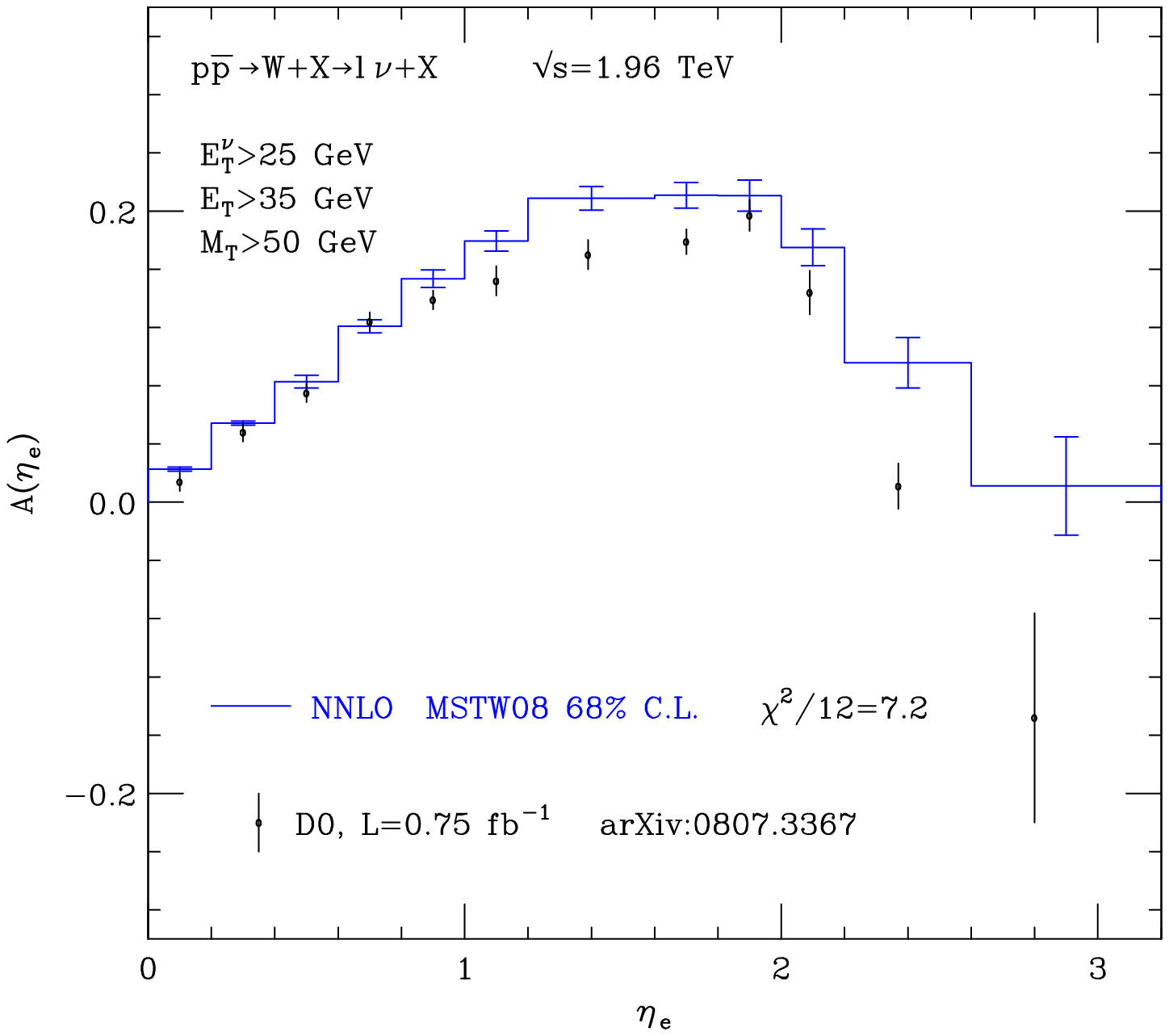}\\
\end{tabular}
\end{center}
\caption{\label{fig:asyD0err}
{\em Electron charge asymmetry at NNLO with PDF uncertainties: $E_T>25$ GeV
(left), $E_T>35$ GeV (right).}}
\end{figure}

In Fig.~\ref{fig:asyD0b} we present the result of the QCD calculation 
performed
by convoluting the NLO partonic cross sections with the NNLO PDFs.
This result is compared with the customary NLO and NNLO results (the same
results as in Fig.~\ref{fig:asyD0}). The comparison shows that the NNLO
corrections to the partonic cross sections are not negligible. Their effect
is quantified by the variation of the $\chi^2$ values, which are reported in 
the plots of Fig.~\ref{fig:asyD0b}. In the region 
$E_T>25$~GeV (plot on the left) the NNLO partonic corrections tend to increase
the value of the asymmetry, while in the region 
$E_T>35$~GeV (plot on the right) the 
corrections tend to decrease the asymmetry. This different behaviour in
different $E_T$ regions
is consistent with our physical expectation about the $E_T$ dependence
of the QCD radiative corrections (see, e.g., our related comments on the CDF
electron asymmetry).

We have computed the scale dependence of the NLO and NNLO results presented 
in Fig.~\ref{fig:asyD0}. This scale dependence is very similar to that 
documented in Fig.~\ref{fig:asycdfscale}. In particular, the scale variations
produce effects that are quantitatively smaller than the size of the
PDF errors, which are considered below.

The NNLO results (solid histograms) of Fig.~\ref{fig:asyD0} are reported in 
Fig.~\ref{fig:asyD0err} by including the PDF errors (68\% C.L.) from the 
MSTW2008 parton densities \cite{Martin:2009iq}.
In most of the rapidity bins, the PDF errors are 
comparable to (or, larger than)
the D0 experimental errors. The inclusion of the PDF errors thus reduces
the differences between the NNLO MSTW2008 results and the D0 electron data,
although a substantial disagreement still remains.

\begin{figure}[htb]
\begin{center}
\begin{tabular}{cc}
\includegraphics[width=0.47\textwidth]{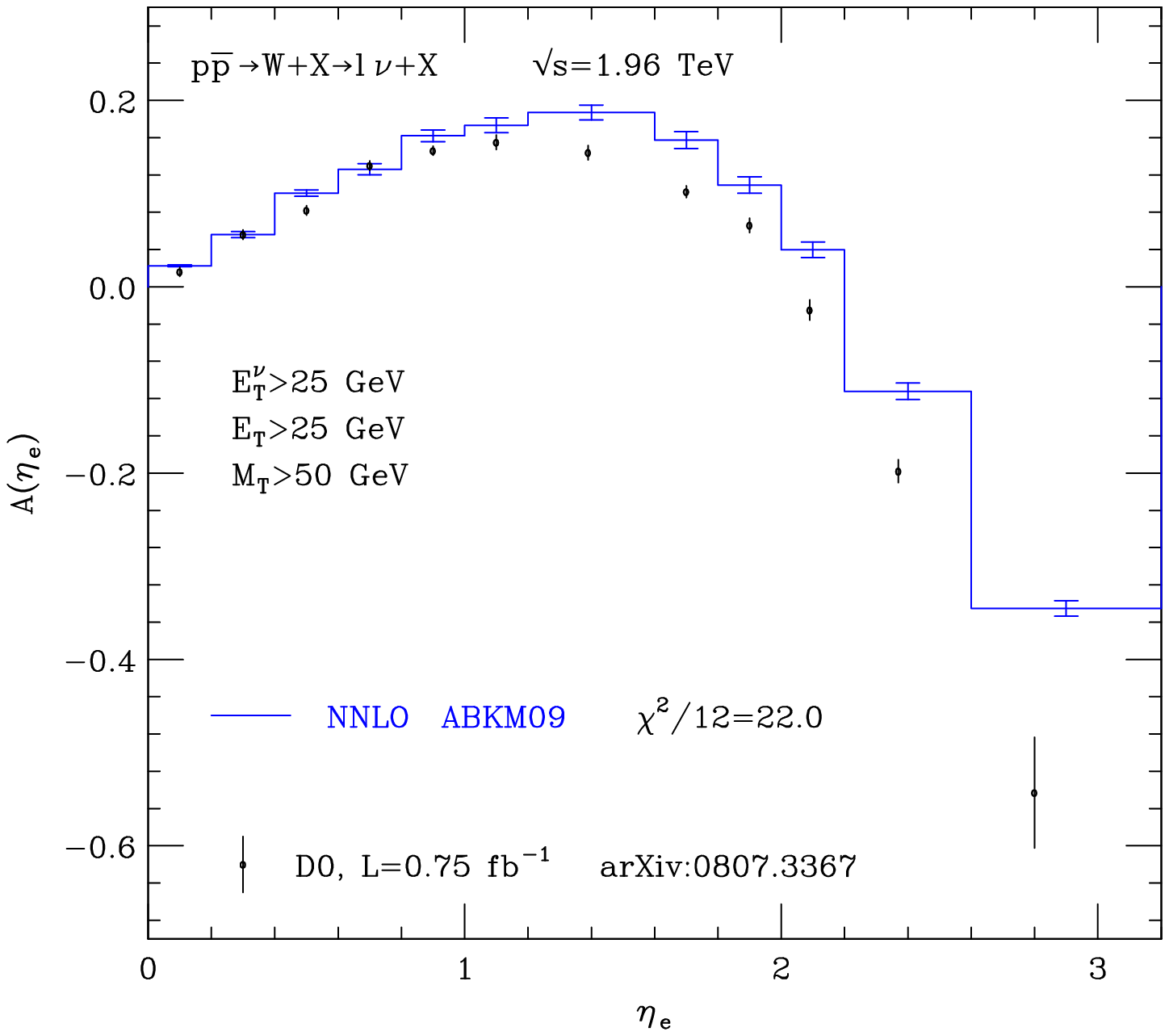} & 
\includegraphics[width=0.47\textwidth]{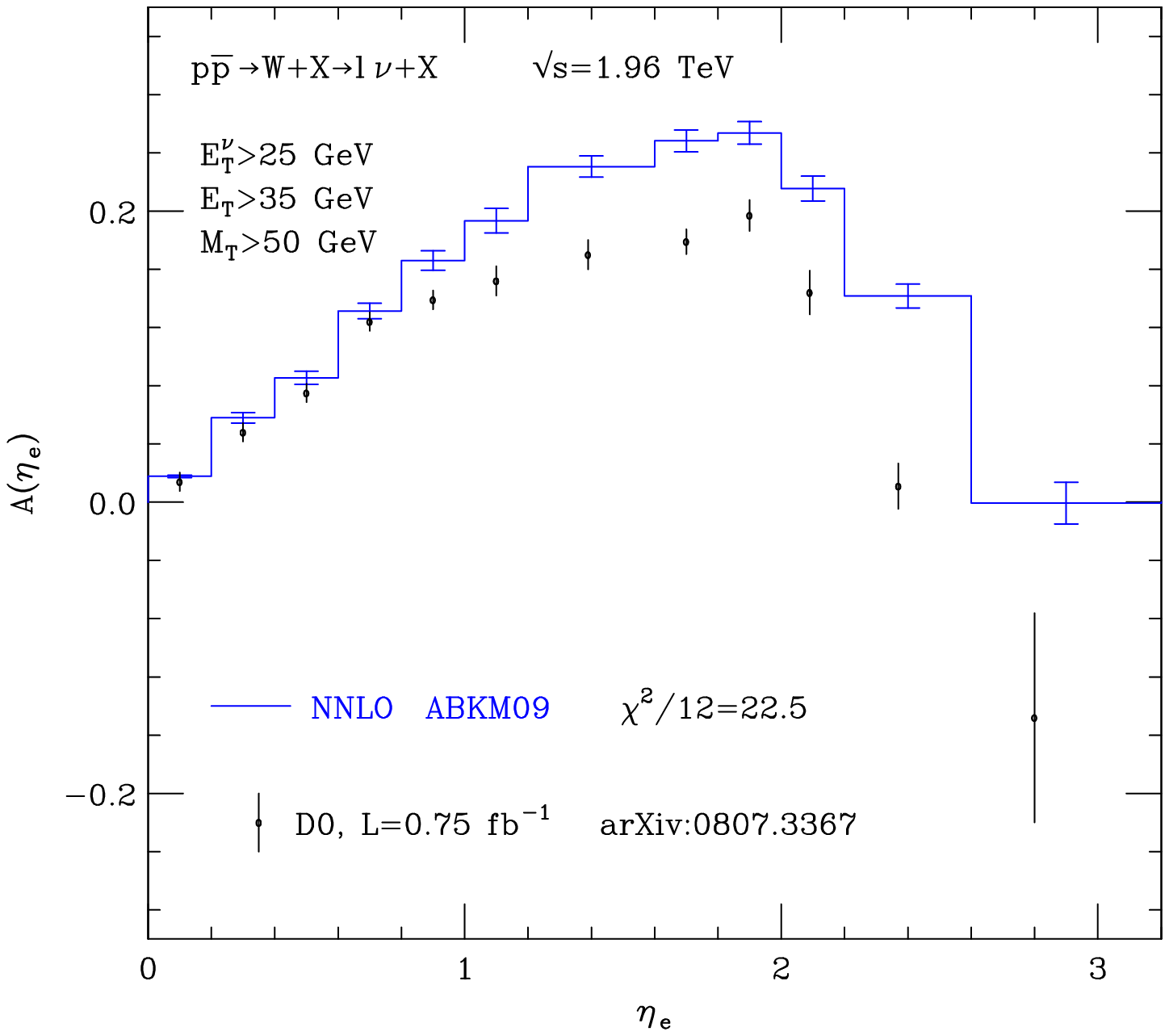}\\
\end{tabular}
\end{center}
\caption{\label{fig:asyD0ale}
{\em Electron charge asymmetry at NNLO with the PDFs of
Ref.~\cite{Alekhin:2009ni}: $E_T>25$ GeV (left), $E_T>35$ GeV (right).}}
\end{figure}

In Fig.~\ref{fig:asyD0ale} we present the NNLO electron asymmetry
computed with the ABKM09 set of PDFs \cite{Alekhin:2009ni}, 
and we include the corresponding PDF errors. These PDF errors are slightly
smaller than the PDF errors of the MSTW2008 set. In the case of the
inclusive-$W$ asymmetry, the ABKM09 result tends to overshoot the 
MSTW2008 result (see Figs.~\ref{fig:wasycdfaljr} and \ref{fig:wmstw})
in the rapidity region where $0.5 \ltap y_W \ltap 2.3$.
This effect appears also in the case of the electron asymmetry, as can be seen
by comparing the results in Fig.~\ref{fig:asyD0ale} with those in 
Fig.~\ref{fig:asyD0err}.
In the region $E_T>25$~GeV (plots on the left), the ABKM09 result
evidently overshoots the MSTW2008 result at large rapidities $(|\eta_e| \gtap
1)$. The overshooting effect is even more evident (with the exception of the
last rapidity bin at high $|\eta_e|$) in the region where $E_T>35$~GeV 
(plots on the right). The use of the ABKM09 parton densities does not reduce
the disagreement with the D0 data, and the corresponding values of
$\chi^2$ (see Fig.~\ref{fig:asyD0ale}) are definitely larger than the 
MSTW2008 values.

\begin{figure}[htb]
\begin{center}
\begin{tabular}{cc}
\includegraphics[width=0.47\textwidth]{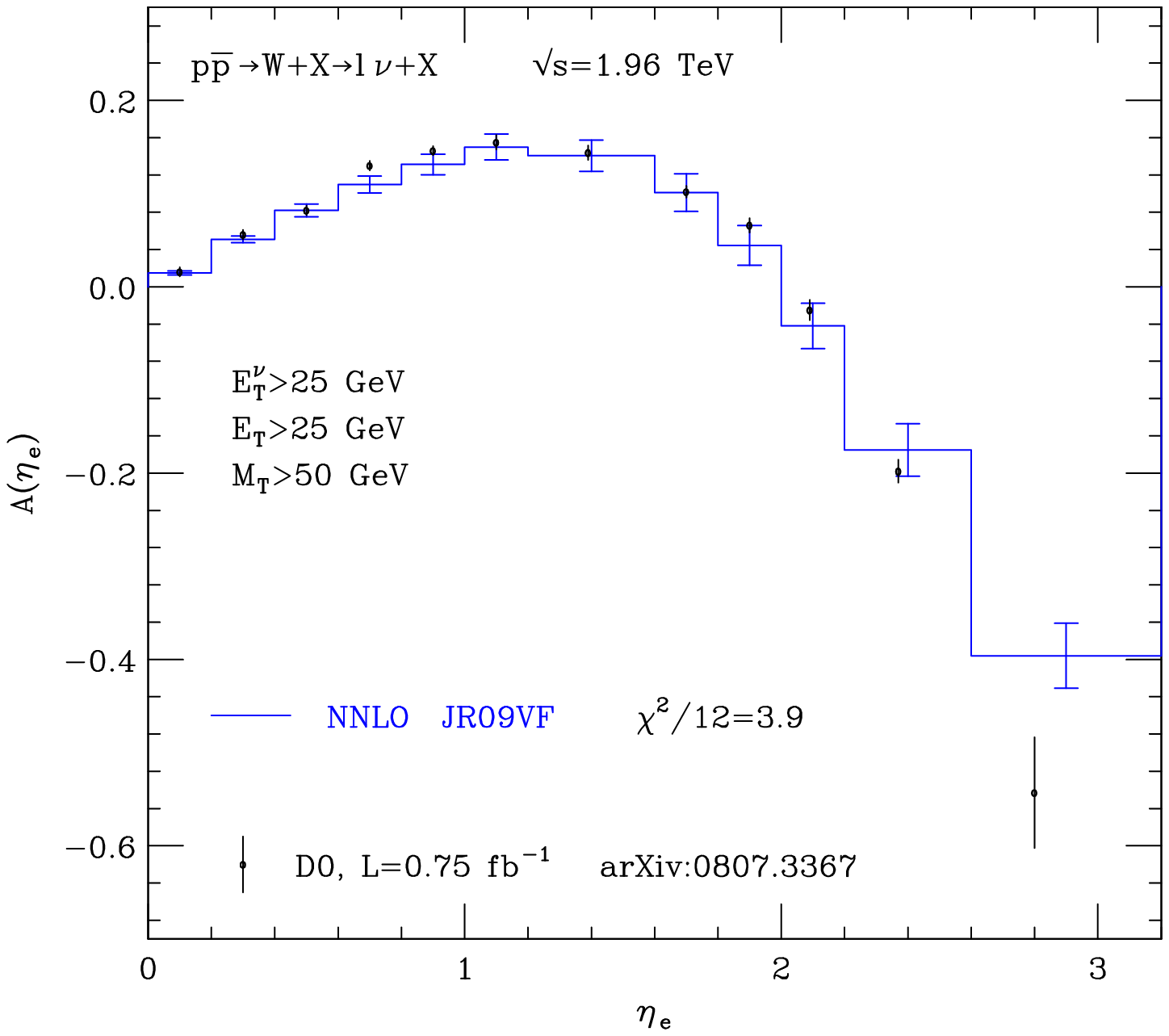} & 
\includegraphics[width=0.47\textwidth]{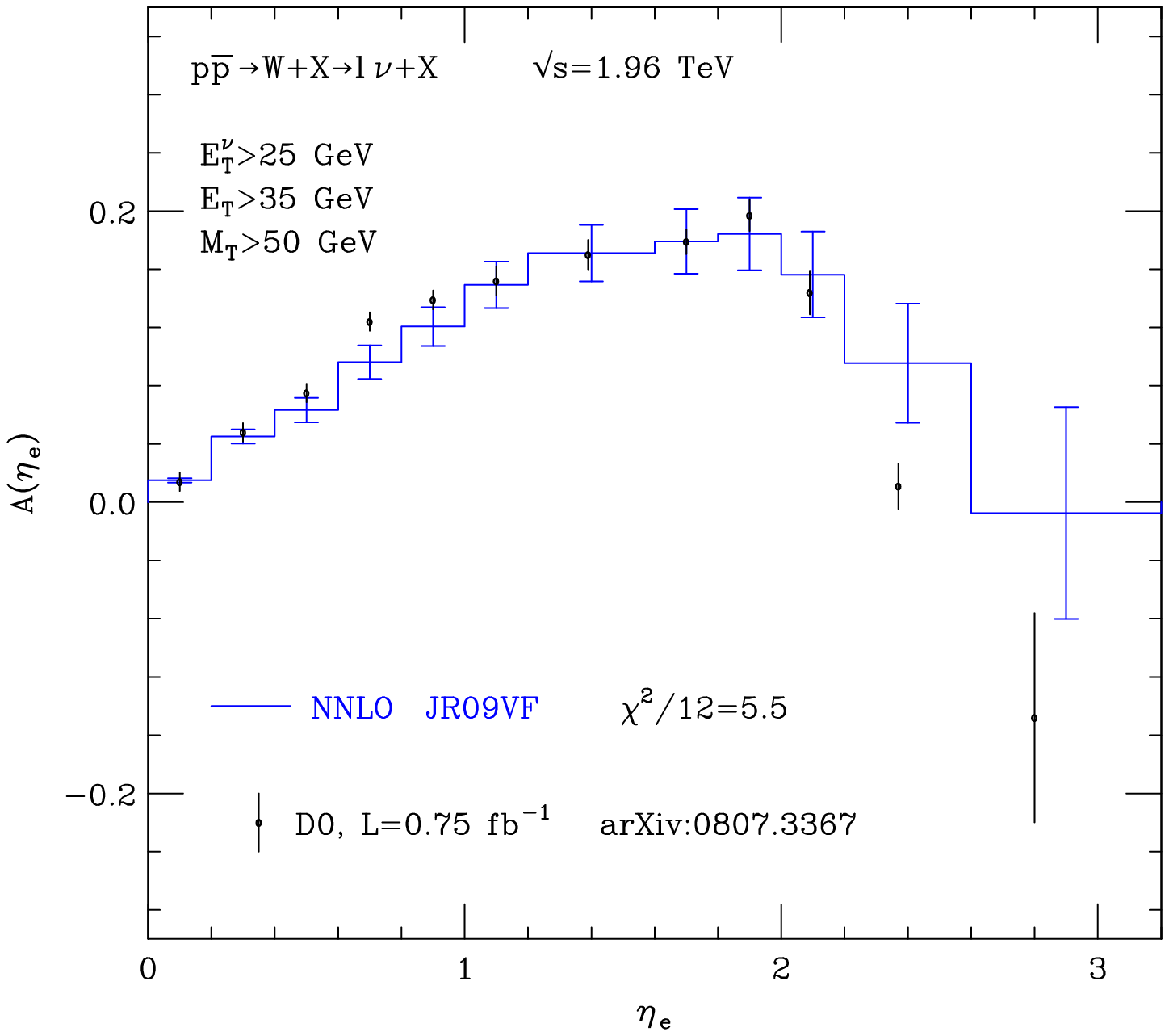}\\
\end{tabular}
\end{center}
\caption{\label{fig:asyD0re}
{\em Electron charge asymmetry at NNLO with the PDFs of 
Ref.~\cite{JimenezDelgado:2009tv}: $E_T>25$ GeV (left), $E_T>35$ GeV (right).}}
\end{figure}

In Fig.~\ref{fig:asyD0re} we present the NNLO electron asymmetry
computed with the JR09VF set of PDFs \cite{JimenezDelgado:2009tv}, 
and we include the corresponding PDF errors.
These PDF errors are larger
than the PDF errors of the MSTW2008 set. 
In the case of the
inclusive-$W$ asymmetry, the JR09VF result tends to undershoot the 
MSTW2008 result (see Figs.~\ref{fig:wasycdfaljr} and \ref{fig:wmstw})
in the rapidity region where $0.2 \ltap y_W \ltap 1.8$.
Comparing the electron asymmetry results in Fig.~\ref{fig:asyD0re} 
with those in Fig.~\ref{fig:asyD0err}, we see that the JR09VF result
undershoots the MSTW2008 result even at high rapidities.
In the region where $E_T>25$~GeV (Fig.~\ref{fig:asyD0re}, left), 
the JR09VF result is (relatively) consistent with 
the D0 data.
In the region where $E_T>35$~GeV (Fig.~\ref{fig:asyD0re}, right), 
the $\chi^2$ value of the MSTW2008 result is reduced by considering the
JR09VF parton densities; the main deviations of the JR09VF result from the
D0 data occur in the rapidity region around $\eta_e \sim 0.8$ and at high
rapidities ($\eta_e \gtap 2.2$).

\clearpage
\section{Rapidity cross section and asymmetry at the LHC}
\label{sec:lhc}

\setcounter{footnote}{1}

In this section we consider $pp$ collisions at LHC energies.
We recall the main features of $W^\pm$ (and~$l^{\pm}$) production, 
and we present results of our QCD calculations at the centre--of--mass energies
$\sqrt{s}=10$~TeV and $7$~TeV.

\subsection{W rapidity distribution and asymmetry}
\label{sec:wlhc}

\begin{figure}[htb]
\begin{center}
\begin{tabular}{c}
\epsfxsize=14truecm
\epsffile{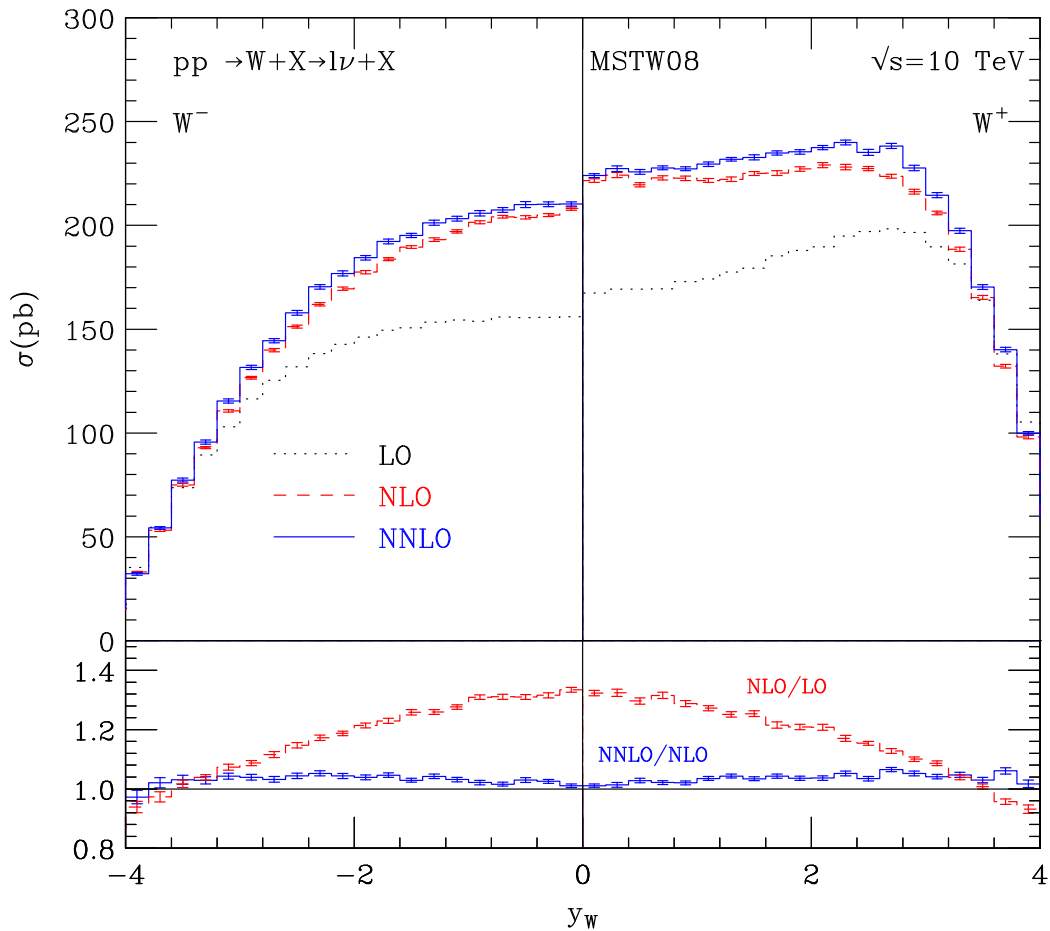}\\
\end{tabular}
\end{center}
\caption{\label{fig:etaWlhc}
{\em Rapidity distribution of an on-shell $W^-$ (left panel) and $W^+$ (right
panel) at the LHC ($\sqrt{s}=10$ TeV),
in LO (black dotted), NLO (red dashed) and NNLO (blue solid) QCD. 
No cuts are applied on the leptons and on their accompanying final state.
The lower panel shows the ratios NLO/LO (red dashed) and 
NNLO/NLO (blue solid) of the cross section results in the upper panel.
}}
\end{figure}

All the numerical results presented in this subsection refer to $W$ production 
at the centre--of--mass energy
$\sqrt{s}=10$~TeV.
In Fig.~\ref{fig:etaWlhc} we present the rapidity 
cross sections
of on-shell
$W^-$ (left panel) and $W^+$ (right panel) bosons.
The histograms in the upper panels
give the values of our LO, NLO and NNLO calculation 
of the cross sections $\sigma=\sigma(W^{\pm}) BR(W\to l{\nu})$ in each
rapidity bin.
Owing to CP invariance of the QCD cross section, both the $W^-$ 
and $W^+$ rapidity distributions are forward--backward symmetric
in $pp$ collisions, and thus only half of the rapidity range 
needs be shown.

The LO cross sections are controlled by the partonic suprocesses in 
Eqs.~(\ref{pro+}) and (\ref{pro-}). The $W^+$ boson is mainly produced
by $u \,{\bar d}$ collisions, while
the $W^-$ boson is mainly produced by $d \,{\bar u}$ collisions. 
The antiquark parton densities ${\bar d}(x)$ and ${\bar u}(x)$ 
of the proton are relatively similar, especially at small values
of the parton momentum fraction $x$ (at LHC energies the typical values of $x$
are smaller than at Tevatron energies). The $u$ quarks carry, on average,
more proton momentum fraction than $d$ quarks and, moreover $u(x)$ is larger
than $d(x)$. As a consequence, the $W^+$ cross section is larger than the 
$W^-$ cross section, and the $W^+$ tends to be produced 
at larger rapidities with respect to the $W^-$.
In Fig.~\ref{fig:etaWlhc} we see that
the $W^+$ and $W^-$ rapidity distributions are (smoothly) peaked 
at $|y_W|\sim 2.5$ and $y_W\sim 0$, respectively.

As in the case of $W$ production at the Tevatron,
having computed the rapidity cross section up to NNLO, we can examine the
quantitative convergence of the perturbative expansion. We
use the K factors defined in Eq.~(\ref{wkfators});
their values for $W^+$ and $W^-$ production at the LHC
are shown in the lower panels of Fig.~\ref{fig:etaWlhc}.
As already found at the Tevatron, the NLO effects are bigger than the NNLO effects on both 
the normalization and the shape of the rapidity cross section.
We note that the K factors for $W^+$ and $W^-$ production are quantitatively
very similar and they also have a very similar
dependence on $y_W$. 
In the rapidity interval $|y_W|\ltap 3$, the NLO K factors vary in the range
$K_{NLO}(y_W)\sim$~1.1--1.3, while the NNLO K factors vary in the range
$K_{NNLO}(y_W)\sim$~1.01--1.05.
At NLO the K factors computed from the ratio of the total 
(i.e. integrated over $y_W$) cross sections are
$K^{W^+}_{NLO}=1.17$ and $K^{W^-}_{NLO}=1.21$, respectively,
whereas at NNLO we find $K_{NNLO}=1.03$ for both $W^+$ and $W^-$
production. Our value of $K_{NNLO}$ is consistent with the
one given in Ref.~\cite{Martin:2009iq}.
We note that, at NLO, the K factors at the LHC are slightly smaller than those at
the Tevatron; at NNLO, the K factors are comparable.
The smallness of $K_{NNLO}-1$ indicates the good quality of the truncated
perturbative expansion.

\begin{figure}[htb]
\begin{center}
\begin{tabular}{c}
\epsfxsize=15truecm
\epsffile{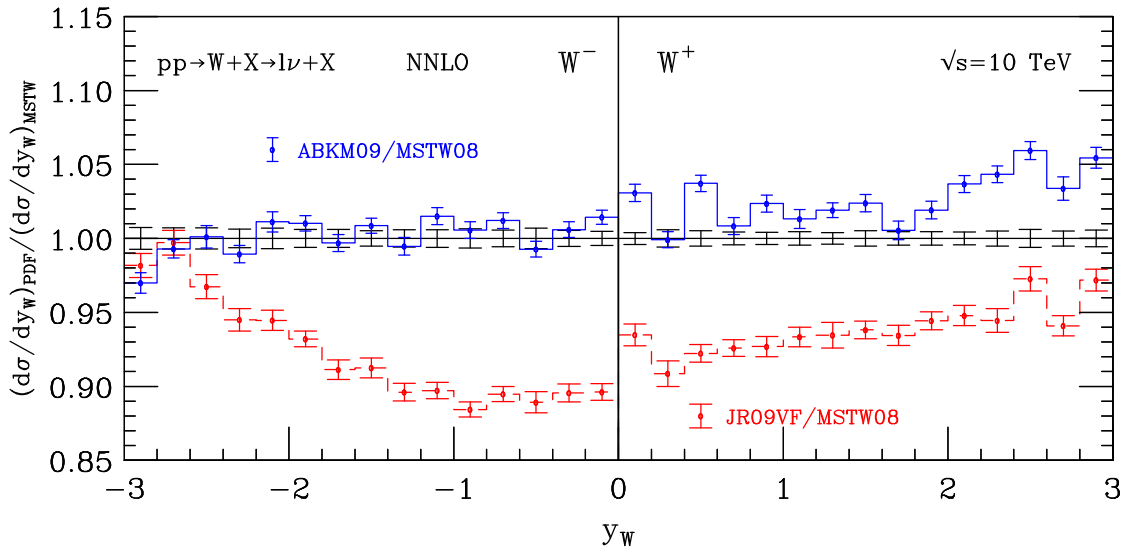}\\
\end{tabular}
\end{center}
\caption{\label{fig:KWlhc}
{\em
On-shell $W^-$ (left panel) and $W^+$ (right panel) production at the LHC ($\sqrt{s}=10$ TeV) in NNLO QCD.
The rapidity cross sections computed with the ABKM09 (blue solid)
and JR09VF (red dashed)
parton densities are rescaled by the corresponding MSTW2008 result.
}}
\end{figure}

As in Sect.~\ref{sec:wtev}, we have repeated our QCD calculation of 
the $W^+$ and $W^-$ rapidity
cross sections by using the PDFs of the
ABKM Collaboration
\cite{Alekhin:2005gq, Alekhin:2009ni} and of the Dortmund Group
\cite{GJRpdf, JimenezDelgado:2009tv}. 
In Fig.~\ref{fig:KWlhc} we show the NNLO
ratios $(d\sigma/dy_W)_{PDF}/(d\sigma/dy_W)_{MSTW}$, where
the cross section in the numerator is computed by using either the
ABKM09 set \cite{Alekhin:2009ni} or the 
JR09VF set \cite{JimenezDelgado:2009tv}, while the cross section in the 
denominator uses the MSTW2008 set.
In the case of $W^-$ production (left panel of Fig.~\ref{fig:KWlhc}) 
and considering the  rapidity region where $|y_W|\ltap 2$,
we see that the
ABKM09 result is consistent
with the MSTW result, within the numerical uncertainties of 
our NNLO computation;
the JR09VF result is instead smaller than the MSTW result, the difference
ranging between about 6 and 11\%.
In the case of $W^+$ production (right panel of Fig.~\ref{fig:KWlhc}),
the ABKM09 result is larger than the MSTW result, the difference ranging
from about 1 to 4\% in the region where $|y_W|\ltap 2$;
the JR09VF result is instead smaller than the MSTW result, the difference
ranging between about 6 and 9\%.
At fixed values of $y_W$, the typical values of parton momentum fractions
probed by the cross section ratios in Fig.~\ref{fig:KWlhc} are about a factor of
five (e.g., using (10~TeV)/(1.96~TeV)$\simeq 5.1$) smaller than those in
Fig.~\ref{fig:KW}; the differences between the ratios in Figs.~\ref{fig:KW} and 
\ref{fig:KWlhc} are due to the underlying differences between the 
ABKM09, MSTW2008 and JR09VF partons in different regions of parton momentum
fraction.
Considering the NNLO total cross sections at the LHC, in the case of
$W^+$ production, the ABKM09 (JR09VF) result  is about 4\% higher (lower) 
than the MSTW2008 result.
In the case of $W^-$ production, the ABKM09 and MSTW2008 results are 
comparable,
whereas the JR09VF result is smaller than the others by about 7\%.
For $W^+$ production, we find the values\footnote{The errors on the 
values of the cross sections
are those from the Monte Carlo integration in our calculation.}
$\sigma_{NNLO}^{W^+}=9.15 \pm 0.03$~nb,
$8.77\pm 0.03$~nb and
$8.38 \pm 0.03$~nb;
for $W^-$ production, we find 
$\sigma_{NNLO}^{W^-}=6.40\pm 0.02$~nb,
$6.40\pm 0.02$ and
$5.94\pm 0.02$.
These values are consistent with
the results of 
Refs.~\cite{Alekhin:2009ni}, 
\cite{Martin:2009iq}
and \cite{JimenezDelgado:2009tv}, once the leptonic branching ratios are taken
into account.

\begin{figure}[htb]
\begin{center}
\begin{tabular}{c}
\epsfxsize=14truecm
\epsffile{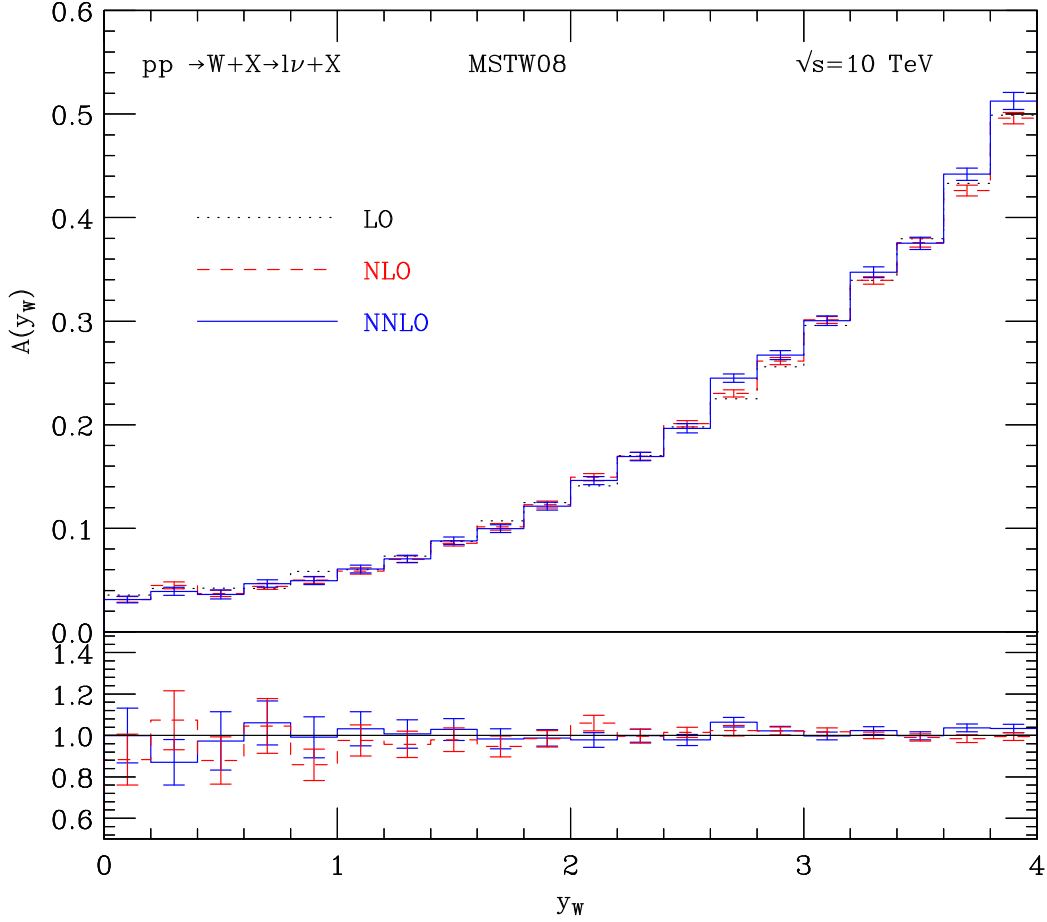}\\
\end{tabular}
\end{center}
\caption{\label{fig:wasymmetrylhc}
{\em The charge asymmetry for on-shell W production at the LHC ($\sqrt{s}=10$
TeV).
Upper panel: LO (dotted), NLO (red dashes), NNLO (blue solid) predictions.
The lower panel shows the ratios NLO/LO (red dashed) and 
NNLO/NLO (blue solid) of the results in the upper panel.
}}
\end{figure}

We now consider the $W$ charge asymmetry. 
In Fig.~\ref{fig:wasymmetrylhc} we
present our results for the asymmetry of on-shell $W$ bosons.
The charge asymmetry is computed up to NNLO in QCD
perturbation theory.
From the histograms in the upper panel,
we see 
that the perturbative result for the $W$ asymmetry at the LHC
is very stable against the effect of QCD radiative corrections.
This is confirmed by the lower panel in the figure, where we show the 
asymmetry K factors at NLO and NNLO (they are computed analogously 
to Eq.~(\ref{wkfators})). The values of the K factors are consistent 
with unity within the numerical uncertainties of the Monte Carlo computation
that we have carried out.

Since the $W^+$ and $W^-$ cross
sections are different, the asymmetry does not vanish at $y_W=0$ where its value
is about 0.04.
As $y_W$ increases, the asymmetry increases and reaches the values
$A(y_W)\sim 0.08$ at $y_W \sim 1.5$, $A(y_W)\sim 0.3$ at $y_W \sim 3$
and $A(y_W)\sim 0.5$ at $y_W\sim 4$.
It is interesting to compare this result with the corresponding
result for the $W$ asymmetry at the Tevatron (see Fig.~\ref{fig:asyW-nocuts}).
At $y_W=0$ the Tevatron asymmetry is constrained to vanish by CP invariance.
However, as $y_W$ increases the Tevatron asymmetry increases much faster than
the LHC asymmetry. The $W$ asymmetry at the Tevatron reaches the values 
$A(y_W)\sim 0.08$ in the region around $y_W \sim 0.5$, and 
$A(y_W)\sim 0.3$ in the region around $y_W \sim 1.5$; the value
$A(y_W)\sim 0.5$ is already reached at $y_W \ltap 2.5$.
This different behaviour of the $W$ asymmetry at the Tevatron and at the LHC
is essentially due to the sensitivity to parton densities with different
values of parton momentum fraction. The increase of the
$W$ charge asymmetry at large rapidities is mainly driven by the difference
(actually, by the ratio)
of the parton densities, $u(x)$ and $d(x)$, of $u$ and $d$ quarks.
Since $u(x)$ and $d(x)$ tend to become similar as their momentum fraction 
$x$ decreases, at large rapidities the $W$ charge asymmetry tends to become
smaller as $\sqrt s$ increases (e.g. going from the Tevatron to the LHC).

The main features of $W$ production at the LHC depend on the PDFs of the 
proton. Increasing the centre--of--mass energy of the colliding protons,
$W$ production probes smaller values of parton momentum fractions $x$.
At smaller values of $x$, all the parton densities (expecially the gluon density)
are larger, and all the quark and antiquark (independently of their flavour)
parton densities tend to become similar. Therefore, the main features
of $W$ production at ${\sqrt s}=10$~TeV (which we have illustrated in this
subsection) are intermediate between those at ${\sqrt s}=7$~TeV and 
those at ${\sqrt s}=14$~TeV.
Increasing ${\sqrt s}$, both the $W^+$ and $W^-$ cross sections increase;
the difference between the absolute values of the $W^+$ and $W^-$
rapidity cross sections, at fixed $y_W$, is reduced; the shapes of 
the $W^+$ and $W^-$rapidity distributions become more similar
(e.g., the peak of the $W^+$ distribution moves toward $y_W \sim 0$);
the $W$ charge asymmetry, at fixed $y_W$, decreases.

\subsection{Charged lepton rapidity distribution and asymmetry}

In this subsection we study
the rapidity distributions of the charged lepton from $W$
decay. 
Owing to EW dynamical correlations (see Eq.~(\ref{stardist}) and the discussion
at the beginning of Sect.~\ref{sec:ltev}),
at the partonic level 
the charged
lepton tends to be produced in the direction of the colliding `down-type' 
quark or antiquark (or, equivalently, in the opposite direction with respect to
the `up-type' quark or antiquark).
As discussed in Sect.~\ref{sec:wlhc}, in $pp$ collisions, the shape 
of the rapidity distribution of the $W^+$ boson is mainly controlled by 
$u \,{\bar d}$ collisions, and the peak of the distribution is due to the larger
momentum fraction carried by $u$ quarks.
Therefore, the parton level EW correlations 
shift the rapidity distribution of the positively-charged lepton 
to be more central than the distribution of the parent $W^+$.
On the contrary, the $W^-$ is mostly produced at central rapidities 
by $d \,{\bar u}$ collisions, and, therefore the EW
correlations shift the rapidity distribution
of the negatively-charged lepton at higher rapidities.
We thus anticipate that, in comparison with the $W$ charge asymmetry,
the lepton charge asymmetry is larger at low rapidities and smaller at high
rapidities. Eventually (as in $p{\bar p}$ collisions at the Tevatron),
at large values of the rapidity the partonic EW asymmetry dominates over the
asymmetry produced by the PDFs, and the lepton charge asymmetry becomes negative.

To the purpose of presenting some quantitative, though
illustrative, results on
the lepton rapidity distributions and asymmetry at the LHC, we refer to
the framework considered in a recent 'Physics Analysis' \cite{CMSnote}
of the CMS Collaboration. The study of Ref.~\cite{CMSnote}, based on data sets
from Monte Carlo simulations, regards the muon rapidity cross sections and 
charge asymmetry that can be measured with the CMS detector at the LHC.

In our  calculations, we use the lepton selection cuts applied in 
Ref.~\cite{CMSnote}. The transverse mass 
and the missing transverse energy are required  
to be $M_T>50$~GeV and  $E_T^\nu>20$~GeV, respectively.
The muons are isolated: the hadronic (partonic) transverse energy 
$E_T^{\rm iso}$ in a cone of radius $R=0.3$ is required to fulfil the 
constraints\footnote{We note that the two constraints also imply
a lower bound on the lepton transverse energy, namely
$E_T>E_{T}^{\rm max}(1-z)=23.75$~GeV.} 
$E_T^{\rm iso}/E_T < z/(1-z)$ with $z=0.05$, and 
$E_T+E_T^{\rm iso}>E_T^{\rm max}=25$~GeV.

\begin{figure}[htb]
\begin{center}
\begin{tabular}{c}
\epsfxsize=14truecm
\epsffile{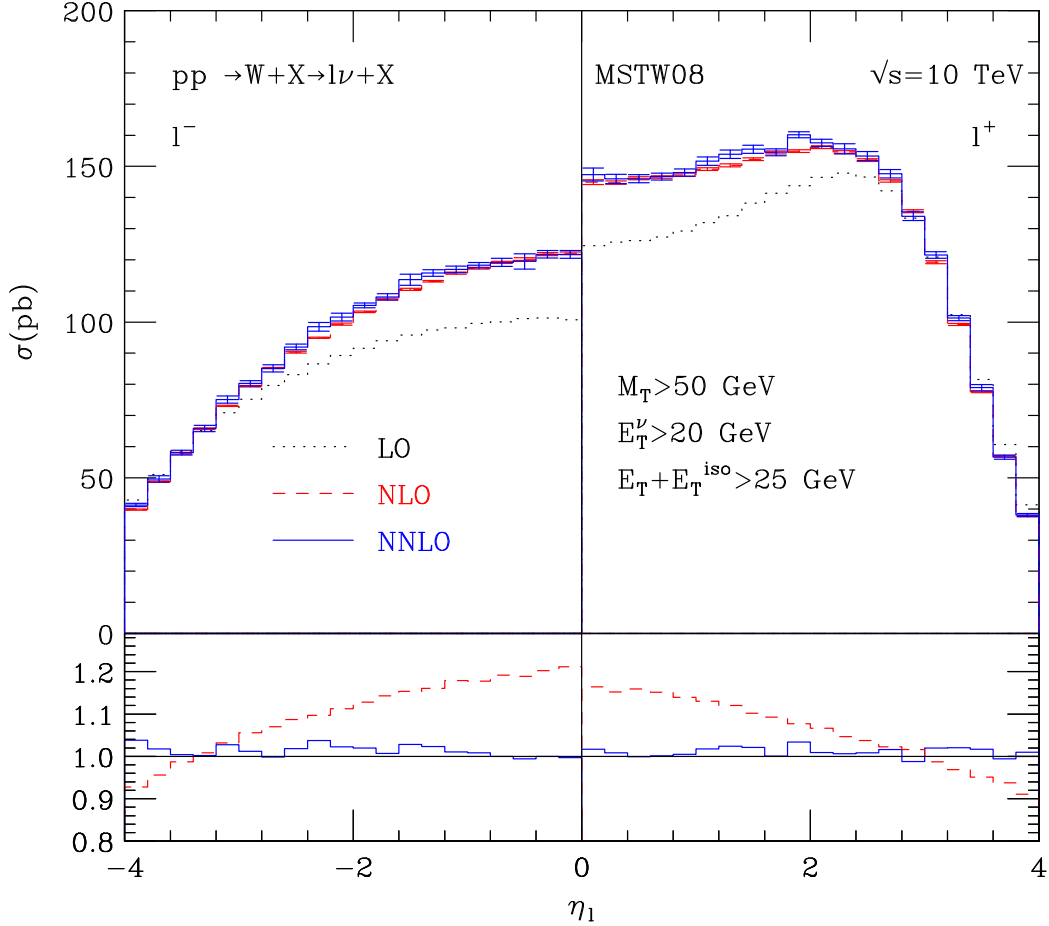}\\
\end{tabular}
\end{center}
\caption{\label{fig:etaplusminuscms}
{\em Rapidity distribution of the $l^-$ (left panel) and $l^+$ (right panel) at the LHC ($\sqrt{s}=10$ TeV),
in LO (black dotted), NLO (red dashed) and NNLO (blue solid) QCD. 
The cuts applied are described in the text.
The lower panel shows the ratios NLO/LO (red dashed) and 
NNLO/NLO (blue solid) of the cross section results in the upper panel.
}}
\end{figure}

The rapidity distributions of the charged leptons at the centre--of-mass energy
${\sqrt s}=10$~TeV are shown in Fig.~\ref{fig:etaplusminuscms}. We present the
results of our QCD calculation up to NNLO.
As in the case of Fig.~\ref{fig:etaWlhc}, we show only
half of the rapidity range, since
the lepton rapidity distributions are invariant under the
replacement $\eta_l \leftrightarrow -\eta_l$.
The qualitative expectations (discussed at the beginning of this subsection)
about the differences between the $W$ and lepton rapidity distributions are
confirmed by the comparison of Fig.~\ref{fig:etaWlhc} with
Fig.~\ref{fig:etaplusminuscms}.
The rapidity distribution of
the $l^+$ is peaked in the rapidity region around $\eta_l\sim 2$, 
and the rapidity distribution
of the $l^-$, though still peaked at $\eta_l\sim 0$, is broader and flatter
than the $W^-$ rapidity distribution.
Moreover, the difference of the rapidity cross sections in the region around
$\eta_l\sim 0$ is larger in Fig.~\ref{fig:etaplusminuscms} than in
Fig.~\ref{fig:etaWlhc}.

We comment on the impact of the QCD radiative corrections. 
The rapidity dependent K factors, defined as
in Eq.~(\ref{wkfators}), are shown in the lower panels 
of Fig.~\ref{fig:etaplusminuscms}.
In the rapidity region where $|\eta_l| \ltap 3$, the NLO K factor of the $l^+$
($l^-$) rapidity cross section
decreases from about 1.16 (1.21) to about 1.01 (1.03). 
The NNLO K factor is very close to unity for both the $l^+$ and $l^-$
rapidity cross sections.
We recall that an equal (rapidity dependent) K factor for $l^+$ and $l^-$ 
production
would imply no radiative
correction to the charge asymmetry.
Since the NLO K factor for $l^-$ production
is slightly larger
than the other, we anticipate a reduction of the lepton asymmetry in going
from LO to NLO. Owing to the similarity of the NNLO K factor, we also 
anticipate very little NNLO effects on the lepton asymmetry.

\begin{figure}[htb]
\begin{center}
\begin{tabular}{c}
\epsfxsize=11truecm
\epsffile{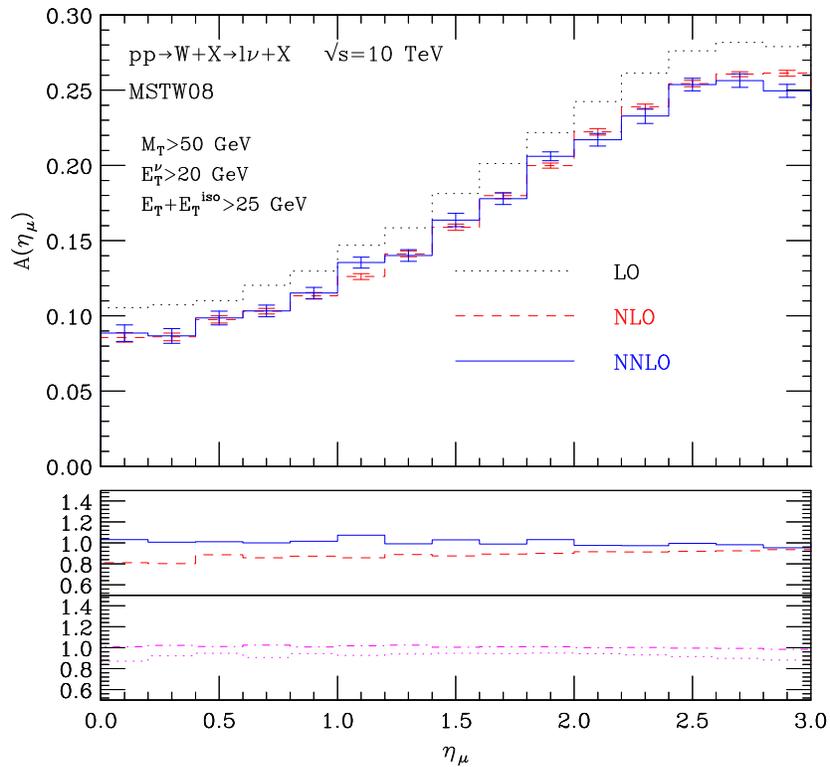}\\
\end{tabular}
\end{center}
\caption{\label{fig:asyCMS}
{\em Muon charge asymmetry at the LHC: $\sqrt{s}=10$ TeV.}}
\end{figure}

The results for the lepton charge asymmetry
are presented in Fig.~\ref{fig:asyCMS}.
The three histograms in the upper panel show the LO (dotted), NLO (dashed) 
and NNLO (solid) results of our calculation. Comparing Fig.~\ref{fig:asyCMS}
with Fig.~\ref{fig:wasymmetrylhc} we see that, as expected, 
the lepton asymmetry is larger (slightly smaller) than the $W$ asymmetry 
at small (large, i.e. $\eta_l \sim 2.5$) rapidities.
Considering the effect of the QCD radiative corrections, we see that, 
in going
from LO to NLO, the QCD corrections tend to decrease the asymmetry, as expected
from the results in Fig.~\ref{fig:etaplusminuscms}. 
The NNLO corrections do not change
the asymmetry in a significant way.
The impact
of the radiative corrections is quantified by the NLO 
(dashed histogram) and NNLO (solid histogram) asymmetry K factors, 
which are shown in
the lower panel of Fig.~\ref{fig:asyCMS}. In the panel we also show
the K factors
computed with the corresponding NLO (dotted) and NNLO (dot-dashed) PDFs by
using the LO partonic cross sections.
The NLO and NNLO K factors vary in the range 
$K_{NLO}(\eta_l) \sim$~0.80--0.94 and $K_{NNLO}(\eta_l) \sim$~0.96--1.07.
Comparing the NLO K factor (dashed histogram)
with the `PDF driven' NLO K factor (dotted histogram), we see that
the variation of the PDFs (in going from the LO to the NLO set)
produce a sizeable contribution to the radiative corrections.
However, a careful inspection of the differences between the
dashed and dotted histograms shows that the
NLO corrections to the partonic cross sections are not negligible.
For instance, the value of the `PDF driven' NLO K factor is 
$K_{NLO}(\eta_l) \sim$~0.88 at both $\eta_l \simeq 0$ and $\eta_l \simeq 3$.
This implies that the NLO corrections to the partonic cross sections
are negative (i.e. they decrease the lepton asymmetry) at small rapidities 
and positive at large rapidities ($\eta_l \sim 2.5$). Therefore, these
corrections tend to slightly reduce the differences between the lepton asymmetry
and the asymmetry of the parent $W$. This effect is qualitatively consistent 
with the analogous effect observed at the Tevatron, for instance, in the
low-$E_T$ bin of the CDF electron asymmetry (see the results in the left panels
of Fig.~\ref{fig:asycdf} and the related comments).

We have repeated our calculations of $W$ and lepton rapidity cross sections and
asymmetries at the centre--of--mass energy $\sqrt{s}=7$~TeV.
The overall features of our results do not change significantly, apart from
the quantitative differences that are expected from the decreased value of
$\sqrt{s}$.
Here we limit ourselves to presenting the results for the lepton charge
asymmetry, considering the same lepton kinematical configuration \cite{CMSnote}
as in Fig.~\ref{fig:asyCMS}. 

\begin{figure}[htb]
\begin{center}
\begin{tabular}{c}
\epsfxsize=11truecm
\epsffile{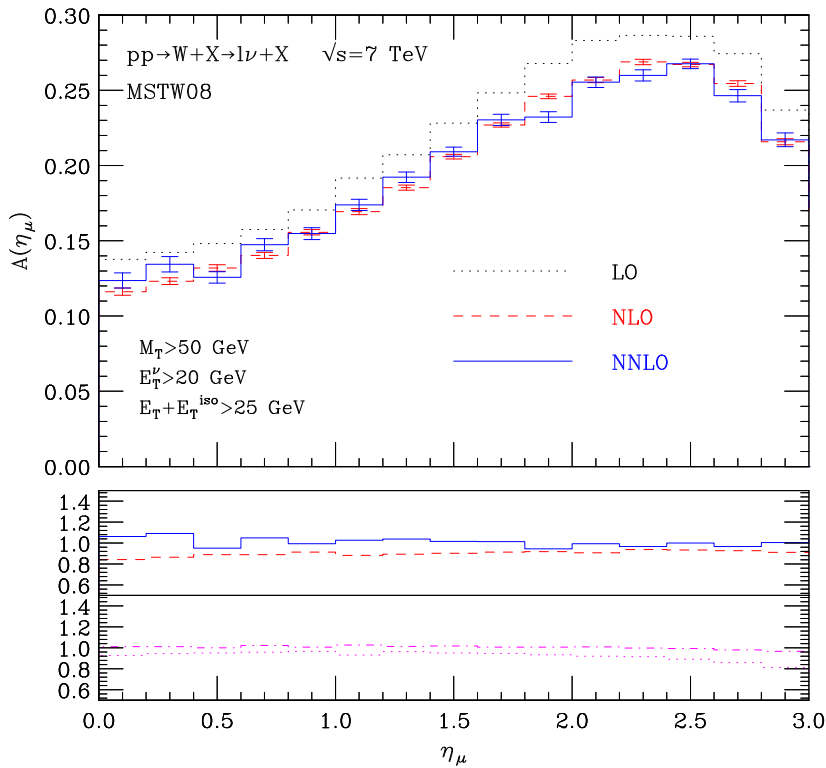}\\
\end{tabular}
\end{center}
\caption{\label{fig:asyCMS_7TeV.ps}
{\em Muon charge asymmetry at the LHC: $\sqrt{s}=7$ TeV.}}
\end{figure}

The lepton asymmetry at $\sqrt{s}=7$~TeV
is shown in Fig.~\ref{fig:asyCMS_7TeV.ps}.
We first note that, at low and medium rapidities ($\eta_l \ltap 2.5$),
the asymmetry is larger at $\sqrt{s}=7$~TeV than
at $\sqrt{s}=10$~TeV. This is not unexpected, since at smaller energies
larger values of parton momentum fraction $x$ are probed, 
where flavour asymmetries of the PDFs are more sizeable.
At high rapidities ($\eta_l \gtap 2.5$), the shape of the asymmetry changes:
$A(\eta_l)$ decreases as $\eta_l$ increases. This behaviour (as we have already
seen at the Tevatron and commented about in Sect.~\ref{sec:ltev}) is a
distinctive effect of the impact of the parton level EW asymmetry on the 
flavour asymmetries of the PDFs. The effect takes place when the
flavour asymmetries are more sizeable, i.e. when larger parton momentum fractions
are probed. Therefore, at fixed $\sqrt s$, the effect sets in in the
high-rapidity region; correspondingly, at higher $\sqrt s$, the effect sets in
at higher rapidities. Indeed, at $\sqrt s= 10$~TeV, the decreasing behaviour
of the lepton asymmetry starts to be visible 
at $\eta_l \sim 3$ (Fig.~\ref{fig:asyCMS}).

As for the impact of the QCD radiative corrections on the lepton charge asymmetry
at $\sqrt{s}=7$~TeV, the effects are very similar to those 
in Fig.~\ref{fig:asyCMS}:
the NLO corrections tend to decrease the asymmetry,
and the NNLO corrections do not significantly change the NLO result. 
Analogously to Fig.~\ref{fig:asyCMS}, in the lower panel of 
Fig.~\ref{fig:asyCMS_7TeV.ps} we present the asymmetry K factors 
at NLO and NNLO,
and the corresponding `PDF driven' K factors. The similarity between the 
K factors and their `PDF driven' versions shows that a sizeable part of the
radiative corrections to the asymmetry is due to the variation of the PDFs
(in particular, going from the LO to the NLO set).
The contribution of the radiative corrections to the partonic cross sections
is nonetheless clearly visible. Comparing the NLO K factor (dashed histogram)
with its PDF driven' version (dotted histogram), we can see that the NLO
corrections to the partonic cross sections are negative at small rapidities
and positive at large rapidities. The same effect has been observed in the lepton
asymmetry at $\sqrt{s}=10$~TeV (see the lower panel in Fig.~\ref{fig:asyCMS}
and related comments).

\section{Summary}
\label{sec:sum}

In this paper we have considered $W$ boson production at hadron colliders,
and the subsequent leptonic decay $W \to l \nu_l$. We have performed QCD studies
of the boson and lepton rapidity distributions, by including the contributions of
higher-order radiative corrections. The main part of our work regards the study
of the lepton charge asymmetry. We have presented the results of the calculation 
of the lepton charge asymmetry up to NNLO in QCD perturbation theory.
Our calculation is carried out by
using the parton level Monte Carlo program
of Ref.~\cite{Catani:2009sm}, which allows us 
to take into account the 
lepton kinematical cuts that are typically applied to perform measurements 
of the charge asymmetry in experiments at hadron colliders.

The lepton charge asymmetry is the result of two different physical mechanisms:
the asymmetry generated by the parton content of the colliding protons and/or
antiprotons and the asymmetry of the partonic (hard-scattering) production
mechanism. The effect of the latter is almost completely cancelled
by calculations or measurements that are highly inclusive over the kinematics of
the leptons from $W$ decay. Therefore, 
the inclusive $W$ asymmetry is mostly sensitive
to the PDFs of the colliding hadrons, and, in particular, 
to the differences between the momentum fraction distributions of $u$ and $d$ 
quarks in the proton.
What is primarily and directly measured is, however, the lepton charge asymmetry. 
The lepton charge asymmetry has an additional sensitivity to 
the underlying partonic scattering, whose asymmetry mostly originates from  
EW dynamics. Moreover, the study the lepton
asymmetry at different values of the lepton transverse energy gives a finer probe
of the momentum fraction dependence of the PDFs of the colliding hadrons.

In the paper we have first considered the case of $p{\bar p}$ collisions at the 
Tevatron Run II. We have 
examined
the charge asymmetry of 
inclusive $W$ production and confirmed earlier findings 
\cite{Anastasiou:2003ds}: the QCD radiative corrections to this quantity
are very small. We have also shown the $W$ asymmetry results obtained by using
different sets of NNLO PDFs. Then, we have
considered the lepton charge asymmetry and computed this observable in
kinematical configurations with various lepton selection cuts, as those
applied by the CDF and D0 Collaborations. We have shown that the QCD radiative
corrections to the lepton asymmetry are small, but they are not as small 
as those to the $W$ asymmetry. This difference is not fully unexpected.
Since the lepton asymmetry also depends on the EW asymmetry generated by the
partonic subprocesses, it is more sensitive to the QCD radiative corrections
to these partonic subprocesses. Moreover, the effect of the QCD corrections
to the lepton charge asymmetry has an impact that
qualitatively and quantitatively depends on the lepton kinematics and the 
specific lepton selection cuts. 
We have performed NNLO calculations of the lepton charge asymmetry by
using different sets of PDFs and presented the comparison with some
of the available Tevatron data. 

We have finally considered the charge asymmetry in $pp$ collisions.
At LHC energies, the lepton charge asymmetry is sensitive to PDFs with
momentum fractions that are smaller than those probed at the Tevatron.
We have presented some illustrative results at the centre--of--mass energies
$\sqrt{s}=10$~TeV and 7~TeV. 
These specific results shown that (as in $p{\bar p}$ collisions at the Tevatron)
the QCD radiative corrections to the lepton charge asymmetry are small,
particularly at NNLO,
though larger than those to the $W$ charge asymmetry.

\noindent {\bf Acknowledgements.}
We would like to thank James Stirling for helpful discussions and comments
on the manuscript. We also thank Gregorio Bernardi, Giorgio Chiarelli, 
Cigdem Issever, Sandra Leone and Junjie Zhu for useful correspondence.


\begin{thebibliography}{99}

\bibitem{Drell:1970wh}
  S.~D.~Drell and T.~M.~Yan,
  Phys.\ Rev.\ Lett.\  {\bf 25} (1970) 316
  [Erratum-ibid.\  {\bf 25} (1970) 902].



\bibitem{Hamberg:1990np}
  R.~Hamberg, W.~L.~van Neerven and T.~Matsuura,
  Nucl.\ Phys.\  B {\bf 359} (1991) 343
  [Erratum-ibid.\  B {\bf 644} (2002) 403];
  R.~V.~Harlander and W.~B.~Kilgore,
  Phys.\ Rev.\ Lett.\  {\bf 88} (2002) 201801.

\bibitem{Anastasiou:2003yy}
  C.~Anastasiou, L.~J.~Dixon, K.~Melnikov and F.~Petriello,
  Phys.\ Rev.\ Lett.\  {\bf 91} (2003) 182002.

\bibitem{Anastasiou:2003ds}
C.~Anastasiou, L.~J.~Dixon, K.~Melnikov and F.~Petriello,
Phys.\ Rev.\ D {\bf 69} (2004) 094008.


\bibitem{Melnikov:2006di}
K.~Melnikov and F.~Petriello,
Phys.\ Rev.\ Lett.\  {\bf 96} (2006) 231803,
Phys.\ Rev.\  D {\bf 74} (2006) 114017.



\bibitem{Catani:2009sm}
  S.~Catani, L.~Cieri, G.~Ferrera, D.~de Florian and M.~Grazzini,
  Phys.\ Rev.\ Lett.\  {\bf 103} (2009) 082001.




\bibitem{Kosower}
  D.~A.~Kosower,
  Phys.\ Rev.\ D {\bf 57} (1998) 5410,
Phys.\ Rev.\ D {\bf 67} (2003) 116003,
Phys.\ Rev.\ D {\bf 71} (2005) 045016.

\bibitem{Weinzierl}
S.~Weinzierl,
JHEP {\bf 0303} (2003) 062,
JHEP {\bf 0307} (2003) 052,
  Phys.\ Rev.\  D {\bf 74} (2006) 014020.

\bibitem{GGG}
  A.~Gehrmann-De Ridder, T.~Gehrmann and E.~W.~N.~Glover,
  Nucl.\ Phys.\  B {\bf 691} (2004) 195,
Phys.\ Lett.\ B {\bf 612} (2005) 36,
Phys.\ Lett.\ B {\bf 612} (2005) 49,
JHEP {\bf 0509} (2005) 056;
A.~Daleo, T.~Gehrmann and D.~Maitre,
JHEP {\bf 0704} (2007) 016.


\bibitem{Frixione:2004is}
S.~Frixione and M.~Grazzini,
JHEP {\bf 0506} (2005) 010.

\bibitem{ST}
  G.~Somogyi, Z.~Trocsanyi and V.~Del Duca,
  JHEP {\bf 0506} (2005) 024,
JHEP {\bf 0701} (2007) 070;
  G.~Somogyi and Z.~Trocsanyi,
  JHEP {\bf 0701} (2007) 052,
  JHEP {\bf 0808} (2008) 042;
  U.~Aglietti, V.~Del Duca, C.~Duhr, G.~Somogyi and Z.~Trocsanyi,
  JHEP {\bf 0809} (2008) 107.




\bibitem{Anastasiou:2004qd}
  C.~Anastasiou, K.~Melnikov and F.~Petriello,
  Phys.\ Rev.\ Lett.\  {\bf 93} (2004) 032002.



\bibitem{Weinzierl:2006ij}
  S.~Weinzierl,
  Phys.\ Rev.\  D {\bf 74} (2006) 014020.


\bibitem{threejets}
  A.~Gehrmann-De Ridder, T.~Gehrmann, E.~W.~N.~Glover and G.~Heinrich,
  Phys.\ Rev.\ Lett.\  {\bf 99} (2007) 132002,
  JHEP {\bf 0711} (2007) 058,
JHEP {\bf 0712} (2007) 094,
Phys.\ Rev.\ Lett.\  {\bf 100} (2008) 172001;

\bibitem{Weinzierl:2008iv}
  S.~Weinzierl,
  Phys.\ Rev.\ Lett.\  {\bf 101} (2008) 162001.



\bibitem{Anastasiou:2004xq}
  C.~Anastasiou, K.~Melnikov and F.~Petriello,
  Phys.\ Rev.\ Lett.\  {\bf 93} (2004) 262002,
Nucl.\ Phys.\ B {\bf 724} (2005) 197;
C.~Anastasiou, G.~Dissertori and F.~Stockli,
JHEP {\bf 0709} (2007) 018.


\bibitem{Catani:2007vq}
  S.~Catani and M.~Grazzini,
  Phys.\ Rev.\ Lett.\  {\bf 98} (2007) 222002;
  M.~Grazzini,
  JHEP {\bf 0802} (2008) 043.





\bibitem{earlyth}
  E.~L.~Berger, F.~Halzen, C.~S.~Kim and S.~Willenbrock,
  Phys.\ Rev.\  D {\bf 40} (1989) 83
  [Erratum-ibid.\  D {\bf 40} (1989) 3789];
  A.~D.~Martin, R.~G.~Roberts and W.~J.~Stirling,
  Mod.\ Phys.\ Lett.\  A {\bf 4} (1989) 1135.



\bibitem{cdfw}
  T.~Aaltonen {\it et al.}  [CDF Collaboration],
  Phys.\ Rev.\ Lett.\  {\bf 102} (2009) 181801.


\bibitem{cdfpre}  
  F.~Abe {\it et al.}  [CDF Collaboration],
  Phys.\ Rev.\ Lett.\  {\bf 68} (1992) 1458,
  Phys.\ Rev.\ Lett.\  {\bf 74} (1995) 850.

\bibitem{cdfrunI}
  F.~Abe {\it et al.}  [CDF Collaboration],
  Phys.\ Rev.\ Lett.\  {\bf 81} (1998) 5754.


\bibitem{cdfe}
  D.~E.~Acosta {\it et al.}  [CDF Collaboration],
  Phys.\ Rev.\  D {\bf 71} (2005) 051104.

\bibitem{d0m}
  V.~M.~Abazov {\it et al.}  [D0 Collaboration],
  Phys.\ Rev.\  D {\bf 77} (2008) 011106.

\bibitem{d0e}
  V.~M.~Abazov {\it et al.}  [D0 Collaboration],
  Phys.\ Rev.\ Lett.\  {\bf 101} (2008) 211801.




\bibitem{Martin:1998sq}
  A.~D.~Martin, R.~G.~Roberts, W.~J.~Stirling and R.~S.~Thorne,
  Eur.\ Phys.\ J.\  C {\bf 4} (1998) 463.


\bibitem{Lai:1999wy}
  H.~L.~Lai {\it et al.}  [CTEQ Collaboration],
  Eur.\ Phys.\ J.\  C {\bf 12} (2000) 375.


\bibitem{Giele:1993dj}
  W.~T.~Giele, E.~W.~N.~Glover and D.~A.~Kosower,
  Nucl.\ Phys.\  B {\bf 403}, 633 (1993).



\bibitem{Balazs:1997xd}
  C.~Balazs and C.~P.~Yuan,
  Phys.\ Rev.\  D {\bf 56} (1997) 5558.






\bibitem{Martin:2009iq}
  A.~D.~Martin, W.~J.~Stirling, R.~S.~Thorne and G.~Watt,
  Eur.\ Phys.\ J.\  C {\bf 63} (2009) 189.

\bibitem{Martin:2009bu}
  A.~D.~Martin, W.~J.~Stirling, R.~S.~Thorne and G.~Watt,
  report IPPP/09/33 (arXiv:0905.3531 [hep-ph]).

\bibitem{CMSnote}
CMS Collaboration, report CMS PAS Note EWK-09-003 (2009).

\bibitem{cteqpdf}
  P.~M.~Nadolsky {\it et al.},
  Phys.\ Rev.\  D {\bf 78} (2008) 013004.


\bibitem{nnpdf}
L.~Del Debbio, S.~Forte, J.~I.~Latorre, A.~Piccione and J.~Rojo  [NNPDF Collaboration],
  JHEP {\bf 0703} (2007) 039;
R.~D.~Ball {\it et al.}  [NNPDF Collaboration],
Nucl.\ Phys.\  B {\bf 809} (2009) 1
[Erratum-ibid.\  B {\bf 816} (2009) 293],
Nucl.\ Phys.\  B {\bf 823} (2009) 195.




\bibitem{Alekhin:2005gq}
  S.~Alekhin,
  JETP Lett.\  {\bf 82} (2005) 628
  [Pisma Zh.\ Eksp.\ Teor.\ Fiz.\  {\bf 82} (2005) 710].


\bibitem{Alekhin:2009ni}
  S.~Alekhin, J.~Blumlein, S.~Klein and S.~Moch,
  report DESY 09-102 (arXiv:0908.2766 [hep-ph]).



\bibitem{GJRpdf}
  M.~Gluck, P.~Jimenez-Delgado and E.~Reya,
  Eur.\ Phys.\ J.\  C {\bf 53} (2008) 355;
  M.~Gluck, P.~Jimenez-Delgado, E.~Reya and C.~Schuck,
  Phys.\ Lett.\  B {\bf 664} (2008) 133;
  P.~Jimenez-Delgado and E.~Reya,
  Phys.\ Rev.\  D {\bf 79} (2009) 074023.

\bibitem{JimenezDelgado:2009tv}
  P.~Jimenez-Delgado and E.~Reya,
  report DOTH-0913 (arXiv:0909.1711 [hep-ph]).

\bibitem{herapdf}
  S.~Chekanov {\it et al.}  [ZEUS Collaboration],
  Eur.\ Phys.\ J.\  C {\bf 42} (2005) 1;
  F.~D.~Aaron {\it et al.}  [H1 Collaboration],
  report DESY 09-005 (arXiv:0904.3513 [hep-ex]);
  H1 and ZEUS Collaborations,
  report DESY 09-158 (arXiv:0911.0884 [hep-ex]).

\bibitem{Amsler:2008zzb}
  C.~Amsler {\it et al.}  [Particle Data Group ],
  Phys.\ Lett.\  B {\bf 667} (2008) 1.


\bibitem{Bodek:2007cz}
  A.~Bodek, Y.~Chung, B.~Y.~Han, K.~S.~McFarland and E.~Halkiadakis,
  Phys.\ Rev.\  D {\bf 77} (2008) 111301.

\bibitem{mcnlo}
  S.~Frixione and B.~R.~Webber,
  JHEP {\bf 0206} (2002) 029;
  S.~Frixione, P.~Nason and B.~R.~Webber,
  JHEP {\bf 0308} (2003) 007.

\bibitem{MRST2006}
  A.~D.~Martin, W.~J.~Stirling, R.~S.~Thorne and G.~Watt,
  Phys.\ Lett.\  B {\bf 652} (2007) 292.

\bibitem{CTEQ6.1}
  D.~Stump, J.~Huston, J.~Pumplin, W.~K.~Tung, H.~L.~Lai, S.~Kuhlmann 
  and J.~F.~Owens,
  JHEP {\bf 0310} (2003) 046.

\bibitem{james}
W.~J.~Stirling, private communication.



\bibitem{book}
R.K.~Ellis, J.~Stirling and B.R.~Webber, {\em QCD and Collider
Physics},
Cambridge University Press, Cambridge, 1996.





\end{thebibliography}
\end{document}